\documentclass[prb,aps,twocolumn,floatfix,showpacs]{revtex4}
\usepackage{graphics}
\begin{document}
\title{Nonzero orbital angular momentum superfluidity in ultracold Fermi gases}
\author{M. Iskin and C. A. R. S{\'a} de Melo}
\affiliation{School of Physics, Georgia Institute of Technology, Atlanta, Georgia 30332, USA}
\date{\today}

\begin{abstract}
We analyze the evolution of superfluidity for nonzero orbital angular momentum channels
from the Bardeen-Cooper-Schrieffer (BCS) to the Bose-Einstein condensation (BEC) limit
in three dimensions.
First, we analyze the low energy scattering properties of finite range interactions
for all possible angular momentum channels.
Second, we discuss ground state ($T = 0$) superfluid properties including
the order parameter, chemical potential, quasiparticle excitation spectrum,
momentum distribution, atomic compressibility, ground state energy and low energy
collective excitations.
We show that a quantum phase transition occurs for nonzero angular momentum pairing,
unlike the $s$-wave case where the BCS to BEC evolution is just a crossover.
Third, we present a gaussian fluctuation theory near the critical temperature ($T = T_{\rm c}$), 
and we analyze the number of bound, scattering and unbound fermions as well as the chemical potential.
Finally, we derive the time-dependent Ginzburg-Landau functional near $T_{\rm c}$, 
and compare the Ginzburg-Landau coherence length with the zero temperature average Cooper pair size.

\end{abstract}
\maketitle

\section{Introduction}
\label{sec:introduction}

Experimental advances involving atomic Fermi gases enabled the
control of interactions between atoms in different hyperfine states
by using Feshbach resonances~\cite{regal2,greiner,hulet,litium1,litium2,litium3,kinast}.
These resonances can be tuned via an external magnetic field and allow the study 
of dilute many body systems with fixed density, but varying interaction
strength characterized by the scattering parameter $a_\ell$.
This technique allows for the study of new phases of strongly interacting fermions.
For instance, the recent experiments from the MIT group~\cite{mitvortex} marked the first
observation of vortices in atomic Fermi gases corresponding to a strong
signature of superfluidity in the $s$-wave ($\ell = 0$) channel.
These studies combined~\cite{regal2,greiner,hulet,litium1,litium2,litium3,kinast,mitvortex}
correspond to the experimental realization of the theoretically
proposed Bardeen-Cooper-Schrieffer (BCS) to Bose-Einstein condensation (BEC)
crossover~\cite{eagles,leggett,nsr,carlos,jan} in three dimensional (3D) $s$-wave superfluids.
Recent extensions of these ideas include trapped fermions~\cite{perali,griffin-trap} and
fermion-boson models~\cite{holland,timmermans,griffin-fb}.

Arguably one of the next frontiers of exploration in ultracold Fermi systems 
is the search for superfluidity in higher angular momentum states ($\ell \ne 0$).
Substantial experimental progress has been made 
recently~\cite{regal3,ticknor,zhang,schunck,gunter}
in connection to $p$-wave ($\ell = 1$) cold Fermi gases, 
making them ideal candidates for the observation of novel triplet 
superfluid phases. These phases may be present not only in atomic, 
but also in nuclear (pairing in nuclei), astrophysics (neutron stars),
and condensed matter (organic superconductors) systems.

The tuning of $p$-wave interactions in ultracold Fermi gases was initially explored
via $p$-wave Feshbach resonances in trap geometries
for $^{40} {\rm K}$ in Ref.~\cite{regal3,ticknor} and $^6{\rm Li}$ in Ref.~\cite{zhang,schunck}.
Finding and sweeping through these resonances is difficult
since they are much narrower than the $s$-wave case,
because atoms interacting via higher angular momentum channels have to
tunnel through a centrifugal barrier to couple to the bound state~\cite{ticknor}.
Furthermore, while losses due to two body dipolar~{\cite{zhang,john}}
or three-body~\cite{regal3,ticknor} processes
challenged earlier $p$-wave experiments, these losses were still present but were 
less dramatic in the very recent optical lattice experiment involving $^{40} {\rm K}$
and $p$-wave Feshbach resonances~\cite{gunter}.

Furthermore, due to the magnetic dipole-dipole interaction between valence electrons of alkali atoms,
the nonzero angular momentum Feshbach resonances corresponding to projections
of angular momentum $\ell$ [$m_\ell = \pm \ell, \pm (\ell -1), ..., 0$]
are nondegenerate (separated from each other) with total number of $\ell + 1$ resonances~\cite{ticknor}.
Therefore, in principle, these resonances can be tuned and studied
independently if the separation between them is larger
than the experimental resolution. Since the ground state is highly dependent on
the separation and detuning of these resonances, it is possible that
$p$-wave superfluid phases can be studied from the BCS to the BEC regime.
For sufficiently large splittings, it has been proposed~\cite{gurarie,skyip} 
that pairing occurs only in $m_\ell = 0$ and does not occur in the $m_\ell = \pm 1$ states.
However, for small splittings, pairing occurs via a linear 
combination of the $m_\ell = 0$ and $m_\ell = \pm 1$ states.
Thus, the $m_\ell = 0$ or $m_\ell = \pm 1$ resonances may be tuned 
and studied independently if the splitting is large enough in 
comparison to the experimental resolution.

The BCS to BEC evolution of $d$-wave ($\ell = 2$) superfluidity
was discussed previously in the literature using continuum~\cite{borkowski,duncan,annett}
and lattice~\cite{hertog,andrenacci} descriptions in connection to high-$T_{\rm c}$ superconductivity.
More recently, $p$-wave superfluidity was analyzed at $T = 0$
for two hyperfine state (THS) systems in 3D~\cite{tlho}, and for single hyperfine state
(SHS) systems in two dimensions (2D)~\cite{botelho1,botelho-pwave,iskin-lattice}, 
using fermion-only models. Furthermore, fermion-boson models were proposed 
to describe $p$-wave superfluidity at zero~\cite{gurarie,skyip} and 
finite temperature~\cite{ohashi} in 3D.

In this manuscript, we present a generalization of the zero and finite temperature analysis
of both THS pseudo-spin singlet and SHS pseudo-spin triplet~\cite{iskinprl} superfluidity in 3D
within a fermion-only description. Our main results are as follows.

Through an analysis of the low energy scattering amplitude within a T-matrix approach,
we find that bound states occur only when the scattering parameter $a_\ell > 0$ for any $\ell$.
The energy of the bound states $E_{{\rm b},\ell}$ involves only 
the scattering length $a_0$ for $\ell = 0$.
However, another parameter $r_\ell$ related to the interaction range $1/k_0$
is necessary to characterize $E_{{\rm b},\ell}$ for $\ell \ne 0$.
Therefore, all superfluid properties for $\ell \ne 0$ depend strongly on $k_0$
and $a_\ell$, while for $\ell = 0$ they depend strongly only on $a_0$ but weakly on $k_0$.

At zero temperature ($T = 0$), we study the possibility of a topological quantum
phase transition in $\ell \ne 0$ atomic Fermi gases during the evolution from BCS to BEC
regime~\cite{borkowski,duncan,gurarie,botelho1,botelho-pwave,iskin-lattice,iskinprl,volovik}.
We show that there is a fundamental difference between the $\ell = 0$ and $\ell \ne 0$ cases. 
In the $s$-wave ($\ell = 0$) case, there is no phase transition as the magnetic field is tuned 
through the Feshbach resonance from the BCS to the BEC limit. 
That is, the zero temperature thermodynamic properties are analytic
functions of the scattering length $a_0$ when the Feshbach resonance is crossed.
In this case, the superfluid ground state does not change in any 
fundamental way as $a_0$ is varied. This has been noted in the condensed 
matter literature long ago~\cite{leggett,nsr,carlos,jan} and it is referred to
as the BCS-BEC crossover problem. However, for $\ell \ne 0$, 
we show that there is a phase transition as the magnetic field is 
swept through the $\ell \ne 0$ Feshbach resonance.
The phase transition does not occur when two body bound states are first formed, 
but occurs when the many body chemical potential crosses a critical value.

To show that such a zero temperature (quantum) phase transition occurs in $\ell \ne 0$, 
we calculate the order parameter, chemical potential, quasiparticle excitation spectrum,
momentum distribution, atomic compressibility, low energy
collective excitations and average Cooper pair size as a function 
of $a_\ell$, and show that they are non-analytic at $T = 0$ 
when the chemical potential $\mu_\ell$ crosses a critical value.
The symmetry of the order parameter remains unchanged through the transition,
as the ground state wavefunction experiences a major rearrangement
of its analytic structure. In addition, the elementary excitations of
the superfluid also change from gapless in the BCS side to fully gapped in the BEC side
leading to qualitatively different thermodynamic properties in both sides.   
Thus, we conclude that there is a potentially observable BCS-BEC phase 
transition in $\ell \ne 0$ atomic Fermi gases in contrast to the
BCS-BEC crossover already found in $s$-wave ($\ell = 0$) gases.

At finite temperatures, we develop a gaussian fluctuation theory 
near the critical temperature ($T \approx T_{{\rm c},\ell}$) 
to analyze the number of unbound, scattering and bound fermions 
as well as the chemical potential. 
We show that while the saddle point number equation
is sufficient in weak coupling where all fermions are unbound, 
the fluctuation contributions have to be taken into account in order to 
recover the BEC physics in strong couplings where all fermions are bound. 

We also derive the time-dependent Ginzburg-Landau (TDGL) functional 
near $T_{{\rm c},\ell}$ and extract the Ginzburg-Landau (GL) coherence length and time.
We recover the usual TDGL equation for BCS superfluids in weak coupling,
whereas in strong coupling we recover the Gross-Pitaevskii (GP) equation
for a weakly interacting dilute Bose gas.
The TDGL equation exhibits anisotropic coherence lengths for $\ell \ne 0$ which
become isotropic only in the BEC limit, in sharp contrast to the $\ell = 0$ case,
where the coherence length is isotropic for all couplings.
Furthermore, for any $\ell$, the GL time is a complex number with a larger 
imaginary component for $\mu_\ell > 0$ reflecting the decay of Cooper pairs 
into the two-particle continuum with short lifetimes. 
However, the imaginary component vanishes for $\mu_\ell \le 0$ and Cooper pairs
become stable with long lifetimes above $T_{{\rm c},\ell}$.

The rest of the paper is organized as follows. In Sec.~\ref{sec:hamiltonian}, 
we analyze the interaction potential in both real and momentum space for nonzero
orbital momentum channels. We introduce the imaginary-time functional integration 
formalism in Sec.~\ref{sec:functional}, and obtain the self-consistency
(order parameter and number) equations. 
There we also discuss the low energy scattering amplitude of a finite range interaction
for all possible angular momentum channels, and relate the self-consistency
equations to scattering parameters.
In Sec.~\ref{sec:ground-state}, we discuss the evolution from
BCS to BEC superfluidity at zero temperature.
There we analyze the order parameter, chemical potential, quasiparticle excitation spectrum,
momentum distribution, atomic compressibility and ground state energy as a function of
scattering parameters.
We also discuss gaussian fluctuations and low energy collective excitations
at zero temperature in Sec.~\ref{sec:gaussian.0}.
In Sec~\ref{sec:normal-state}, we present the evolution of superfluidity 
from the BCS to the BEC regimes near the critical temperature.
There we discuss the importance of gaussian fluctuations, and
analyze the number of unbound, scattering and bound fermions, 
critical temperature and chemical potential as a function of scattering parameters.
In Sec.~\ref{sec:tdgl}, we derive TDGL equation and extract the GL coherence length and time.
There, we recover the GL equation in the BCS and the GP equation
in the BEC limit. A short summary of our conclusions is given in Sec.~\ref{sec:conclusions}.
Finally, we present in Appendices~\ref{sec:appa} and~\ref{sec:appb}
the coefficients for the low frequency and long wavelength expansion of the action 
at zero and finite temperatures, respectively.

\section{Generalized Hamiltonian}
\label{sec:hamiltonian}

The Hamiltonian for a dilute Fermi gas is given by
\begin{eqnarray}
\label{eqn:hamiltonian.pwave}
H = \sum_{\mathbf{k}, s_1}\xi(\mathbf{k})a_{\mathbf{k}, s_1}^\dagger a_{\mathbf{k}, s_1}
&+& \frac{1}{2{\cal V}}\sum_{\mathbf{k},\mathbf{k'},\mathbf{q}} \sum_{s_1,s_2,s_3,s_4}
V_{s_1,s_2}^{s_3,s_4}(\mathbf{k},\mathbf{k'}) \nonumber \\
&&
b_{s_1,s_2}^\dagger (\mathbf{k},\mathbf{q}) b_{s_3,s_4} (\mathbf{k'},\mathbf{q}), 
\end{eqnarray}
where $s_n$ labels the pseudo-spins corresponding to 
trapped hyperfine states and ${\cal V}$ is the volume. 
These states are represented by 
the creation operator $ a_{\mathbf{k}, s_1}^\dagger$, and
$b_{s_1,s_2}^\dagger (\mathbf{k},\mathbf{q}) = a_{\mathbf{k}+\mathbf{q}/2, s_1}^\dagger 
a_{-\mathbf{k}+\mathbf{q}/2, s_2}^\dagger$.
Here, $\xi(\mathbf{k})= \epsilon(\mathbf{k}) - \mu$ where 
$\epsilon(\mathbf{k}) = k^2/(2M)$ is the energy and
$\mu$ is the chemical potential of fermions.

The interaction term can be written in a separable form
$
V_{s_1,s_2}^{s_3,s_4}(\mathbf{k},\mathbf{k'}) = \Gamma_{s_1,s_2}^{s_3,s_4} V(\mathbf{k},\mathbf{k'}),
$
where $\Gamma_{s_1,s_2}^{s_3,s_4}$ is the spin and $V(\mathbf{k},\mathbf{k'})$ is
the spatial part, respectively.
In the case of THS case, where $s_n \equiv (\uparrow, \downarrow)$, 
both pseudo-spin singlet and pseudo-spin triplet pairings are allowed.
However, we concentrate on the pseudo-spin singlet THS state with 
$
\Gamma_{s_1,s_2}^{s_3,s_4} = \Gamma_{s_1,s_2}^{s_3,s_4} \delta_{s_1,-s_2} \delta_{s_2,s_3} \delta_{s_3,-s_4}.
$
In addition, we discuss the SHS case ($s_n \equiv \, \uparrow$),
where only pseudo-spin triplet pairing is allowed, and the interaction is given by
$
\Gamma_{s_1,s_2}^{s_3,s_4} = \Gamma_{s_1,s_2}^{s_3,s_4} \delta_{s_1,s_2} \delta_{s_3,s_4} \delta_{s_4,\uparrow}.
$
In this manuscript, we analyze THS singlet and
SHS triplet cases for all allowable angular momentum channels.
THS triplet pairing is more involved due to the more complex nature of 
the vector order parameters, and therefore, we postpone this discussion for a future manuscript.

The two fermion interaction can be expanded as
\begin{equation}
V(\mathbf{k}, \mathbf{k'}) = \int d^3\mathbf{r} V(r)e^{i(\mathbf{k} - \mathbf{k'})\cdot \mathbf{r}},
\label{eqn:Vreal}
\end{equation}
and should have the necessary symmetry under the Parity operation, 
where the transformation $\mathbf{k} \to -\mathbf{k}$ or $\mathbf{k'} \to -\mathbf{k'}$
leads to $V(\mathbf{k}, \mathbf{k'})$ for singlet, and $-V(\mathbf{k}, \mathbf{k'})$ 
for triplet pairing.
Furthermore, $V(\mathbf{k}, \mathbf{k'})$ is invariant under the 
transformation $(\mathbf{k},\mathbf{k'}) \to (-\mathbf{k},-\mathbf{k'})$, and 
$V(\mathbf{k}, \mathbf{k'})$ reflects the Pauli exclusion principle.

In order to obtain an approximate expression for the atomic interaction potential,
we use the Fourier expansion of a plane wave in 3D
\begin{equation}
e^{i\mathbf{k} \cdot \mathbf{r}} = 4\pi \sum_{\ell,m_\ell} i^\ell j_\ell(kr) 
Y_{\ell,m_\ell}^*(\widehat{\mathbf{r}}) Y_{\ell,m_\ell}(\widehat{\mathbf{k}}) ,
\end{equation}
where $j_\ell(kr)$ is the spherical Bessel function of order $\ell$ and
$Y_{\ell,m_\ell}(\widehat{\mathbf{k}})$ is the spherical harmonic
of order $(\ell,m_\ell)$, in Eq.~\ref{eqn:Vreal} to evaluate the matrix elements of the
interaction potential in $\mathbf{k}$-space
\begin{equation}
V(\mathbf{k},\mathbf{k'}) = 
4\pi\sum_{\ell,m_\ell}V_\ell(k,k')Y_{\ell,m_\ell}(\widehat{\mathbf{k}}) Y_{\ell,m_\ell}^*(\widehat{\mathbf{k'}}).
\label{eqn:Vsh}
\end{equation}
Here, $\sum_{\ell,m_\ell} = \sum_{\ell = 0}^{\infty} \sum_{m_\ell = - \ell}^{\ell}$, and
$\widehat{\mathbf{k}}$ denotes the angular dependence $(\theta_{\mathbf{k}}, \phi_{\mathbf{k}})$.
The $(k,k')$ dependent coefficients $V_\ell(k,k')$ are related to the real space potential
$V(r)$ through the relation
\begin{equation}
V_\ell(k,k') = 4\pi\int_0^\infty dr r^2 j_\ell(kr) j_\ell(k'r) V(r).
\end{equation}
The index $\ell$ labels angular momentum states in 3D, with 
$\ell = 0,1,2, ...$ corresponding to $s,p,d, ...$ channels, respectively.

In the long wavelength limit ($k \to 0$), one can show that the $k$ dependence 
of this potential becomes exactly separable. In fact, for $kr \ll 1$ and $k'r \ll 1$,
the asymptotic expression of the spherical Bessel function for small arguments can be used,
giving $V_\ell(k,k') = C_\ell k^\ell k'^\ell$, with the coefficient $C_\ell$ dependent on
the particular choice of the real space potential. In the opposite limit, where $kr \gg 1$ and 
$k'r \gg 1$, the potential is not separable. In this case,
$V_\ell(k,k')$ mixes different $k$ and $k'$, and shows an oscillatory behavior 
(which is dependent on the exact form of $V(r)$) with a decaying envelope that is
proportional to $1/(kk')$.

Under these circumstances, we choose to study a model potential that contains most of the
features described above. One possibility is to retain only one of the $\ell$ terms
in Eq.~(\ref{eqn:Vsh}), by assuming that the dominant contribution 
to the scattering process between fermionic atoms occurs in the $\ell$th angular momentum channel.
This assumption may be experimentally relevant since atom-atom dipole interactions
split different angular momentum channels such that they may be tuned independently.
Using the properties discussed above, we write
\begin{equation}
V_\ell(k,k') = - \lambda_\ell \Gamma_\ell(k) \Gamma_\ell(k'),
\end{equation}
where $\lambda_\ell > 0$ is the interaction strength, and the function
\begin{equation}
\Gamma_\ell(k) = \frac{(k/k_0)^\ell} {(1 + k^2/k_0^2)^{\frac{\ell + 1}{2}}}
\end{equation}
describes the momentum dependence. 
Here, $k_0 \sim R_0^{-1}$ plays the role of the interaction range in real space
and sets the scale at small and large momenta.
In addition, the diluteness condition ($n_\ell R_0^3 \ll 1$) requires
$(k_0/k_{\rm F})^3 \gg 1$, where $n_\ell$ is the density of atoms and 
$k_{\rm F}$ is the Fermi momentum. This function reduces to $\Gamma_\ell(k) \sim k^\ell$
for small $k$, and behaves as $\Gamma_\ell(k) \sim 1/k$ for large $k$,
which guarantees the correct qualitative behavior expected for $V_\ell(k,k')$ according
to the analysis above.

\section{Functional integral formalism}
\label{sec:functional}

In this section, we describe in detail the THS singlet case for even angular momentum states.
A similar approach for the SHS triplet case for odd angular momentum 
states can be found in Ref.~\cite{iskin-lattice}, and therefore, 
we do not repeat the same analysis here. However, we point out the main 
differences between the two cases whenever it is necessary.

\subsection{THS Singlet Effective Action}
\label{sec:effective-action}

In the imaginary-time functional integration formalism ($\hbar=k_{\rm B}=1$, and $\beta=1/T$), 
the partition function for the THS singlet case can be written as
\begin{equation}
Z_\ell=\int D(a^\dagger,a)e^{-S_\ell}
\end{equation}
with action
\begin{equation}
S_\ell=\int_0^\beta d\tau\left[ \sum_{\mathbf{k}, s} a_{\mathbf{k}, s}^\dagger(\tau)
(\partial_\tau) a_{\mathbf{k}, s}(\tau) + H_\ell(\tau) \right]
\end{equation}
where the Hamiltonian for the $\ell$th angular momentum channel is
\begin{eqnarray}
H_\ell(\tau) &=& \sum_{\mathbf{k}, s} \xi_\ell(\mathbf{k}) 
a_{\mathbf{k}, s}^\dagger(\tau) a_{\mathbf{k}, s}(\tau) \nonumber \\
&-& \frac{4\pi\lambda_\ell}{{\cal V}} \sum_{\mathbf{q},m_\ell} 
b_{\ell, m_\ell}^\dagger(\mathbf{q},\tau) b_{\ell, m_\ell}(\mathbf{q},\tau).
\end{eqnarray}
Here,
$
b_{\ell, m_\ell} (\mathbf{q},\tau) = \sum_{\mathbf{k}} 
\Gamma_\ell(k) Y_{\ell,m_\ell}(\widehat{\mathbf{k}})
a_{\mathbf{k}+\mathbf{q}/2,\uparrow} a_{\mathbf{k}-\mathbf{q}/2,\downarrow}
$
and 
$
\xi_\ell(\mathbf{k}) = \epsilon(\mathbf{k}) - \mu_\ell.
$
We first introduce the Nambu spinor 
$\psi^\dagger(p)=( a_{p, \uparrow}^\dagger , a_{-p, \downarrow} )$, 
where $p=(\mathbf{k}, iw_j)$ 
denotes both momentum and fermionic Matsubara frequency $w_j=(2j+1)\pi/\beta$, 
and use a Hubbard-Stratonovich transformation 
\begin{eqnarray}
e^{-\sum_{q} \lambda |b(q)|^2}
&=& \int D[\Phi^\dagger,\Phi] \nonumber \\
&&e^{\sum_{q}\left[\frac{|\Phi(q)|^2}{\lambda} + b^\dagger(q)\Phi(q) + \Phi^\dagger(q) b(q) \right]}
\end{eqnarray}
to decouple fermionic and bosonic degrees of freedom.
Integration over the fermionic part [$D(\psi^\dagger,\psi)$] leads to the action
\begin{eqnarray}
S_\ell^{\rm eff} &=& \beta \sum_{q,m_\ell} \frac{|\Phi_{\ell,m_\ell}(q)|^2}{4\pi {\cal V}^{-1} \lambda_\ell} \nonumber \\
&+& \sum_{p,p'} \left[ \beta \xi_\ell(\mathbf{k}) \delta_{p,p'}
- {\rm Tr} \ln \left( \mathbf{G}_\ell/\beta \right)^{-1} \right] ,
\label{eqn:s_eff}
\end{eqnarray}
where $q=(\mathbf{q}, iv_j)$, with bosonic Matsubara frequency $v_j=2\pi j/\beta$.
Here,
\begin{eqnarray}
\mathbf{G}_\ell^{-1} &=& \Phi_\ell^*(q)\Gamma_\ell(\frac{p+p'}{2})\sigma_- 
+ \Phi_\ell(-q)\Gamma_\ell(\frac{p+p'}{2})\sigma_+ \nonumber \\
&& \hskip 1cm + \left[iw_j \sigma_0-\xi_\ell(\mathbf{k})\sigma_3\right]\delta_{p,p'}
\end{eqnarray}
is the inverse Nambu propagator,
$\Phi_\ell(q) = \sum_{m_\ell}\Phi_{\ell,m_\ell}(q)Y_{\ell,m_\ell}(\widehat{\frac{\mathbf{k}+\mathbf{k'}}{2}})$ 
is the bosonic field,
and $\sigma_{\pm}=(\sigma_1\pm\sigma_2)/2$ and $\sigma_{i}$ is the Pauli spin matrix.
The bosonic field
\begin{equation}
\Phi_{\ell,m_\ell}(q) = \Delta_{\ell,m_\ell}\delta_{q,0}+ \Lambda_{\ell,m_\ell}(q)
\end{equation}
has $\tau$-independent $\Delta_{\ell,m_\ell}$ and $\tau$-dependent $\Lambda_{\ell,m_\ell}(q)$ parts.

Performing an expansion in $S_\ell^{\rm {eff}}$ to quadratic order in $\Lambda_{\ell,m_\ell}(q)$ leads to
\begin{equation}
S_\ell^{\rm{gauss}} = S_\ell^{\rm{sp}} + \frac{\beta}{2}\sum_{q,m_\ell,m'_\ell} 
\widetilde{\Lambda}_{\ell,m_\ell}^\dagger(q) 
\mathbf{F}^{-1}_{\ell,m_\ell,m'_\ell}(q) \widetilde{\Lambda}_{\ell,m'_\ell}(q),
\label{eqn:gaction}
\end{equation}
where the vector $\widetilde{\Lambda}_{\ell,m_\ell}^\dagger(q)$ is such that
$\widetilde{\Lambda}_{\ell,m_\ell}^\dagger(q) = [\Lambda_{\ell,m_\ell}^\dagger(q), \Lambda_{\ell,m_\ell}(-q)]$,
and $\mathbf{F}^{-1}_{\ell,m_\ell,m'_\ell}(q)$ are the matrix elements 
of the inverse fluctuation propagator matrix $\mathbf{F}^{-1}_{\ell}(q)$. 
Furthermore, $S_\ell^{\rm{sp}}$ is the saddle point action given by
\begin{eqnarray}
S_\ell^{\rm sp} &=& \beta \sum_{m_\ell}\frac{|\Delta_{\ell,m_\ell}|^2}{4\pi {\cal V}^{-1} \lambda_\ell} \nonumber \\
&+& \sum_{p} \left[ \beta\xi_\ell(\mathbf{k})
- {\rm Tr} \ln \left( \mathbf{G}_\ell^{\rm sp}/\beta \right)^{-1} \right], 
\end{eqnarray}
and the saddle point inverse Nambu propagator is
\begin{equation}
(\mathbf{G}_\ell^{\rm sp})^{-1} = iw_j \sigma_0-\xi_\ell(\mathbf{k})\sigma_3 +
\Delta_\ell^*(\mathbf{k})\sigma_- + \Delta_\ell(\mathbf{k}) \sigma_+,
\end{equation}
with saddle point order parameter
\begin{equation}
\Delta_\ell(\mathbf{k}) = \Gamma_\ell(k) \sum_{m_\ell} \Delta_{\ell,m_\ell} Y_{\ell,m_\ell}(\widehat{\mathbf{k}}).
\end{equation}
Notice that, $\Delta_\ell(\mathbf{k})$ may involve several different $m_\ell$ 
for a given angular momentum channel $\ell$.

The matrix elements of the inverse fluctuation matrix $\mathbf{F}_{\ell}^{-1}(q)$ are given by
\begin{eqnarray}
(\mathbf{F}_{\ell,m_\ell,m_\ell'}^{-1})_{11} 
&=& \frac{1}{\beta} \sum_p (\mathbf{G}_\ell^{\rm sp})_{11}(\frac{q}{2} + k) 
(\mathbf{G}_\ell^{\rm sp})_{11}(\frac{q}{2} - k) \nonumber \\
&&
\Gamma_\ell^2 (p) Y_{\ell,m_\ell}(\widehat{\mathbf{k}}) 
Y_{\ell,m_\ell'}^*(\widehat{\mathbf{k}}) - \frac{\delta_{m_\ell,m_\ell'}{\cal V}}{4\pi\lambda_\ell}, 
\label{eqn:fluct.F11}
\\
(\mathbf{F}_{\ell,m_\ell,m_\ell'}^{-1})_{12} 
&=& 
\frac{1}{\beta} \sum_p (\mathbf{G}_\ell^{\rm sp})_{12} (\frac{q}{2} + k) 
(\mathbf{G}_\ell^{\rm sp})_{12}(\frac{q}{2} - k) \nonumber \\
&&
\Gamma_\ell^2(p) Y_{\ell,m_\ell}(\widehat{\mathbf{k}}) Y_{\ell,m_\ell'}^*(\widehat{\mathbf{k}}).
\label{eqn:fluct.F12}
\end{eqnarray}
Notice that while 
$(\mathbf{F}_{\ell,m_\ell,m_\ell'}^{-1})_{12}(q) = $ 
$(\mathbf{F}_{\ell,m_\ell,m_\ell'}^{-1})_{21}(q)$ 
are even under the transformations
$\mathbf{q}\rightarrow -\mathbf{q}$ and $iv_j\rightarrow -iv_j$; 
$(\mathbf{F}_{\ell,m_\ell,m_\ell'}^{-1})_{11}(q) =$  
$(\mathbf{F}_{\ell,m_\ell,m_\ell'}^{-1})_{22}(-q)$ 
are even only under $\mathbf{q}\rightarrow -\mathbf{q}$, having
no defined parity in $iv_j$.
 
The Gaussian action Eq.~(\ref{eqn:gaction}) leads to the thermodynamic potential
$\Omega_\ell^{\rm{gauss}} = \Omega_\ell^{\rm{sp}} + \Omega_\ell^{\rm{fluct}}$, where 
\begin{eqnarray}
\Omega_\ell^{\rm{sp}} &=& \sum_{m_\ell} \frac{|\Delta_{\ell,m_\ell}|^2}{4\pi {\cal V}^{-1} \lambda_\ell} + 
\sum_{\mathbf{k}}\big \lbrace\xi_\ell(\mathbf{k}) - E_\ell(\mathbf{k}) \nonumber \\
&-& \frac{2}{\beta}\ln\left[1 + \exp(-\beta E_\ell(\mathbf{k}))\right] \big\rbrace, 
\label{eqn:sptp} \\
\Omega_\ell^{\rm{fluct}} &=& \frac{1}{\beta} \sum_{q}\ln\det[\mathbf{F}^{-1}_{\ell}(q)/(2\beta)]
\label{eqn:flucttp}
\end{eqnarray}
are the saddle point and fluctuation contributions, respectively.
Here,
\begin{equation}
E_\ell(\mathbf{k}) = \left[\xi_\ell^2(\mathbf{k})+|\Delta_\ell(\mathbf{k})|^2 \right]^{\frac{1}{2}},
\end{equation}
is the quasiparticle energy spectrum.
Having completed the presentation of the functional integral formalism,
we discuss next the self-consistency equations for the order parameter and
the chemical potential.

\subsection{Self-consistency Equations}
\label{sec:self-consistency}
The saddle point condition $\delta S_\ell^{\rm{sp}} /\delta \Delta_{\ell,m_\ell}^* = 0$ 
leads to the order parameter equation
\begin{equation}
\frac{\Delta_{\ell,m_\ell}}{4\pi \lambda_\ell} = \frac{1}{{\cal V}}
\sum_{\mathbf{k}}\frac{\Delta_{\ell}(\mathbf{k})\Gamma_\ell(k)
Y_{\ell,m_\ell}^*(\widehat{\mathbf{k}})}{2E_\ell(\mathbf{k})} 
\tanh\frac{\beta E_\ell(\mathbf{k})}{2},
\end{equation}
which can be expressed in terms of experimentally relevant 
parameters via the $T$-matrix approach~\cite{tlho}.

The low energy two body scattering amplitude between a pair of fermions
in the $\ell$th angular momentum channel is given by~\cite{landau}
\begin{equation}
f_\ell(k) = - \frac{k^{2\ell}} {1/a_\ell - r_\ell k^2 + ik^{2\ell + 1}},
\end{equation}
where $r_\ell < 0$ and $a_\ell$ are the effective range and scattering parameter, 
respectively. 
Here $r_\ell$ has dimensions of $L^{2\ell - 1}$ and
$a_\ell$ has dimensions of $L^{2\ell + 1}$, where $L$ is the size of the system.
The energy of the two body bound state is determined from the poles of $f_\ell(k \to i\kappa_\ell)$,
and is given by $E_{{\rm b},\ell} = - \kappa_\ell^2/(2M)$.
Bound states occur when $a_0 > 0$ for $\ell = 0$, and
$a_{\ell \ne 0} r_{\ell \ne 0} < 0$ for $\ell \ne 0$. Since $r_{\ell} < 0$,
bound states occur only when $a_\ell > 0$ for all $\ell$,
in which case the binding energies are given by
\begin{eqnarray}
E_{{\rm b}, 0} &=& - \frac{1}{M a_0^2}, \\
E_{{\rm b},\ell \ne 0} &=& \frac{1}{M a_\ell r_\ell}.
\end{eqnarray}
Notice that, only a single parameter ($a_0$) is sufficient to describe 
the low energy two body problem for $\ell = 0$, while two parameters $(a_\ell, r_\ell)$
are necessary to describe the same problem for $\ell \ne 0$.
The point at which $1/(k_{\rm F}^{2\ell + 1} a_\ell) = 0$ corresponds 
to the threshold for the formation of a two body bound state in vacuum. 
Beyond this threshold, $a_0$ for $\ell = 0$ and 
$|a_{\ell \ne 0} r_{\ell \ne 0}|$ for $\ell \ne 0$ 
are the size of the bound states.

For any $\ell$, the two body scattering amplitude is related to the $T$-matrix via
\begin{eqnarray}
f_\ell(k) = -\frac{M}{4\pi} T_\ell[k,k;2\epsilon(\mathbf{k}) + i0^+],
\end{eqnarray}
where the $T$-matrix is given by
\begin{eqnarray}
T(\mathbf{k},\mathbf{k'},E) = V(\mathbf{k},\mathbf{k'}) + 
\frac{1}{{\cal V}}\sum_{\mathbf{k''}} \frac{V(\mathbf{k},{\mathbf{k''}})
T(\mathbf{k''},\mathbf{k'},E)} {E - 2\epsilon(\mathbf{k''}) + i0^+} \nonumber .
\end{eqnarray}
Using the spherical harmonics expansion for both 
$V(\mathbf{k},\mathbf{k'})$ and $T(\mathbf{k},\mathbf{k'},E)$
leads to two coupled equations,
\begin{eqnarray}
\frac{1}{\lambda_\ell} &=& -\frac{M}{4\pi k_0^{2\ell} a_\ell} +
\frac{1}{{\cal V}} \sum_{\mathbf{k}} \frac{\Gamma_\ell^2(k)}{2\epsilon(\mathbf{k})}, 
\label{eqn:a_ell} \\
r_{\ell \ne 0} &=& -\frac{\pi k_0^{2\ell}}{M^2 {\cal V}}\sum_{\mathbf{k}} \frac{\Gamma_\ell^2(k)}{\epsilon^2(\mathbf{k})} -
\frac{\ell + 1}{k_0^2 a_\ell},
\label{eqn:r_ell}
\end{eqnarray}
relating $\lambda_\ell$ and $k_0$ to $a_\ell$ and $r_\ell$.
Except for notational differences, notice that these relations 
are identical to previous results~\cite{tlho}.
After performing momentum integrations we obtain
\begin{eqnarray}
k_0^{2\ell + 1} a_\ell &=& \frac{M k_0\lambda_\ell \sqrt{\pi}}{M k_0\lambda_\ell\widetilde{\phi}_\ell - 4\pi\sqrt{\pi}}, 
\label{eqn:ka} \\
-\frac{1}{a_{\ell \ne 0} r_{\ell \ne 0}} &=& \frac{2 k_0^2 \sqrt{\pi}}{k_0^{2\ell + 1} a_\ell \phi_\ell + 2(\ell + 1)\sqrt{\pi}},
\label{eqn:ar}
\end{eqnarray}
where 
$
\widetilde{\phi}_{\ell} = \Gamma(\ell + 1/2)/\Gamma(\ell + 1)
$ 
and
$
\phi_{\ell} = \Gamma(\ell - 1/2)/\Gamma(\ell + 1).
$
Here $\Gamma(x)$ is the Gamma function. 
Notice that, $k_0^{2\ell + 1} a_\ell$ diverges and changes sign when
$M k_0 \lambda_\ell \widetilde{\phi}_\ell = 4\pi\sqrt{\pi}$,
which corresponds to the critical coupling for Feshbach resonances (the unitarity limit).

\begin{figure} [htb]
\centerline{\scalebox{0.67}{\includegraphics{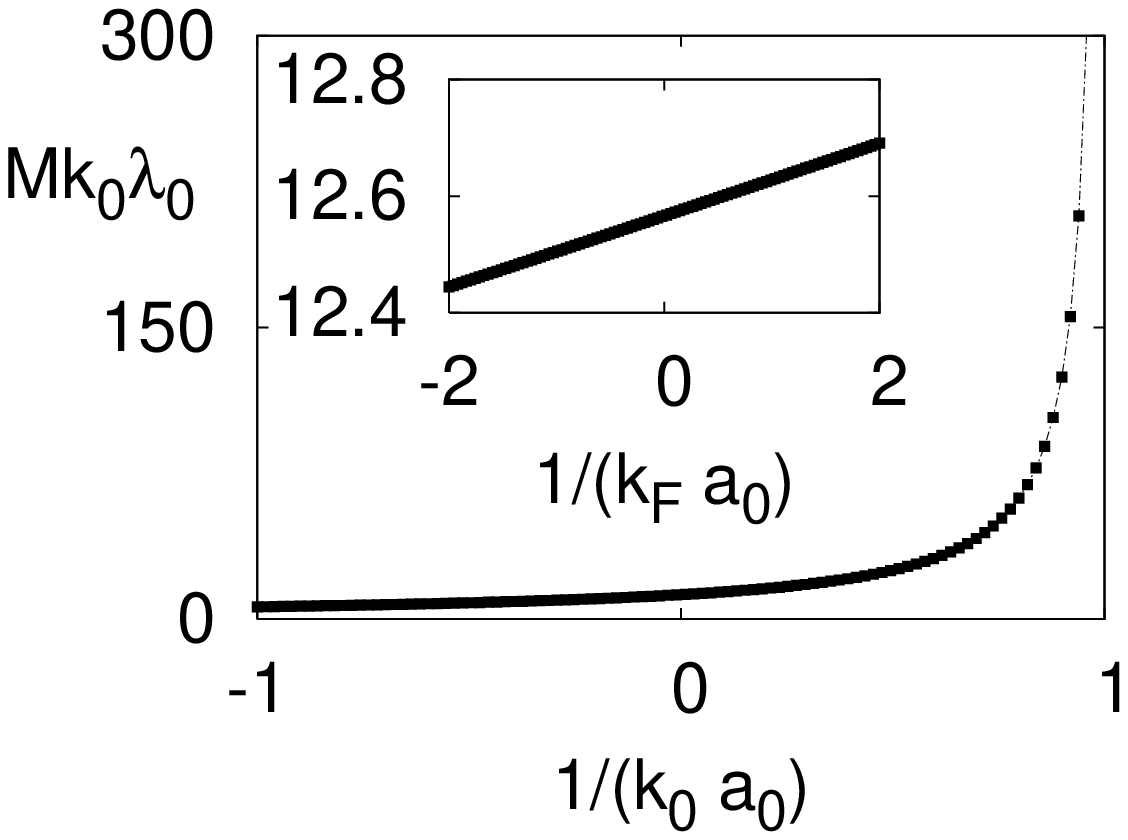}}}
\caption{\label{fig:swave.lambda} Plot of original interaction strength 
$Mk_0 \lambda_0$ versus scattering parameter $1/(k_0 a_0)$. 
The inset shows $Mk_0 \lambda_0$ versus $1/(k_{\rm F} a_0)$ for $k_0 \approx 200 k_{\rm F}$.
}
\end{figure}
\begin{figure} [htb]
\centerline{\scalebox{0.67}{\includegraphics{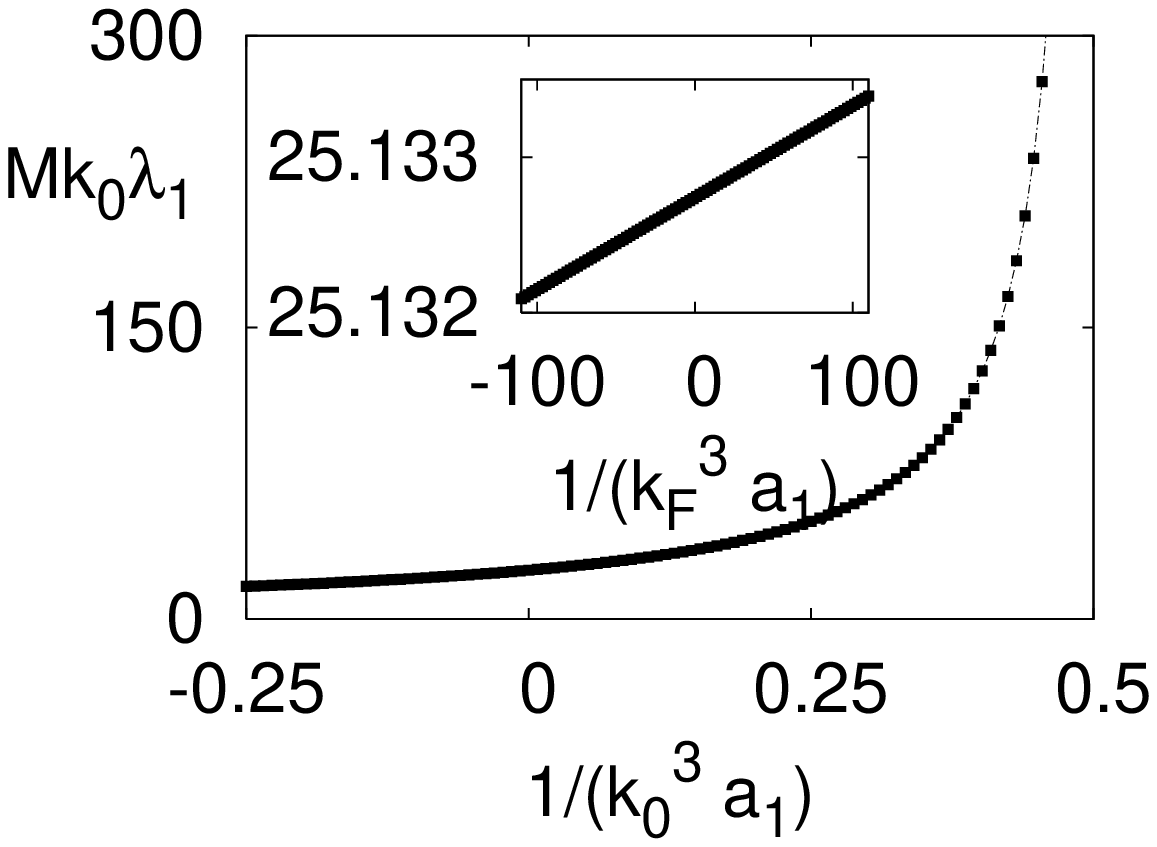}}}
\caption{\label{fig:pwave.lambda} Plot of original interaction strength 
$Mk_0 \lambda_1$ versus scattering parameter $1/(k_0^3 a_1)$.
The inset shows $Mk_0 \lambda_1$ versus $1/(k_{\rm F}^3 a_1)$ for $k_0 \approx 200 k_{\rm F}$.
}
\end{figure}

In addition, the scattering parameter has a maximum value 
in the zero ($\lambda_\ell \to 0$) and a minimum value in the infinite 
($\lambda_\ell \to \infty$) coupling limits given respectively by
\begin{eqnarray}
k_0^{2\ell + 1} a_{\ell \ne 0}^{\rm max} &=& - \frac{2(\ell + 1)\sqrt{\pi}}{\phi_\ell}, 
\,\, (a_\ell < 0), 
\label{eqn:abcs} \\
k_0^{2\ell + 1} a_\ell^{\rm min} &=& \frac{\sqrt{\pi}}{\widetilde{\phi}_\ell},
\,\, (a_\ell > 0).
\label{eqn:abec}
\end{eqnarray}
The first condition Eq.~(\ref{eqn:abcs}) (when $\lambda_\ell \to 0$) follows from Eq.~(\ref{eqn:ar}) 
where $r_{\ell \ne 0} < 0$ has to be satisfied for all possible $a_{\ell \ne 0}$.
However, there is no condition on $r_0$ for $\ell = 0$, and $k_0 a_0^{\rm max} = 0$
in the BCS limit.
The second condition Eq.~(\ref{eqn:abec}) (when $\lambda_\ell \to \infty$) follows from Eq.~(\ref{eqn:ka}),
which is valid for all possible $\ell$.
The minimum $a_\ell$ for a finite range interaction is associated with the Pauli principle, which
prevents two identical fermions to occupy the same state.
Thus, while the scattering parameter can not be arbitrarily small for a finite range
potential, it may go to zero as $k_0 \to \infty$.
Furthermore, the binding energy is given by 
\begin{equation}
E_{{\rm b},\ell \ne 0} = -\frac{2\sqrt{\pi}} {M k_0^{2\ell - 1} a_\ell \phi_\ell},
\end{equation}
when $k_0^{2\ell + 1} a_\ell \phi_\ell \gg 2(\ell + 1)\sqrt{\pi}$.

In Fig.~\ref{fig:swave.lambda}, we plot the original interaction strength $Mk_0\lambda_0$ 
versus the scattering parameter $k_0 a_0$ for the $s$-wave ($\ell = 0$) channel. 
Notice that, $k_0 |a_0| \to 0$ in the BCS and $k_0 a_0 \to 1$ in the BEC limit.
A divergence in $k_0a_0$ corresponds to a $s$-wave Feshbach resonance
occurring at $Mk_0\lambda_0 = 4\pi$.

In Fig.~\ref{fig:pwave.lambda}, we plot the original interaction strength $Mk_0\lambda_1$ 
versus the scattering parameter $k_0^3 a_1$ for the $p$-wave ($\ell = 1$) channel. 
Notice that, $k_0^3 |a_1| \to 4$ in the BCS and $k_0^3 a_1 \to 2$ in the BEC limit.
A divergence in $k_0^3 a_1$ corresponds to a $p$-wave Feshbach resonance
occurring at $Mk_0\lambda_1 = 8\pi$.

Thus, the order parameter equation in terms of the scattering parameter is rewritten as
\begin{eqnarray}
\frac{M{\cal V}\Delta_{\ell,m_\ell}}{16\pi^2 k_0^{2\ell} a_\ell} 
&=& \sum_{\mathbf{k},m'_\ell}
\left[ \frac{1}{2\epsilon(\mathbf{k})} - 
\frac{\tanh[\beta E_\ell(\mathbf{k})/2]} {2E_\ell(\mathbf{k})}
\right] \nonumber \\
&& \hskip 2mm \Delta_{\ell,m'_\ell} \Gamma_\ell^2(k)
Y_{\ell,m_\ell}^*(\widehat{\mathbf{k}})Y_{\ell,m'_\ell}(\widehat{\mathbf{k}}).
\label{eqn:opeqn}
\end{eqnarray}
This equation is valid for both THS pseudo-spin singlet and SHS pseudo-spin triplet states.
However, there is one important difference between pseudo-spin singlet and pseudo-spin triplet states.
For pseudo-spin singlet states, the order parameter is a scalar function of $\mathbf{k}$,
while it is a vector function for pseudo-spin triplet states discussed next.

In general, the triplet order parameter can be written
in the standard form~\cite{leggett-review}
\begin{equation}
\mathbf{O}_\ell (\mathbf{k})
= \left( \begin{array}{cc} -d_\ell^x (\mathbf{k}) + id_\ell^y (\mathbf{k}) & d_\ell^z (\mathbf{k}) 
\\ d_\ell^z (\mathbf{k}) & d_\ell^x (\mathbf{k}) + id_\ell^y (\mathbf{k}) \end{array}\right),
\end{equation}
where the vector 
$
d_\ell (\mathbf{k}) = [d_\ell^x (\mathbf{k}), d_\ell^y (\mathbf{k}), d_\ell^z (\mathbf{k})]
$ 
is an odd function of $\mathbf{k}$.
Thefore, all up-up, down-down and up-down components may exist for a 
THS pseudo-spin triplet interaction.
However, in the SHS pseudo-spin triplet case only the up-up or down-down component
may exist leading to $\Delta_\ell (\mathbf{k}) \propto (\mathbf{O}_\ell)_{s_1 s_1}(\mathbf{k})$.
Thus, for the up-up case $d_\ell^z (\mathbf{k}) = 0$ and $d_\ell^x (\mathbf{k}) = - id_\ell^y (\mathbf{k})$,
leading to $d_\ell (\mathbf{k}) = d_\ell^x (\mathbf{k})(1,i,0)$, which
breaks time reversal symmetry, as expected from a fully spin polarized state.
The corresponding down-down state has $d_\ell (\mathbf{k}) = d_\ell^x (\mathbf{k})(1,-i,0)$.
Furthermore, the simplified form of the SHS triplet order parameter
allows a treatment similar to that of THS singlet states.
However, it is important to mention that the THS triplet case can be investigated
using our approach, but the treatment is more complicated.

The order parameter equation has to be solved self-consistently with the number equation 
$N_\ell = -\partial \Omega_\ell/\partial {\mu_\ell}$ where $\Omega_\ell$ is the full thermodynamic potential. 
In the approximations used, 
\begin{equation}
N_\ell \approx N_\ell^{\rm gauss} = N_{\ell}^{\rm{sp}} + N_\ell^{\rm{fluct}}
\label{eqn:totalN.0}
\end{equation}
has two contributions. The saddle point contribution to the number equation is
\begin{equation}
N_\ell^{\rm{sp}} = \sum_{\mathbf{k}, s} n_\ell(\mathbf{k}),
\label{eqn:spnumbereqn}
\end{equation}
where $n_\ell(\mathbf{k})$ is the momentum distribution given by
\begin{equation}
n_\ell(\mathbf{k})=\frac{1}{2}\left[ 1 - \frac{\xi_\ell(\mathbf{k})}{E_\ell(\mathbf{k})}\tanh\frac{\beta E_\ell(\mathbf{k})}{2} \right].
\label{eqn:nk}
\end{equation}
For the SHS triplet case, the summation over $s$ is not present in $N_\ell^{\rm{sp}}$.
The fluctuation contribution to the number equation is
\begin{equation}
N_\ell^{\rm{fluct}} = -\frac{1}{\beta}\sum_{q} 
\frac{\partial [\det \mathbf{F}^{-1}_{\ell}(q)]/\partial {\mu_\ell}}
{\det \mathbf{F}^{-1}_{\ell}(q)},
\end{equation}
where $\mathbf{F}^{-1}_{\ell}(q)$ is the inverse fluctuation matrix
defined in Eq.~(\ref{eqn:fluct.F11}) and (\ref{eqn:fluct.F12}).

In the rest of the paper, we analyze analytically the superfluid properties 
at zero temperature (ground state) and near the critical temperatures
for THS singlet (only even $\ell$) and SHS triplet (only odd $\ell$) cases.
In addition, we analyze numerically the $s$-wave ($\ell = 0$) channel of 
THS singlet and $p$-wave ($\ell = 1$) channel of
SHS triplet cases, which are currently of intense theoretical and experimental interest
in ultracold Fermi atoms.

\section{Bcs to Bec Evolution at $T = 0$}
\label{sec:ground-state}

At low temperatures, the saddle point self-consistent (order parameter and number) 
equations are sufficient to describe ground state properties in the weak coupling BCS and 
strong coupling BEC limits~\cite{leggett}.
However, fluctuation corrections to the number equation 
may be important in the intermediate regime~\cite{pieri}.

Ground state properties ($T = 0$) are investigated by solving
saddle point self-consistency (order parameter and number) equations
to obtain $\Delta_{\ell,m_\ell}$ and $\mu_\ell$, which are discussed next.

\subsection{Order Parameter and Chemical Potential}
\label{sec:order-parameter}

We discuss in this section $\Delta_{\ell,m_\ell}$ and $\mu_\ell$.
In weak coupling, we first introduce a shell about the Fermi energy
$|\xi_\ell(\mathbf{k})| \le w_{\rm D}$ such that
$\epsilon_{\rm F} \gg w_{\rm D} \gg \Delta_\ell(\mathbf{k}_{\rm F})$, inside of
which one may ignore the 3D density of states factor ($\sqrt{\epsilon/\epsilon_{\rm F}}$)
and outside of which one may ignore $\Delta_{\ell}(\mathbf{k})$.
While in sufficiently strong coupling,
we use $\xi_\ell(\mathbf{k}) \gg |\Delta_\ell (\mathbf{k})|$ to derive the 
analytic results discussed below. 
It is important to notice that, in strictly weak and strong coupling, 
the self-consistency equations Eq.~(\ref{eqn:spnumbereqn}) and (\ref{eqn:opeqn}) 
are decoupled, and play reversed roles. 
In weak (strong) coupling the order parameter
equation determines $\Delta_{\ell,m_\ell}$ ($\mu_\ell$) and the
number equation determines $\mu_\ell$ ($\Delta_{\ell,m_\ell}$).

In weak coupling, the number equation Eq.~(\ref{eqn:spnumbereqn}) leads to
\begin{eqnarray}
\mu_\ell = \epsilon_{\rm F}
\end{eqnarray}
for any $\ell$ where $\epsilon_{\rm F} = k_{\rm F}^2/(2M)$ is the Fermi energy.
In strong coupling, the order parameter equation Eq.~(\ref{eqn:opeqn}) leads to
\begin{eqnarray}
\mu_0 &=& -\frac{1}{2M a_0^2},
\label{eqn:mu_0} \\
\mu_{\ell \ne 0} &=& - \frac{\sqrt{\pi}} {M k_0^{2\ell -1} a_\ell \phi_\ell},
\label{eqn:mu_ell}
\end{eqnarray}
where 
$
\phi_\ell = \Gamma(\ell - 1/2) / \Gamma(\ell + 1)
$
and $\Gamma(x)$ is the Gamma function.
This calculation requires that $a_0 k_0 > 1$ for $\ell = 0$ and that
$k_0^{2\ell + 1} a_\ell \phi_\ell > (\ell + 1) \sqrt{\pi}$ 
for $\ell \ne 0$ for the order parameter equation to have a solution 
with $\mu_\ell < 0$ in the strong coupling limit.
In the BEC limit $\mu_0 = - k_0^2/[2M (k_0 a_0 - 1)^2]$ for $\ell = 0$.
Notice that, $\mu_0 = -1/(2M a_0^2)$ when $k_0 a_0 \gg 1$ 
[or $|\mu_0| \ll \epsilon_0 = k_0^2/(2M)$], and thus, 
we recover the contact potential ($k_0 \to \infty$) result. 
In the same spirit, to obtain the expressions in Eq.~(\ref{eqn:mu_0}) and (\ref{eqn:mu_ell}),
we assumed $|\mu_\ell| \ll \epsilon_0$.
Notice that, $\mu_\ell = E_{{\rm b},\ell}/2$ in this limit for any $\ell$.

On the other hand, the solution of the order parameter equation in the weak coupling limit is
\begin{eqnarray}
|\Delta_{0,0}| &=& 16 \sqrt{\pi} \epsilon_{\rm F} 
\exp \left[ 2 + \frac{\pi}{2} \frac{k_{\rm F}} {k_0} - \frac{\pi}{2k_{\rm F}|a_0|}\right], \\
|\Delta_{\ell \ne 0,m_\ell}| &\sim& \left(\frac{k_0}{k_{\rm F}}\right)^\ell \epsilon_{\rm F} \nonumber \\
&& \exp\left[t_\ell \left(\frac{k_0}{k_{\rm F}}\right)^{2\ell -1} - \frac{\pi}{2k_{\rm F}^{2\ell+1}|a_\ell|}\right],
\end{eqnarray}
where $t_1 = \pi /4$ and $t_{\ell > 1} = \pi 2^{\ell + 1} (2\ell - 3)!! / \ell!$.
These expressions are valid only when the exponential terms are small.
The solution of the number equation in the strong coupling limit is
\begin{eqnarray}
|\Delta_{0,0}| &=& 8\epsilon_{\rm F}\left( \frac{\mu_0}{9\epsilon_{\rm F}} \right)^\frac{1}{4}, \\
\sum_{m_\ell} |\Delta_{\ell \ne 0,m_\ell}|^2 &=& \frac{64 \sqrt{\pi}}{3\phi_\ell} \epsilon_{\rm F} (\epsilon_{\rm F}\epsilon_0)^\frac{1}{2}
\end{eqnarray}
to order $\mu_\ell/\epsilon_0$, 
where we assumed that $\xi_\ell(\mathbf{k}) \gg |\Delta_\ell(\mathbf{k})|$
for sufficiently strong couplings with $|\mu_\ell| \ll \epsilon_0$.

\begin{figure} [htb]
\centerline{\scalebox{0.67}{\includegraphics{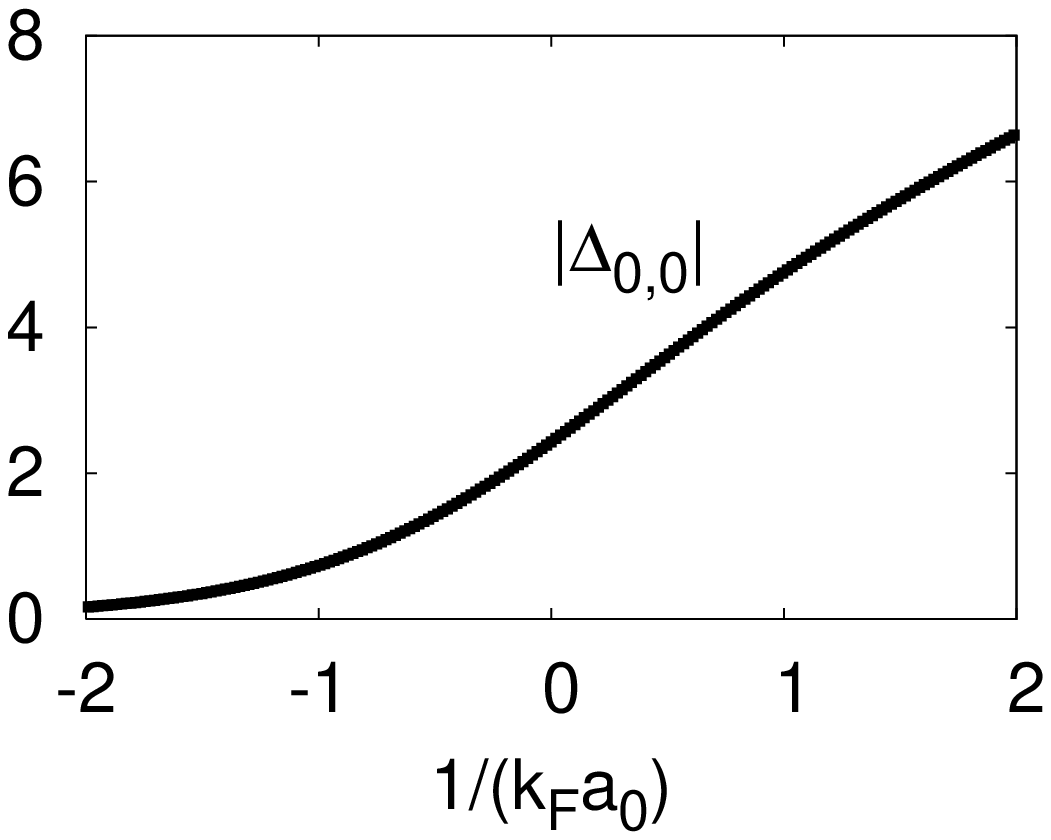}}}
\caption{\label{fig:swave.gap} Plot of 
reduced order parameter $\Delta_{\rm r} = |\Delta_{0,0}|/\epsilon_{\rm F}$
versus interaction strength $1/(k_{\rm F} a_0)$ for $k_0 \approx 200k_{\rm F}$.
}
\end{figure}
\begin{figure} [htb]
\centerline{\scalebox{0.67}{\includegraphics{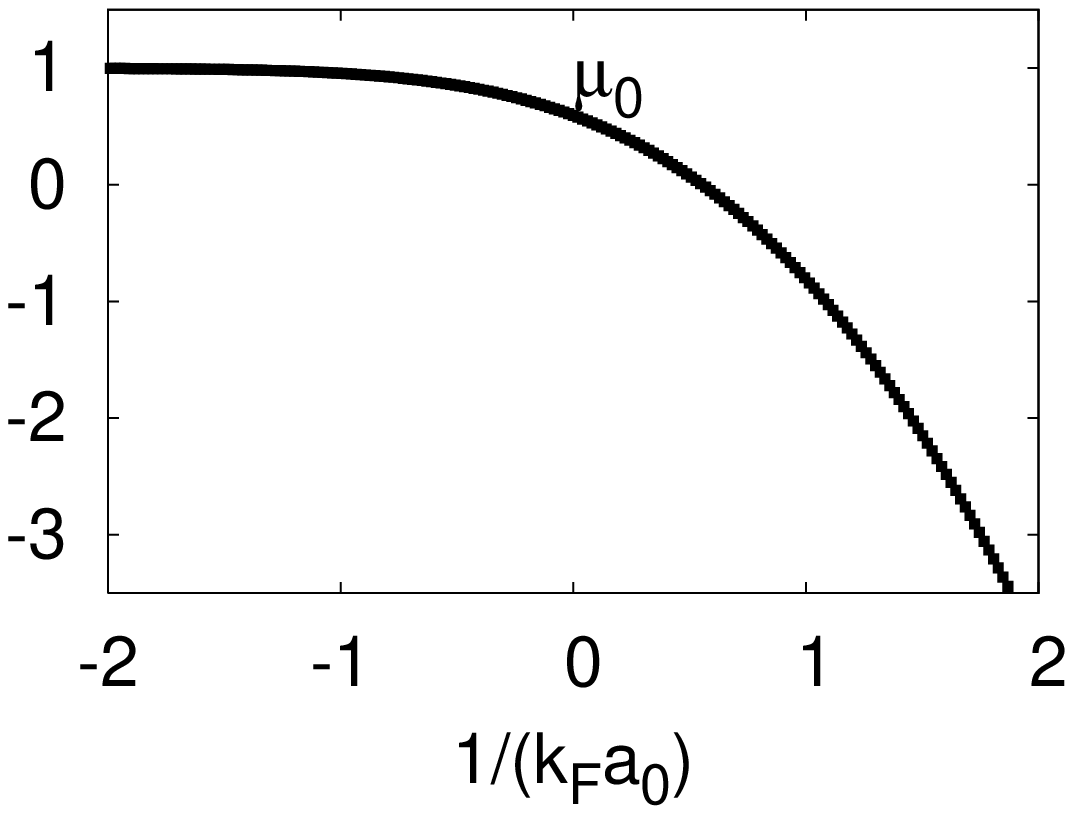}}}
\caption{\label{fig:swave.mu0} Plot of 
reduced chemical potential $\mu_{\rm r} = \mu_0/\epsilon_{\rm F}$ (inset)
versus interaction strength $1/(k_{\rm F} a_0)$ for $k_0 \approx 200k_{\rm F}$.
}
\end{figure}

Next, we present numerical results for two particular states.
First, we analyze the THS $s$-wave ($\ell = 0, m_\ell = 0$) case, where 
$\Delta_0(\mathbf{k}) = \Delta_{0,0} \Gamma_0(k) Y_{0,0}(\widehat{\mathbf{k}})
$
with 
$
Y_{0,0}(\widehat{\mathbf{k}}) = 1/\sqrt{4\pi}.
$
Second, we discuss the SHS $p$-wave ($\ell = 1, m_\ell = 0$) case, where 
$
\Delta_1(\mathbf{k}) = \Delta_{1,0} \Gamma_1(k) Y_{1,0}(\widehat{\mathbf{k}})
$
with
$
Y_{1,0}(\widehat{\mathbf{k}}) = \sqrt{3/(4\pi)} \cos(\theta_\mathbf{k}).
$
In all numerical calculations, we choose $k_0 \approx 200 k_{\rm F}$ to
compare $s$-wave and $p$-wave cases.

In Figs.~\ref{fig:swave.gap} and~\ref{fig:swave.mu0}, 
we show $|\Delta_{0,0}|$ and $\mu_0$ at $T = 0$ for the $s$-wave case. 
Notice that the BCS to BEC evolution range in $1/(k_{\rm F} a_0)$ is of order $1$.
Furthermore, $|\Delta_{0,0}|$ grows continuously without saturation with
increasing coupling, while $\mu_0$ changes from $\epsilon_{\rm F}$ to $E_{{\rm b},0}/2$ 
continuously and decreases as $-1/(2M a_0^2)$ for strong couplings.
Thus, the evolution of $|\Delta_{0,0}|$ and $\mu_0$ as a function of 
$1/(k_{\rm F}a_0)$ is smooth. 
For completeness, it is also possible to obtain analytical values of $a_0$
and $\Delta_{0,0}$ when the chemical potential vanishes.
When $\mu_0 = 0$, we obtain for 
$
|\Delta_{0,0}| = 8\epsilon_{\rm F}[\pi^2\sqrt{\pi}/\Gamma^4(1/4)]^{1/3}
\approx 3.73 \epsilon_{\rm F}
$ 
at 
$
1/(k_{\rm F}a_0) = (2\pi^3\sqrt{\pi} \epsilon_{\rm F} / |\Delta_{0,0}|)^{1/2} / [2\Gamma^2(3/4)] 
\approx 0.554,
$
which also agrees with the numerical results.
Here $\Gamma(x)$ is the Gamma function.

\begin{figure} [htb]
\centerline{\scalebox{0.67}{\includegraphics{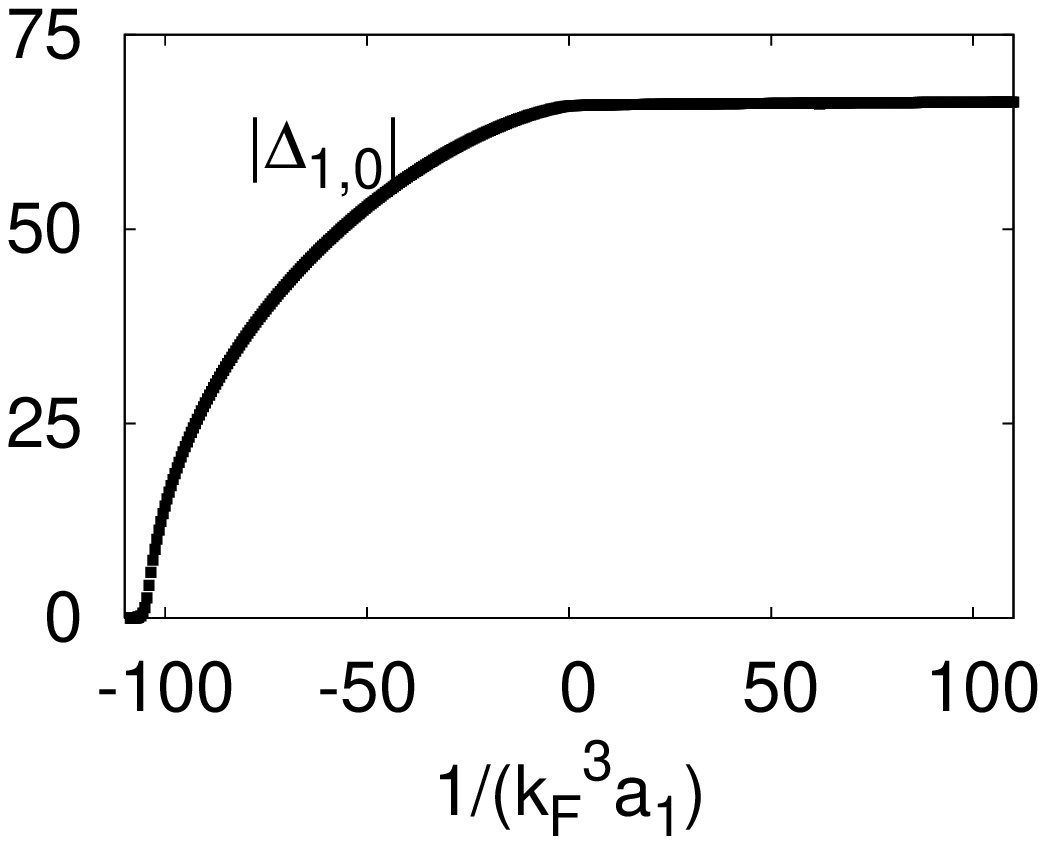}}}
\caption{\label{fig:pwave.gap} Plots of 
reduced order parameter $\Delta_{\rm r} = |\Delta_{1,0}|/\epsilon_{\rm F}$
versus interaction strength $1/(k_{\rm F}^3 a_1)$ for $k_0 \approx 200k_{\rm F}$.
}
\end{figure}
\begin{figure} [htb]
\centerline{\scalebox{0.67}{\includegraphics{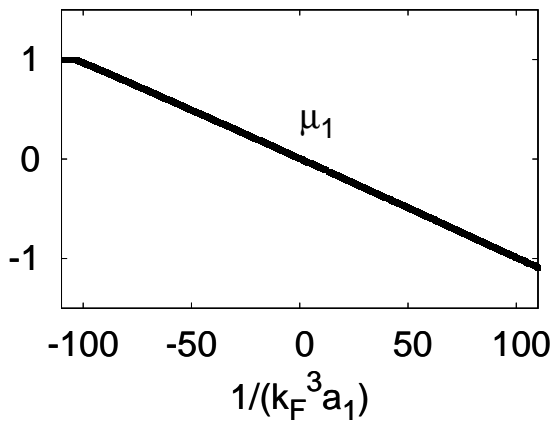}}}
\caption{\label{fig:pwave.mu0} Plots of 
reduced chemical potential $\mu_{\rm r} = \mu_1/\epsilon_{\rm F}$ (inset)
versus interaction strength $1/(k_{\rm F}^3 a_1)$ for $k_0 \approx 200k_{\rm F}$.
}
\end{figure}

In Figs.~\ref{fig:pwave.gap} and~\ref{fig:pwave.mu0}, we show $|\Delta_{1,0}|$ 
and $\mu_1$ at $T = 0$ for the $p$-wave case.
Notice that the BCS to BEC evolution range in $1/(k_{\rm F}^3 a_1)$ is of order $k_0/k_{\rm F}$.
Furthermore, $|\Delta_{1,0}|$ grows with increasing coupling
but saturates for large $1/(k_{\rm F}^3 a_1)$, while $\mu_1$ changes from $\epsilon_{\rm F}$ 
to $E_{{\rm b},1}/2$ continuously and decreases as $-1/(Mk_0a_1)$ for strong couplings.
For completeness, we present the limiting expressions
\begin{eqnarray}
|\Delta_{1,0}| &=& 24 \frac{k_0}{k_{\rm F}} \epsilon_{\rm F} \exp \left[-\frac{8}{3}
+ \frac{\pi k_0}{4k_{\rm F}} - \frac{\pi}{2k_{\rm F}^3 |a_1|}\right], \\
|\Delta_{1,0}| &=& 8\epsilon_{\rm F}\left( \frac{\epsilon_0}{9\epsilon_{\rm F}} \right)^\frac{1}{4},
\end{eqnarray}
in the weak and strong coupling limits, respectively.

The evolution of $|\Delta_{1,0}|$ and $\mu_1$ are qualitatively
similar to recent $T = 0$ results for THS fermion~\cite{tlho} 
and SHS fermion-boson~\cite{skyip} models.
Due to the angular dependence of $\Delta_1(\mathbf{k})$,
the quasiparticle excitation spectrum $E_1(\mathbf{k})$ is gapless for $\mu_1 > 0$,
and fully gapped for $\mu_1 < 0$.
Furthermore, both $\Delta_{1,0}$ and $\mu_1$ are nonanalytic exactly when $\mu_1$
crosses the bottom of the fermion energy band $\mu_1 = 0$ at $1/(k_{\rm F}^3 a_1) \approx 0.48$.
The nonanalyticity does not occur in the first derivative of $\Delta_{1,0}$ or $\mu_1$
as it is the case in 2D~\cite{botelho-pwave}, but occurs in the second and higher derivatives.
Thus, in the $p$-wave case, the BCS to BEC evolution is not a crossover,
but a quantum phase transition occurs, as can be seen in the quasiparticle
excitation spectrum to be discussed next.

\subsection{Quasiparticle Excitations}
\label{sec:quasiparticle}

The quasiparticle excitation spectrum 
$
E_\ell(\mathbf{k}) = [\xi_\ell^2(\mathbf{k}) + |\Delta_\ell(\mathbf{k})|^2]^{1/2}
$
is gapless at $\mathbf{k}$-space regions where the conditions 
$
\Delta_\ell(\mathbf{k}) = 0
$
and 
$
\epsilon(\mathbf{k}) = \mu_\ell
$ 
are both satisfied.
Notice that the second condition is only satisfied
in the BCS side $\mu_\ell \ge 0$, and therefore, 
the excitation spectrum is always gapped in the BEC side ($\mu_\ell < 0$).

For $\ell = 0$, the order parameter is isotropic in $\mathbf{k}$-space
without zeros (nodes) since it does not have any angular dependence.
Therefore, the quasiparticle excitation spectrum 
is fully gapped in both BCS ($\mu_0 > 0$) and BEC ($\mu_0 < 0$) sides, since
\begin{eqnarray}
\min \{ E_0 (\mathbf{k}) \} &=& |\Delta_0(k_{\mu_0})|, \,\, (\mu_0 > 0), 
\label{eqn:swave.qp.bcs} \\
\min \{ E_0 (\mathbf{k}) \} &=& \sqrt{|\Delta_0 (0)|^2 + \mu_0^2}, \,\, (\mu_0 < 0).
\label{eqn:swave.qp.bec}
\end{eqnarray}
Here, $k_{\mu_\ell}  = \sqrt{2M \mu_\ell}$.
This implies that the evolution of the quasiparticle excitation spectrum 
from weak coupling BCS to strong coupling BEC regime is smooth 
when $\mu_0 = 0$ for $\ell = 0$ pairing.

\begin{figure} [htb]
\centerline{\scalebox{0.61}{\includegraphics{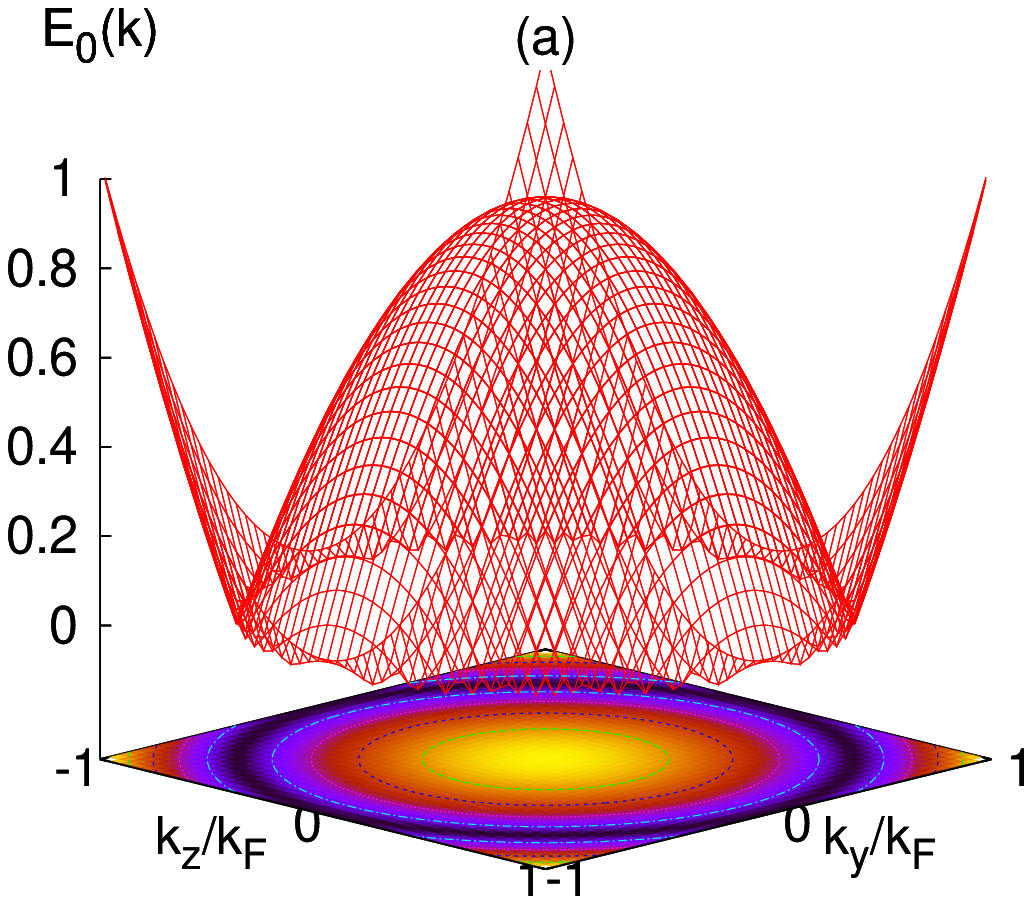}}}
\vskip 8mm
\centerline{\scalebox{0.61}{\includegraphics{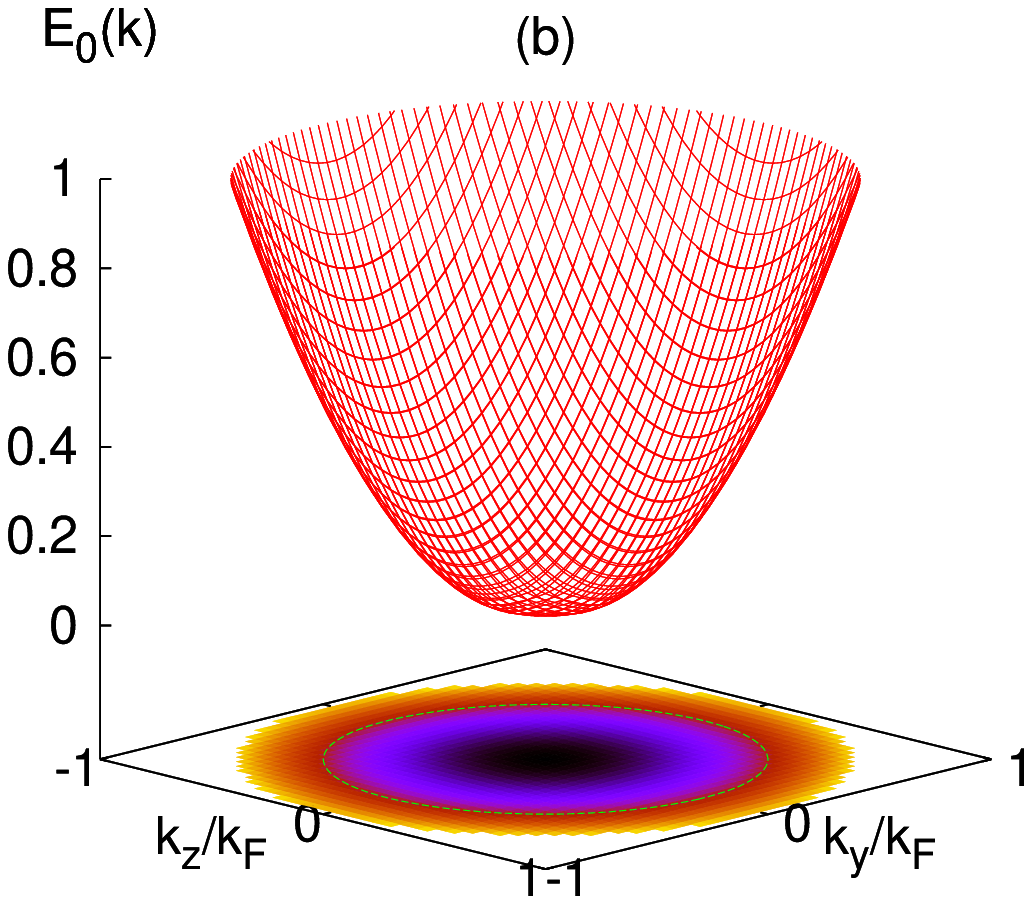}}}
\vskip 8mm
\centerline{\scalebox{0.61}{\includegraphics{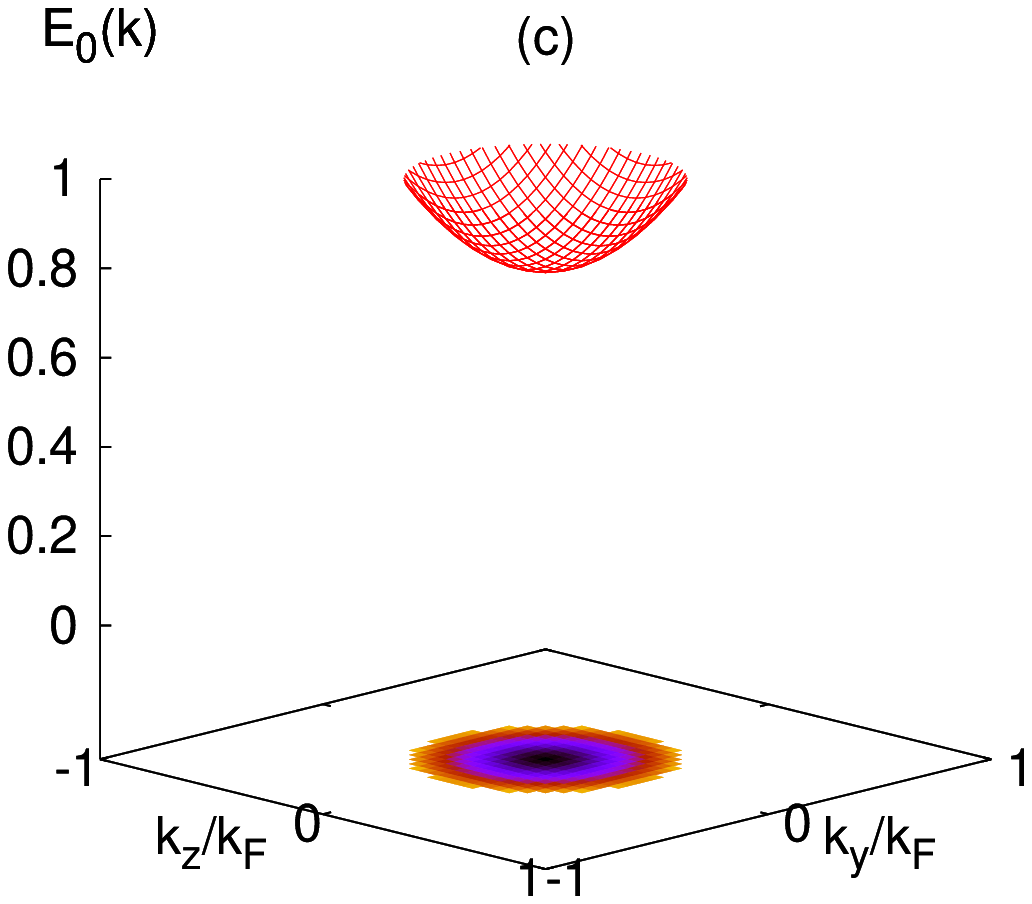}}}
\caption{\label{fig:swave.qp} (Color online) 
Plots of quasiparticle excitation spectrum $E_0(k_x = 0, k_y, k_z)$ when
a) $\mu_0 > 0$ (BCS side) for $1/(k_{\rm F} a_0) = -1$,
b) $\mu_0 = 0$ (intermediate regime) for $1/(k_{\rm F} a_0) \approx 0.55$, and
c) $\mu_0 < 0$ (BEC side) for $1/(k_{\rm F} a_0) = 1$
versus momentum $k_y/k_{\rm F}$ and $k_z/k_{\rm F}$.
}
\end{figure}

In Fig.~\ref{fig:swave.qp}, we show $E_0(k_x = 0, k_y,k_z)$
for an $s$-wave ($\ell = 0, m_\ell = 0$) superfluid when
a) $\mu_0 > 0$ (BCS side) for $1/(k_{\rm F} a_0) = -1$,
b) $\mu_0 = 0$ (intermediate regime) for $1/(k_{\rm F} a_0) \approx 0.55$, and
c) $\mu_0 < 0$ (BEC side) for $1/(k_{\rm F} a_0) = 1$.
Notice that the quasiparticle excitation spectrum is gapped for 
all three cases.
However, the situation for $\ell \ne 0$ is very different as discussed next.

For $\ell \ne 0$, the order parameter is anisotropic in $\mathbf{k}$-space
with zeros (nodes) since it has an angular dependence.
Therefore, while the quasiparticle excitation spectrum
is gapless in the BCS ($\mu_{\ell \ne 0} > 0$) side,
it is fully gapped in the BEC ($\mu_{\ell \ne 0} < 0$) side, since
\begin{eqnarray}
\min \{ E_{\ell \ne 0} (\mathbf{k}) \} &=& 0, \,\, (\mu_\ell > 0), 
\label{eqn:pwave.qp.bcs}\\
\min \{ E_{\ell \ne 0} (\mathbf{k}) \} &=& |\mu_\ell|, \,\, (\mu_\ell < 0).
\label{eqn:pwave.qp.bec}
\end{eqnarray}
This implies that the evolution of quasiparticle excitation spectrum
from weak coupling BCS to strong coupling BEC regime is not smooth 
for $\ell \ne 0$ pairing having a nonanalytic behavior when $\mu_{\ell \ne 0} = 0$.
This signals a quantum phase transition from a gapless to a fully gapped
state exactly when $\mu_{\ell \ne 0}$ drops below the bottom of 
the energy band $\mu_{\ell \ne 0} = 0$.

\begin{figure} [htb]
\centerline{\scalebox{0.61}{\includegraphics{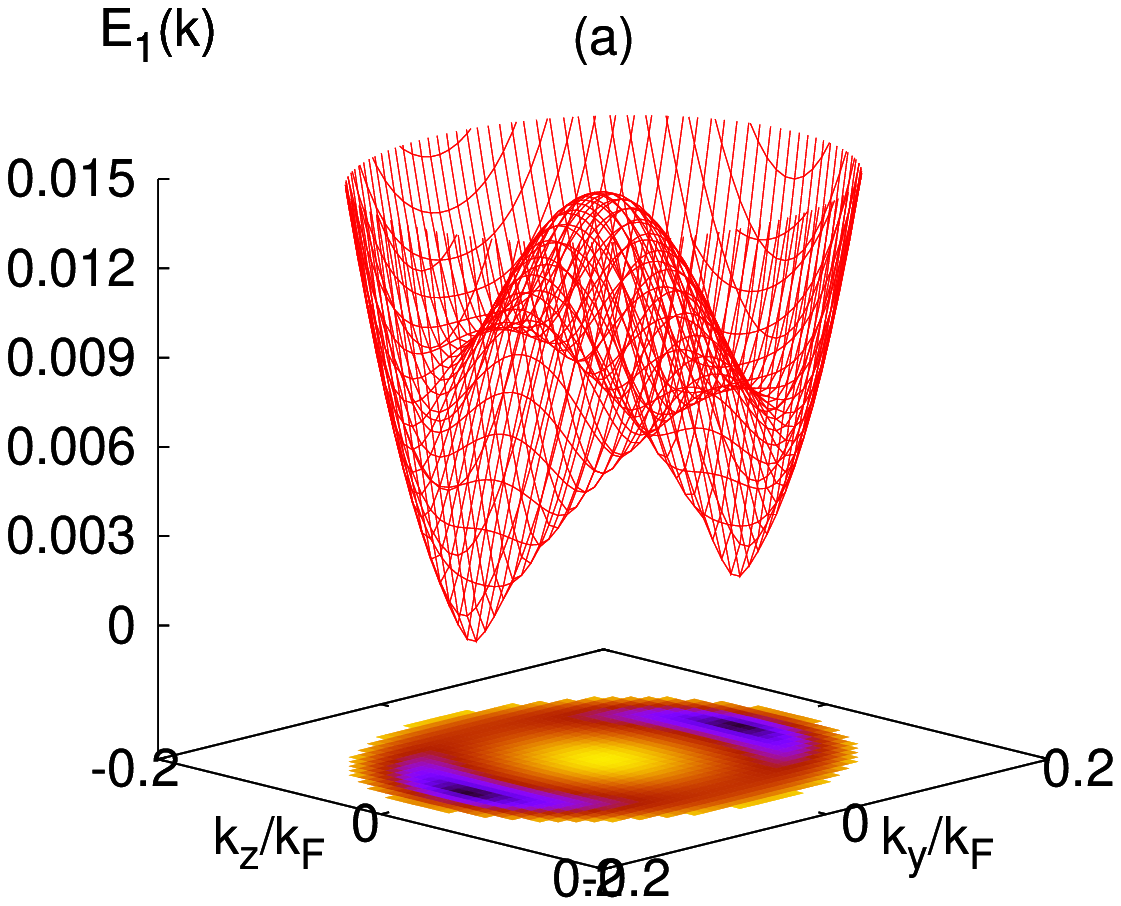}}}
\vskip 8mm
\centerline{\scalebox{0.61}{\includegraphics{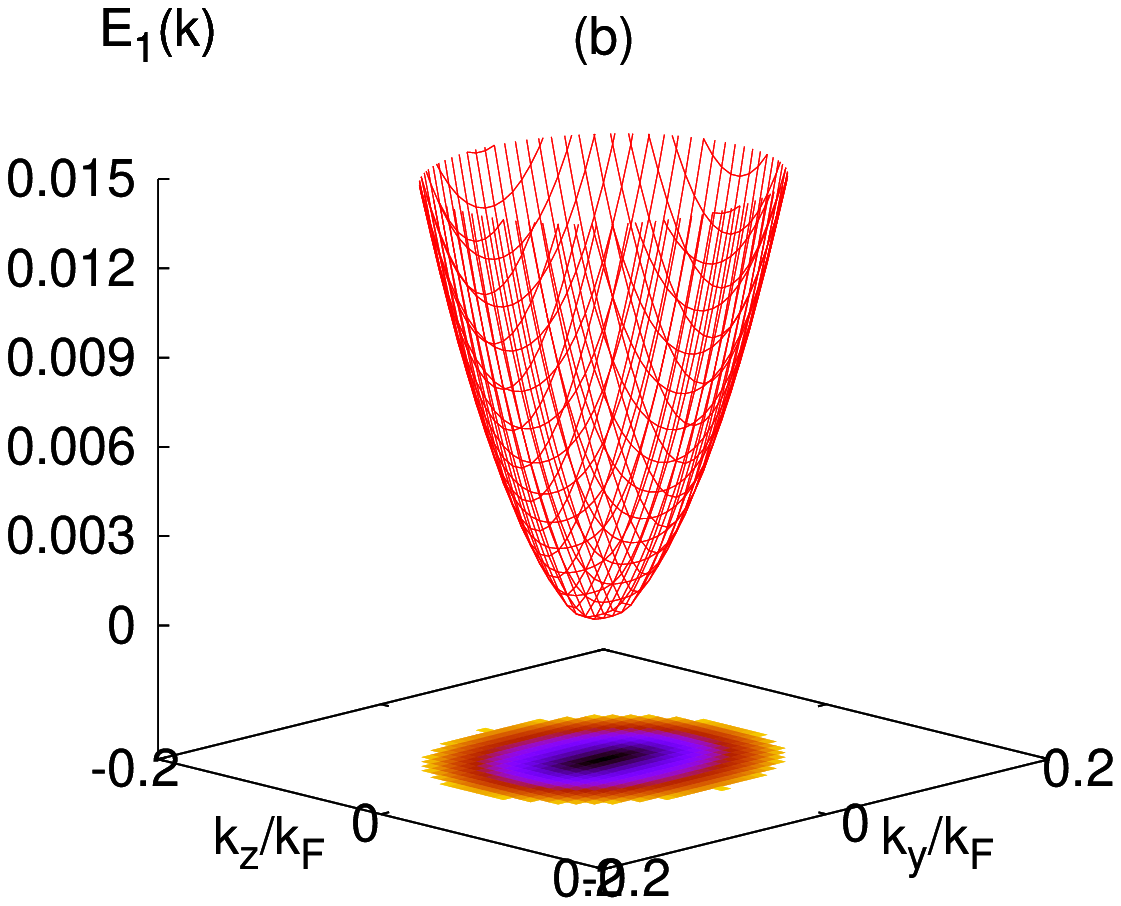}}}
\vskip 8mm
\centerline{\scalebox{0.61}{\includegraphics{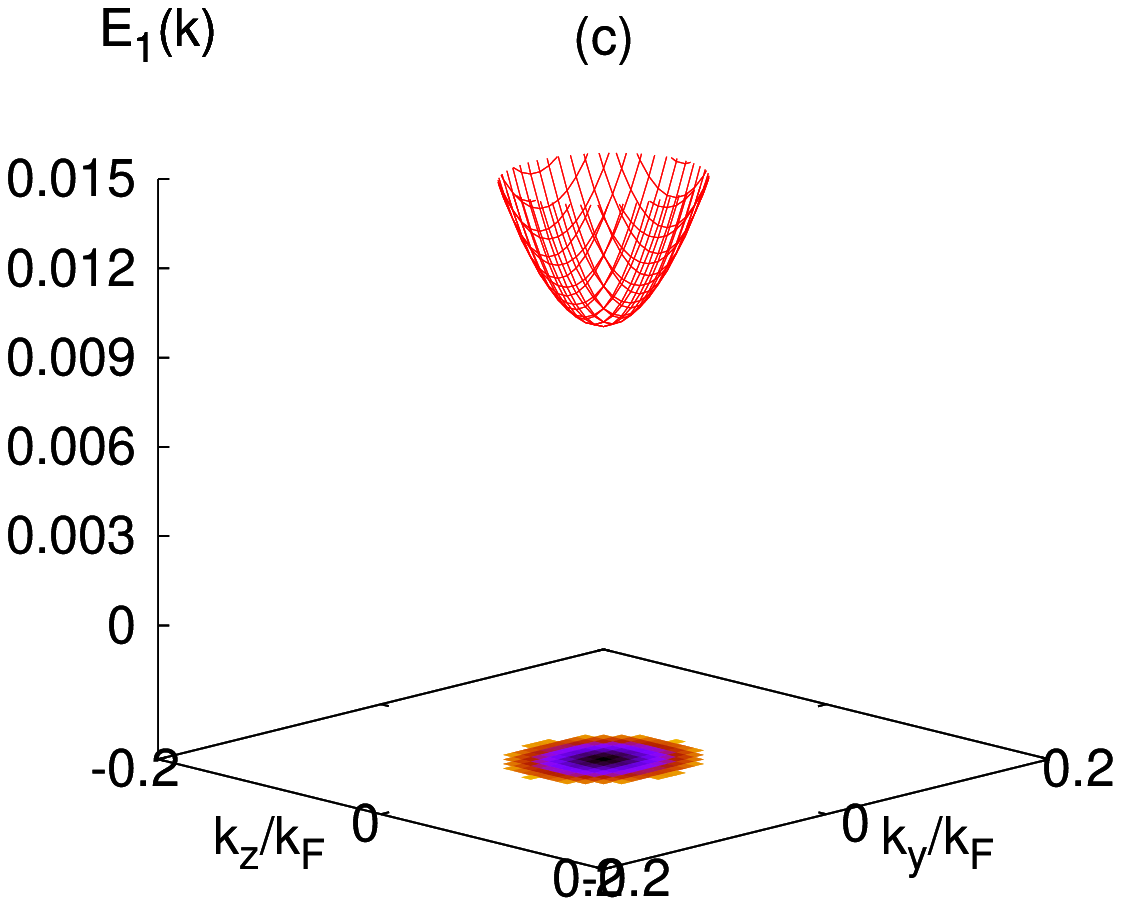}}}
\caption{\label{fig:pwave.qp} (Color online) 
Plots of quasiparticle excitation spectrum $E_1(k_x = 0, k_y, k_z)$ in 
a) $\mu_1 > 0$ (BCS side) for $1/(k_{\rm F}^3 a_1) = -1$,
b) $\mu_1 = 0$ (intermediate regime) for $1/(k_{\rm F}^3 a_1) \approx 0.48$, and
c) $\mu_1 < 0$ (BEC side) for $1/(k_{\rm F}^3 a_1) = 1$
versus momentum $k_y/k_{\rm F}$ and $k_z/k_{\rm F}$.
}
\end{figure}

In Fig.~\ref{fig:pwave.qp}, we show $E_1 (k_x = 0, k_y,k_z)$
for a $p$-wave ($\ell = 1, m_\ell = 0$) superfluid when
a) $\mu_1 > 0$ (BCS side) for $1/(k_{\rm F}^3a_1) = -1$,
b) $\mu_1 = 0$ (intermediate regime) for $1/(k_{\rm F}^3a_1) \approx 0.48$, and 
c) $\mu_1 < 0$ (BEC side) for $1/(k_{\rm F}^3a_1) = 1$.
The quasiparticle excitation spectrum is gapless when
$
\Delta_1(\mathbf{k}) \propto k_z/k_{\rm F} = 0
$
and
$
k_x^2 + k_y^2 + k_z^2 = 2M \mu_1
$
are both satisfied in certain regions of $\mathbf{k}$-space.
For $k_x = 0$, these conditions are met only when 
$k_z = 0$ and $k_y = \pm \sqrt{2M \mu_1}$ for a given $\mu_1$.
Notice that, these points come closer as the interaction ($\mu_1$) increases (decreases),
and when $\mu_1 = 0$ they become degenerate at $\mathbf{k} = \mathbf{0}$. 
For $\mu_1 < 0$, the second condition can not be satisfied,
and thus, a gap opens in the excitation spectrum
of quasiparticles as shown in Fig.~\ref{fig:pwave.qp}c.

The spectrum of quasiparticles plays an important role in the thermodynamic properties
of the evolution from BCS to BEC regime at low temperatures.
For $\ell = 0$, thermodynamic quantities depend exponentially on
$T$ throughout the evolution. Thus, a smooth crossover occurs at $\mu_0 = 0$.
However, for $\ell \ne 0$, thermodynamic quantitites depend exponentially 
on $T$ only in the BEC side, while they have a power law dependence on $T$ 
in the BCS side. Thus, a non-analytic evolution occurs at $\mu_{\ell \ne 0} = 0$.
This can be seen best in the momentum distribution which is discussed next.

\subsection{Momentum Distribution}
\label{sec:md}

In this section, we analyze the momentum distribution 
$
n_\ell(\mathbf{k}) = \left[ 1 - \xi_\ell(\mathbf{k}) / E_\ell(\mathbf{k}) \right]/2
$
in the BCS $(\mu_\ell > 0)$ and BEC sides $(\mu_\ell < 0)$, which reflect
the gapless to gapped phase transition for nonzero
angular momentum superfluids.

\begin{figure} [htb]
\centerline{\scalebox{0.61}{\includegraphics{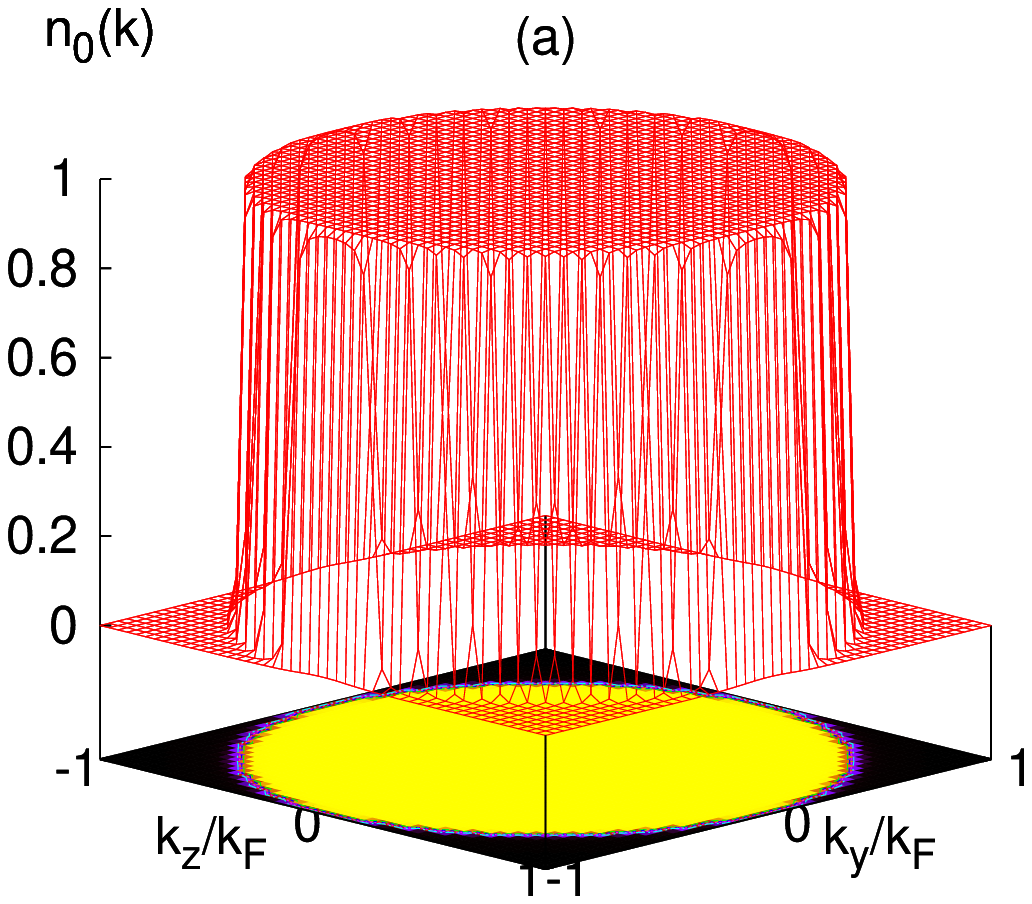}}}
\vskip 8mm
\centerline{\scalebox{0.61}{\includegraphics{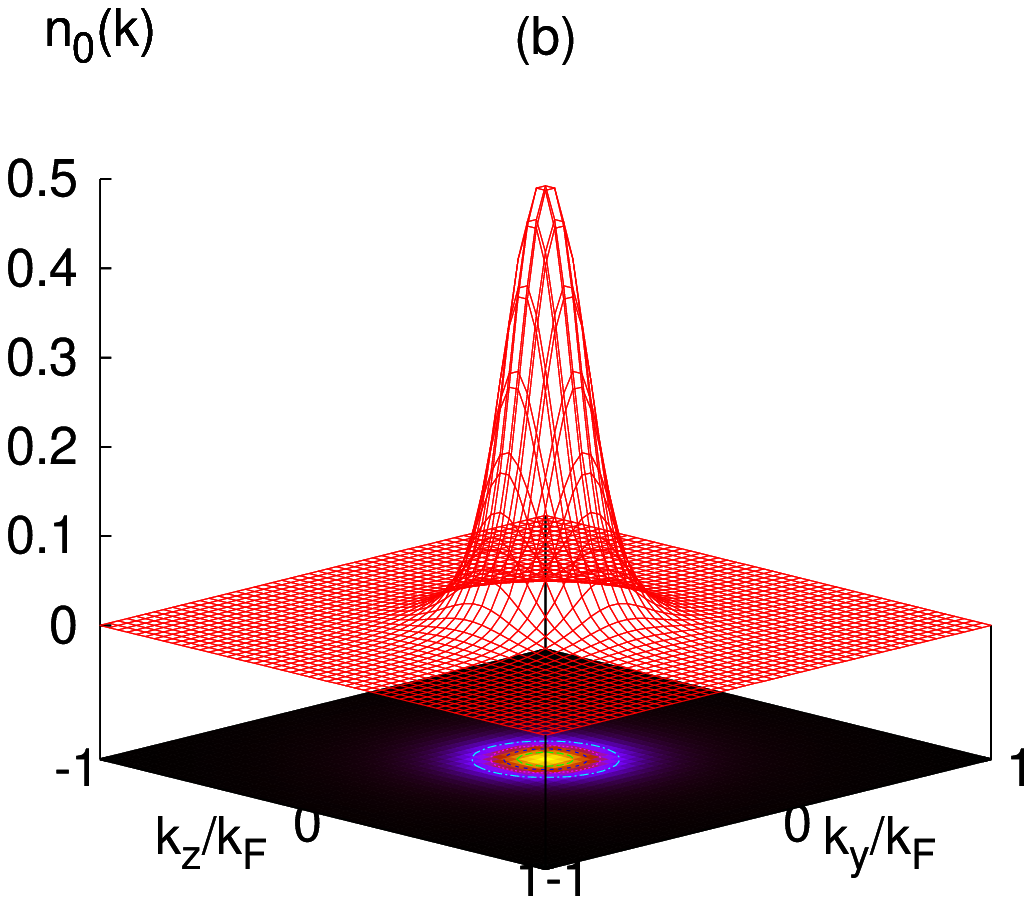}}}
\vskip 8mm
\centerline{\scalebox{0.61}{\includegraphics{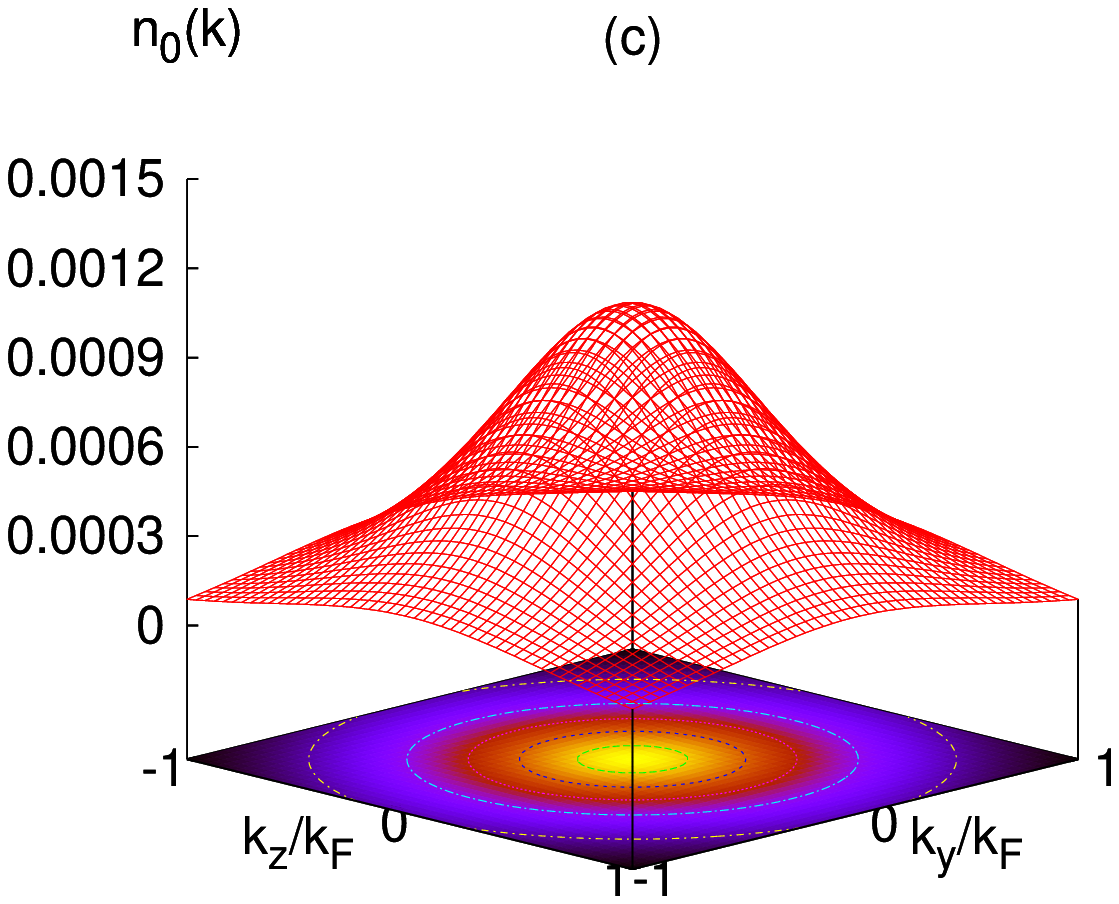}}}
\caption{\label{fig:swave.md} (Color online) 
Contour plots of momentum distribution $n_0(k_x = 0, k_y, k_z)$ when
a) $\mu_0 > 0$ (BCS side) for $1/(k_{\rm F} a_0) = -1$,
b) $\mu_0 = 0$ (intermediate regime) for $1/(k_{\rm F} a_0) \approx 0.55$, and
c) $\mu_0 < 0$ (BEC side) for $1/(k_{\rm F} a_0) = 1$
versus momentum $k_y/k_{\rm F}$ and $k_z/k_{\rm F}$.
}
\end{figure}

In Fig.~\ref{fig:swave.md}, we show $n_0(k_x = 0, k_y,k_z)$ 
for an $s$-wave ($\ell = 0, m_\ell = 0$) superfluid when 
a) $\mu_0 > 0$ (BCS side) for $1/(k_{\rm F} a_0) = -1$,
b) $\mu_0 = 0$ (intermediate regime) for $1/(k_{\rm F} a_0) \approx 0.55$, and
c) $\mu_0 < 0$ (BEC side) for $1/(k_{\rm F} a_0) = 1$.
As the interaction increases the Fermi sea with locus $\xi_0 ({\mathbf k}) = 0$ 
is suppressed, and pairs of atoms with opposite momenta become 
more tightly bound.
As a result, $n_0(\mathbf{k})$ broadens in the BEC side
since fermions with larger momentum participate 
in the formation of bound states. Notice that, the evolution is a
crossover without any qualitative change.
Furthermore, $n_0(k_x,k_y = 0,k_z)$ and $n_0(k_x,k_y,k_z = 0)$ can be
trivially obtained from $n_0(k_x = 0,k_y,k_z)$, since
$n_0(k_x,k_y,k_z)$ is symmetric in $k_x, k_y$ and $k_z$.

\begin{figure} [htb]
\centerline{\scalebox{0.61}{\includegraphics{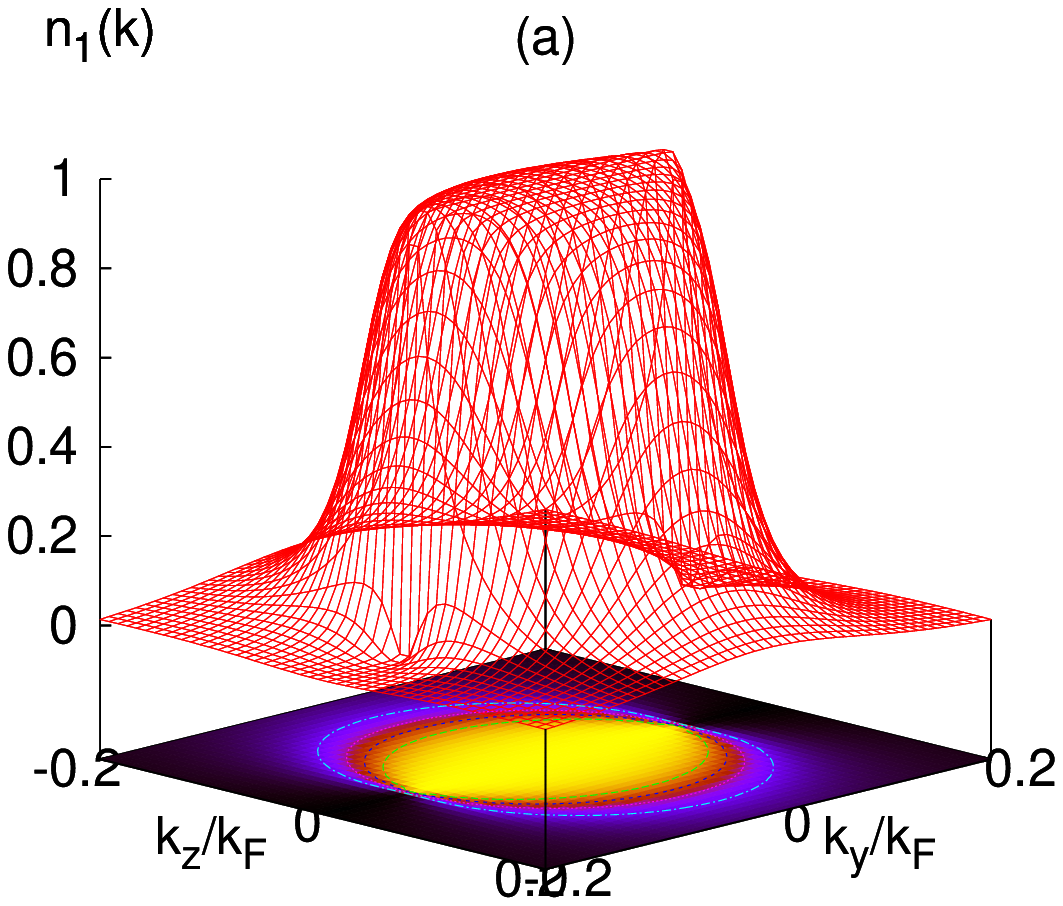}}}
\vskip 8mm
\centerline{\scalebox{0.61}{\includegraphics{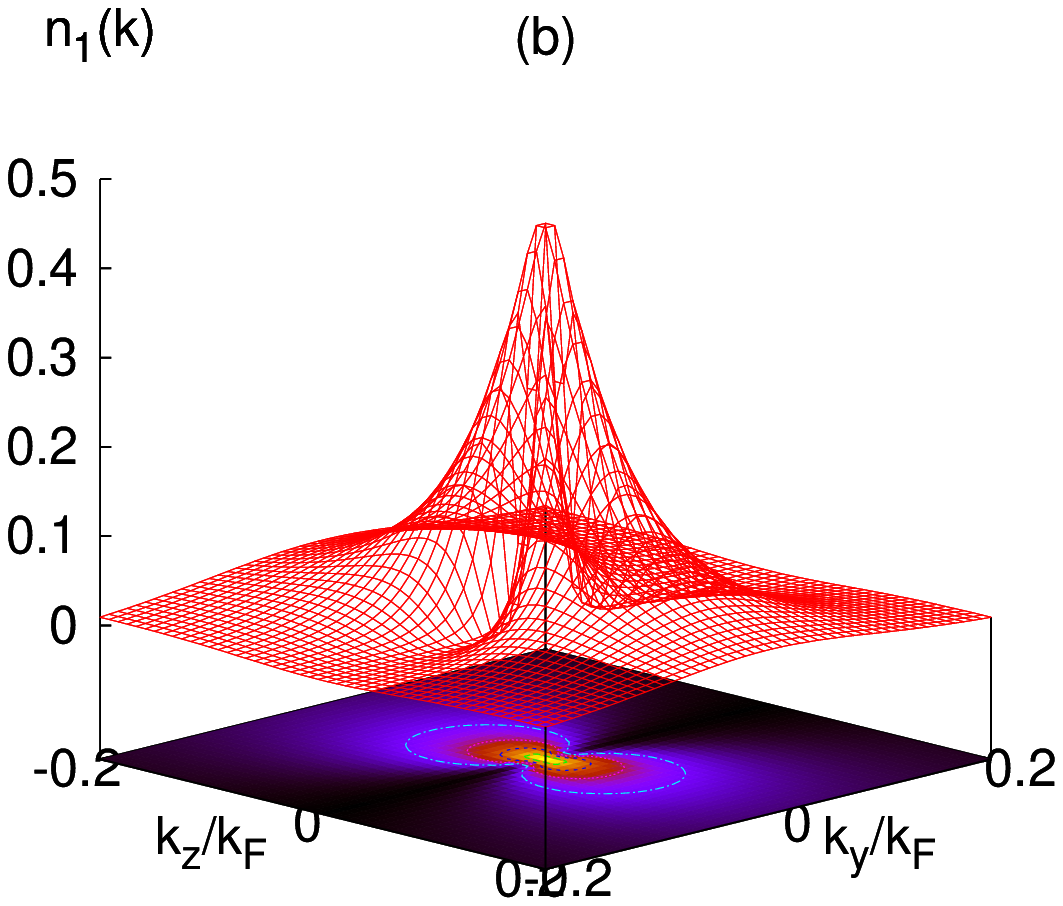}}}
\vskip 8mm
\centerline{\scalebox{0.61}{\includegraphics{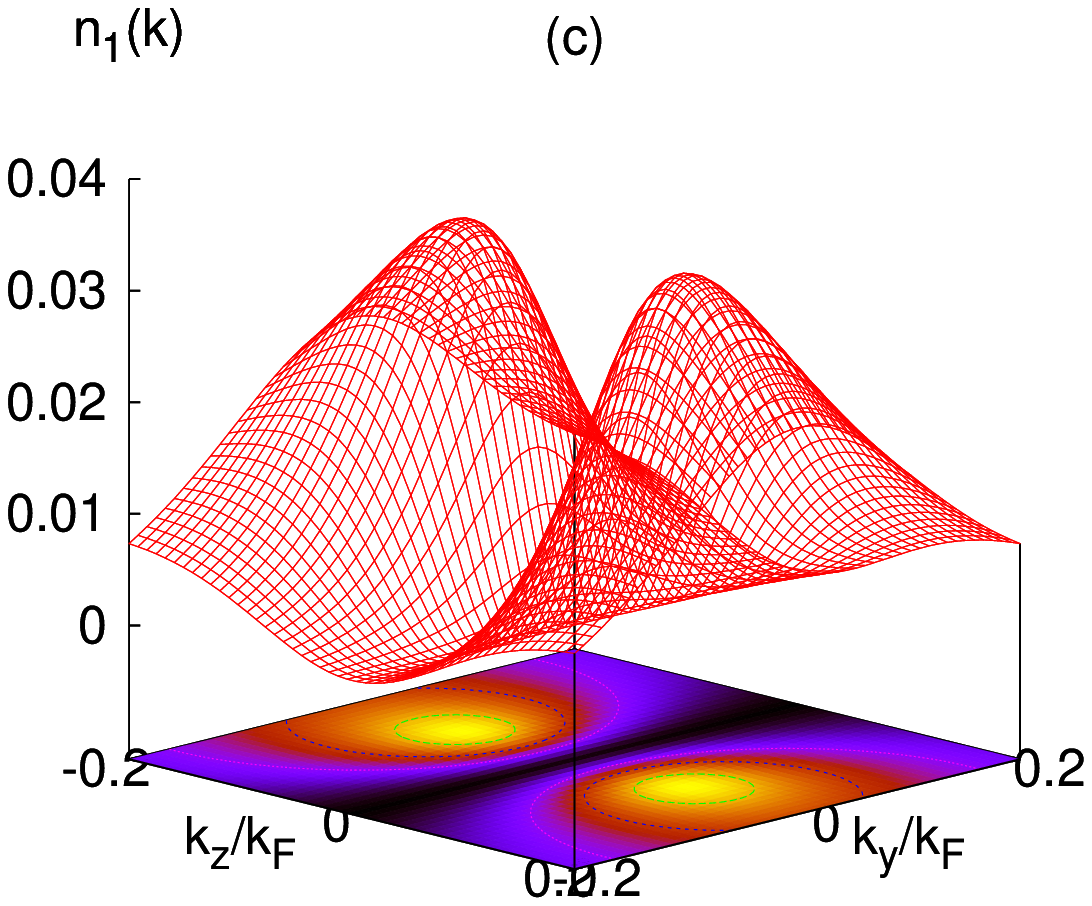}}}
\caption{\label{fig:pwave.md} (Color online) 
Contour plots of momentum distribution $n_1(k_x = 0, k_y, k_z)$ in 
a) $\mu_1 > 0$ (BCS side) for $1/(k_{\rm F}^3 a_1) = -1$,
b) $\mu_1 = 0$ (intermediate regime) for $1/(k_{\rm F}^3 a_1) \approx 0.48$, and
c) $\mu_1 < 0$ (BEC side) for $1/(k_{\rm F}^3 a_1) = 1$
versus momentum $k_y/k_{\rm F}$ and $k_z/k_{\rm F}$.
}
\end{figure}

In Fig.~\ref{fig:pwave.md}, we show $n_1 (k_x = 0, k_y,k_z)$ 
for a $p$-wave ($\ell = 1, m_\ell = 0$) superfluid when
a) $\mu_1 > 0$ (BCS side) for $1/(k_{\rm F}^3a_1) = -1$,
b) $\mu_1 = 0$ (intermediate regime) for $1/(k_{\rm F}^3a_1) \approx 0.48$, and 
c) $\mu_1 < 0$ (BEC side) for $1/(k_{\rm F}^3a_1) = 1$.
Notice that $n_1 (k_x = 0, k_y,k_z)$ is largest in the BCS side
when $k_z/k_{\rm F} = 0$, but it vanishes along $k_z/k_{\rm F} = 0$ for any $k_y/k_{\rm F}$ 
in the BEC side. As the interaction increases the Fermi sea with locus $\xi_1 ({\mathbf k}) = 0$ 
is suppressed, and pairs of atoms with opposite momenta become 
more tightly bound. As a result, the large momentum distribution in the vicinity of
$\mathbf{k} = \mathbf{0}$ splits into two peaks around finite ${\mathbf k}$ 
reflecting the $p$-wave symmetry of these tightly bound states. 
Furthermore, $n_1 (k_x, k_y, k_z = 0) = [1 - {\rm sgn} ( \xi_1 ({\mathbf k}) )]/2$ for any $\mu_1$,
and $n_1 (k_x, k_y = 0, k_z)$ is trivially obtained from 
$n_1 (k_x = 0, k_y,k_z)$, since $n_1(\mathbf{k})$ is symmetric in $k_x, k_y$.
Here, ${\rm sgn}$ is the Sign function.

Thus, $n_1(\mathbf{k})$ for the $p$-wave case has a major rearrangement 
in $\mathbf{k}$-space with increasing interaction,
in sharp contrast to $s$-wave.
This qualitative difference between $p$-wave and $s$-wave symmetries
around $\mathbf{k} = \mathbf{0}$ explicitly shows a direct measurable
consequence of the gapless to gapped quantum phase transition when $\mu_1 = 0$,
since $n_1(\mathbf{k})$ depends explicitly on $E_1({\mathbf k})$.
These quantum phase transitions are present in all nonzero angular momentum states,
and can be further characterized through the atomic compressibility
as discussed in the next section.

\subsection{Atomic Compressibility}
\label{sec:ac}

At finite temperatures, the isothermal atomic compressibility is defined by
$\kappa_\ell^T (T) = -(\partial {\cal V} / \partial {\cal P})_{T, N_{\ell}}/{\cal V}$ 
where ${\cal V}$ is the volume and ${\cal P}$ is the pressure of the gas.
This can be rewritten as
\begin{equation}
\kappa_\ell^T (T)=-\frac{1}{N_\ell^2} \left( \frac{\partial^2\Omega_\ell} {\partial \mu_\ell^2} \right)_{T,{\cal V}}
=\frac{1}{N_\ell^2} \left( \frac{\partial N_\ell}{\partial \mu_\ell} \right)_{T,{\cal V}},
\end{equation}
where the partial derivative 
$
\partial N_\ell / \partial \mu_\ell$ at $T \approx 0
$ 
is given by
\begin{equation}
\label{eqn:dndmu}
\frac{\partial N_\ell}{\partial \mu_\ell} \approx \frac{\partial N_\ell^{\rm{sp}}}{\partial \mu_\ell} = 
\sum_{\mathbf{k},s} \frac{|\Delta_\ell(\mathbf{k})|^2}{2E_\ell^3(\mathbf{k})}.
\end{equation}

The expression above leads to 
$
\kappa_0^T(0) = 2N(\epsilon_{\rm F})/N_0^2
$ 
in weak coupling BCS and 
$
\kappa_0^T(0) = 2 N(\epsilon_{\rm F}) \epsilon_{\rm F}/(3|\mu_0|N_0^2)
$
in strong coupling BEC limit for $\ell = 0$, where 
$
N(\epsilon_{\rm F}) = M {\cal V} k_{\rm F}/(2\pi^2)
$
is the density of states per spin at the Fermi energy.
Notice that $\kappa_0^T(0)$ decreases as $a_0^2$ in strong coupling since 
$|\mu_0| = 1/(2M a_0^2)$.
However, we only present the strong coupling results
for higher angular momentum states since they exhibit
an interesting dependence on $a_\ell$ and $k_0$.
In the case of THS pseudo-spin singlet, we obtain
$
\kappa_{\ell > 1}^T(0) = 4 N(\epsilon_{\rm F}) \epsilon_{\rm F} \bar{\phi}_\ell/( \epsilon_0 \phi_\ell N_\ell^2)
$
for $\ell > 1$, while in the case of SHS states we obtain
$
\kappa_1^T(0) = N(\epsilon_{\rm F}) \epsilon_{\rm F}/(\sqrt{\epsilon_0|\mu_1|} N_\ell^2)
$
for $\ell = 1$ and
$
\kappa_{\ell > 1}^T(0) = 2 N(\epsilon_{\rm F}) \epsilon_{\rm F} \bar{\phi}_\ell/( \epsilon_0 \phi_\ell N_\ell^2)
$
for $\ell > 1$.
Here
$
\phi_\ell = \Gamma(\ell - 1/2) / \Gamma(\ell + 1)
$
and
$
\bar{\phi}_\ell = \Gamma(\ell - 3/2) / \Gamma(\ell + 1),
$
where $\Gamma(x)$ is the Gamma function.
Notice that $\kappa_1^T(0)$ decreases as $\sqrt{a_1}$ 
for $\ell = 1$ since $|\mu_1| = 1(Mk_0a_1)$ and $\kappa_{\ell > 1}^T(0)$ is a 
constant for $\ell > 1$ in strong coupling.

\begin{figure} [htb]
\centerline{\scalebox{0.67}{\includegraphics{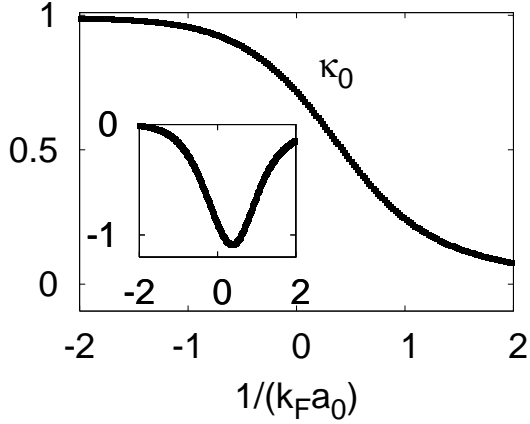} }}
\caption{\label{fig:swave.ac} Plot of reduced isothermal atomic compressibility
$\kappa_{\rm r} = \kappa_0^T (0) / \widetilde{\kappa}_0$ versus interaction strength $1/(k_{\rm F} a_0)$
for $k_0 \approx 200k_{\rm F}$.
The inset shows the numerical derivative of $d \kappa_{\rm r} / d[(k_{\rm F}a_0)^{-1}]$ 
versus $1/(k_{\rm F} a_0)$. Here, $\widetilde{\kappa}_0$ is the weak coupling compressibility.
}
\end{figure}

In Fig.~{\ref{fig:swave.ac}}, we show the evolution of 
$\kappa_0^T (0)$ for a $s$-wave $(\ell = 0, m_\ell = 0)$ superfluid
from the BCS to the BEC regime. 
$\kappa_0^T (0)$ decreases continuously, and thus
the evolution is a crossover (smooth) as can be seen in the inset
where the numerical derivative of $\kappa_0^T(0)$ with respect to
$1/(k_{\rm F} a_0)$ is shown $\{ d\kappa_0^T(0) / d[(k_{\rm F}a_0)^{-1}] \}$. 
This decrease is associated with the increase of the gap of 
the excitation spectrum as a function of $1/(k_{\rm F} a_0)$.
In this approximation, the gas is incompressible [$\kappa_0^T(0) \to 0$] in the extreme BEC limit.

\begin{figure} [htb]
\centerline{\scalebox{0.67}{\includegraphics{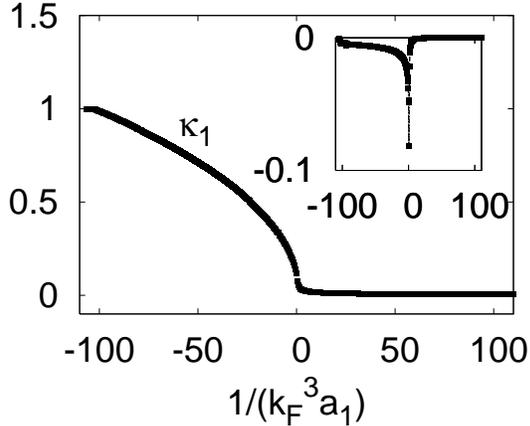} }}
\caption{\label{fig:pwave.ac} Plot of reduced isothermal atomic compressibility
$\kappa_{\rm r} = \kappa_1^T (0) / \widetilde{\kappa}_1$ versus interaction strength $1/(k_{\rm F}^3 a_1)$
for $k_0 \approx 200k_{\rm F}$. 
The inset shows the numerical derivative of $d \kappa_{\rm r} / d[(k_{\rm F}^3 a_1)^{-1}]$ 
versus $1/(k_{\rm F}^3 a_1)$. Here $\widetilde{\kappa}_1$ is the weak coupling compressibility.
}
\end{figure}

In Fig.~{\ref{fig:pwave.ac}}, we show the evolution of 
$\kappa_1^T (0)$ for a $p$-wave $(\ell = 1, m_\ell = 0)$ superfluid 
from the BCS to the BEC regime.
Notice that, there is a change in qualitative behavior when $\mu_1 = 0$ at $1/(k_{\rm F}^3 a_1) \approx 0.48$
as can be seen in the inset where the numerical derivative of $\kappa_1^T(0)$ with respect to
$1/(k_{\rm F}^3 a_1)$ is shown $\{ d\kappa_1^T(0) / d[(k_{\rm F}^3 a_1)^{-1}] \}$.
Thus, the evolution from BCS to BEC is not a crossover, but a quantum phase 
transition occurs when $\mu_1 = 0$~\cite{botelho1,botelho-pwave,iskin-lattice,gurarie}.

The non-analytic behavior occurring when $\mu_{\ell \ne 0} = 0$ 
can be understood from higher derivatives of $\kappa_\ell$ with respect to $\mu_\ell$
\begin{equation}
\left[ \frac{\partial \kappa_\ell^T(T)} {\partial \mu_\ell} \right]_{T,{\cal V}}
= - 2N_\ell [\kappa_\ell^T(T)]^2 
+ \frac{1}{N_\ell^2} \left( \frac{\partial^2 N_\ell} {\partial \mu_\ell^2} \right)_{T,{\cal V}}.
\end{equation}
For instance, the second derivative
$
\partial^2 N_\ell^{\rm sp} / \partial \mu_\ell^2 = 
3\sum_{\mathbf{k},s} |\Delta_\ell(\mathbf{k})|^2 \xi_\ell(\mathbf{k})/[2E_\ell^5(\mathbf{k})]
$
tends to zero in the weak ($\mu_\ell \approx \epsilon_{\rm F} > 0$) and 
strong ($\mu_\ell \approx E_{{\rm b},\ell}/2 < 0 $) coupling limits.
On the other hand, when $\mu_\ell = 0$, 
$\partial^2 N_\ell^{\rm sp} / \partial \mu_\ell^2$ is finite only for $\ell = 0$, 
and it diverges for $\ell \ne 0$.
This divergence is logarithmic for $\ell = 1$, and of higher order
for $\ell > 1$. Thus, we conclude again that higher derivatives of
$N_\ell^{\rm sp}$ are nonanalytic when $\mu_{\ell \ne 0} = 0$,
and that a quantum phase transition occurs for $\ell \ne 0$.

Theoretically, the calculation of the isothermal atomic compressibility 
$\kappa_\ell^T (T)$ is easier than the isentropic atomic compressibility $\kappa_\ell^{\cal S} (T)$.
However, performing measurements of $\kappa_\ell^{\cal S} (T)$ may be simpler in cold
Fermi gases, since the gas expansion upon release from the trap is expected to be 
nearly isentropic. Fortunately, $\kappa_\ell^{\cal S} (T)$ is related to $\kappa_\ell^T (T)$ via the 
thermodynamic relation
\begin{equation}
\kappa_\ell^{\cal S} (T) = \frac{C_\ell^{\cal V} (T)}{C_\ell^{\cal P} (T)} \kappa_\ell^T (T),
\end{equation}
where $\kappa_\ell^T (T) > \kappa_\ell^{\cal S} (T)$ since
specific heat capacitites $C_\ell^{\cal P}(T) > C_\ell^{\cal V} (T)$.
Furthermore, at low temperatures ($T \approx 0$)
the ratio $C_\ell^{\cal P} (T) / C_\ell^{\cal V} (T) \approx \rm{const}$,
and therefore, $\kappa_\ell^S (T \approx 0) \propto \kappa_\ell^T (T \approx 0)$. 
Thus, we expect qualitatively similar behavior
in both the isentropic and isothermal compressibilities at low temperatures $(T \approx 0)$.

The measurement of the atomic compressibility could also be performed via
an analysis of particle density fluctuations~\cite{jin-correlations, ealtman}. 
As it is well know from thermodynamics~\cite{reif},
$\kappa_\ell^T(T)$ is connected to density fluctuations via the relation
\begin{eqnarray}
\langle n_\ell^2 \rangle -\langle n_\ell \rangle^2 = T \langle n_\ell \rangle^2 \kappa_\ell^T (T),
\end{eqnarray}
where $\langle n_\ell \rangle$ is the average density of atoms. 
From the measurement of density fluctuations $\kappa_\ell^{T}(T)$ 
can be extracted at any temperature $T$.

It is important to emphasize that in 
this quantum phase transition at $\mu_{\ell \ne 0} = 0$, the symmetry of the order parameter 
does not change as is typical in the Landau classification of phase transitions.
However, a clear thermodynamic signature occurs in derivatives of
the compressibility suggesting that the phase transition is higher than second order
according to Ehrenfest's classification. Therefore, if the symmetry of
the order parameter does not change when $\mu_\ell$ changes sign,
what is changing? To address this question, the topology of momentum space
is discussed next.

\subsection{Topological quantum phase transitions}
\label{sec:qpt}

In what follows, we discuss the role of momentum space 
topology~\cite{volovik-book,duncan,volovik} in the non-analytic 
behavior of the thermodynamic potential, when $\mu_{\ell \ne 0} = 0$.
To investigate the role of topology, we make an immediate connection 
to the Lifshitz transition~\cite{lifshitz} in the context of 
ordinary metals at $T = 0$ and high pressure. 
In the conventional Lifshitz transition, the Fermi surface 
$\epsilon (\mathbf{k}, {\cal P}) = \epsilon_{\rm F}$ changes its topology as the pressure
${\cal P}$ is changed. For an isotropic pressure ${\cal P}$, the deviation 
$\Delta {\cal P} = {\cal P} - {\cal P}_{\rm c}$ from the critical pressure 
${\cal P}_{\rm c}$ is proportional to $\Delta \mu = \mu - \mu_{\rm c}$
where $\mu_{\rm c}$ is the critical chemical potential at the transition point. 
A typical example of the Lifshitz transition
is the disruption of a neck of the Fermi surface which leads to 
a non-analytic behavior of the number of states ${\cal N} (\mu)$ 
inside the Fermi surface. In this case, ${\cal N} (\mu)$ behaves as
$
{\cal A} (\mu_{\rm c}) + {\cal B} |\mu - \mu_{\rm c}|^{3/2}
$ 
for $\mu < \mu_{\rm c}$, 
and as ${\cal A}(\mu_{\rm c})$ for $\mu > \mu_{\rm c}$, 
in the vicinity of $\mu_{\rm c}$. 
Here, ${\cal K} = (3/2) {\cal B} |\mu - \mu_{\rm c}|^{1/2}/n_{\rm c}^2$
is the electronic compressibility, where $n_{\rm c} = N_{\rm c}/{\cal V}$ 
is the critical density of electrons at the transition point.
Notice that ${\cal K}$ is nonanalytic, but it is not singular.
The quantity that signals a phase transition in this case is not 
${\cal K}$, but the thermopower ${\cal Q}$, which is proportional 
to $-\partial \ln (n^2{\cal K})/\partial \mu$,
thus leading to ${\cal Q} \propto - |\Delta \mu|^{-1/2}$. 
In the conventional Lifshitz transition,
the system lowers its energy by 
$
\Delta {\cal E} \propto  - |\Delta \mu|^{5/2} \propto  - |\Delta {\cal P}|^{5/2},
$
and the transition is said to be of second and half 
order~\cite{abrikosov}.

The topological transition proposed here is analogous to the
Lifshitz transition in the sense that the surface in momentum
space corresponding to $E_{\ell \ne 0}(\mathbf{k}) = E_{\ell \ne 0} (\mathbf{k},\mu_\ell) = 0$
changes from a well defined set of $\mathbf{k}$ points for $\mu_{\ell \ne 0} > 0$
to a null set for $\mu_{\ell \ne 0} < 0$. Here, $E_{\ell \ne 0} (\mathbf{k},\mu_\ell)$
plays the role of $\epsilon(\mathbf{k},{\cal P})$ and $\mu_{\ell \ne 0} = \mu_{\rm c} = 0$
plays the role of the critical pressure ${\cal P}_{\rm c}$.

For the Lifshitz transition in $\ell \ne 0$ superfluids, there is
a non-analytic behavior in $\partial^2 N_{\ell \ne 0} /\partial \mu_\ell^2$, 
and thus in $\partial \kappa_{\ell \ne 0}^T(0) / \partial \mu_\ell$.
This behavior in $\kappa_{\ell \ne 0}^T(0)$ is due to the collapse of 
all order parameter nodes at ${\mathbf{k} = 0}$, which
produce a gap in the excitation spectrum $E_{\ell \ne 0}(\mathbf{k})$ 
and a massive discontinuous rearrangement of the momentum distribution 
$n_{\ell \ne 0}(\mathbf{k})$ in the ground state as $\mu_{\ell \ne 0} \to \mu_{\ell \ne 0}^{\rm c} = 0$.
A direct topological analogy with the standard Lifshitz transition can be 
made by noticing the collapse of locus of zero quasiparticle excitation energy 
at $\mu_{\ell \ne 0} = \mu_{\ell \ne 0}^{\rm c}$
in the excitation spectrum of the system.
Generalized topological invariants can be invented along the lines
of Ref.~\cite{volovik-book,duncan,volovik}, however, we do not 
discuss them here. Instead, we analyze next the phase diagram at zero temperature.

\subsection{Phase diagram}
\label{sec:pd}

To have a full picture of the evolution from the BCS to the BEC limit at $T = 0$,
it is important to analyze thermodynamic quantities at low temperatures.
In particular, it is important to determine the quantum critical region (QCR)
where a qualitative change occurs in quantities such as the specific heat, compressibility
and spin susceptibility. Here, we do not discuss in detail the QCR, but we analyze
the contributions from quasiparticle excitations to thermodynamic properties.
However, the discussion can be extended to include collective excitations~\cite{duncan}
(see Sec.~\ref{sec:gaussian.0}).

Next, we point out a major difference between $\ell = 0$ and $\ell \ne 0$ states
in connection with the spectrum of the quasiparticle excitations 
(see Sec.~\ref{sec:quasiparticle}) and their contribution to low temperature thermodynamics.

\begin{figure} [htb]
\centerline{\scalebox{0.5}{\includegraphics{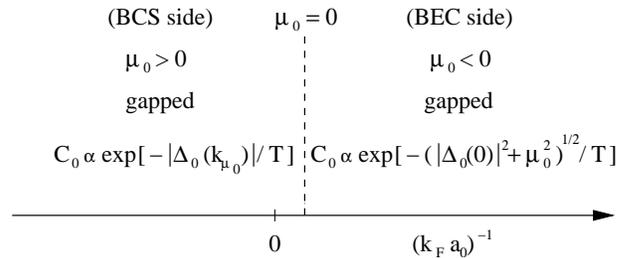} }}
\caption{\label{fig:swave.phase} The phase diagram of $s$-wave superfluids as a function of $1/(k_{\rm F} a_0)$.
}
\end{figure}

For $\ell = 0$, quasiparticle excitations are gapped for all couplings,
and therefore, thermodynamic quantitites such as atomic compressibility, 
specific heat and spin susceptibility have an exponential dependence 
on the temperature and the minimum energy of quasiparticle excitations 
$\sim \exp[-\min \{ E_0 (\mathbf{k}) \}/T]$.
Using Eqs.~(\ref{eqn:swave.qp.bcs}) and~(\ref{eqn:swave.qp.bec})
leads to $\sim \exp[ - |\Delta_0(k_{\mu_0})|/T]$ in the BCS side ($\mu_0 > 0$) and 
$\sim \exp[-\sqrt{|\Delta_0 (0)|^2 + \mu_0^2}/T]$ in the BEC side ($\mu_0 < 0$)
as shown in Fig.~\ref{fig:swave.phase}, where $k_{\mu_\ell}  = \sqrt{2M \mu_\ell}$.
Notice that, there is no qualitative change across $\mu_0 = 0$ at small but finite 
temperatures. This indicates the absence of a QCR and confirms there is only a crossover
for $s$-wave ($\ell = 0$) superfluids at $T = 0$.

\begin{figure} [htb]
\centerline{\scalebox{0.5}{\includegraphics{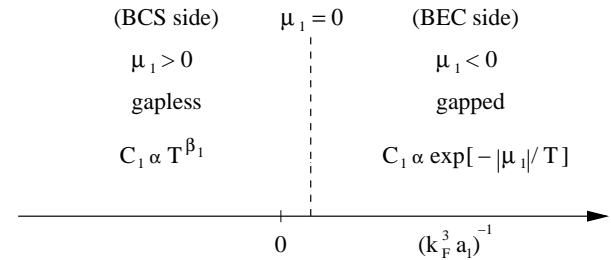} }}
\caption{\label{fig:pwave.phase} The phase diagram of $p$-wave superfluids as a function of $1/(k_{\rm F}^3 a_1)$.
}
\end{figure}

For $\ell \ne 0$, quasiparticle excitations are gapless in the BCS side and are only 
gapped in the BEC side, and therefore, while thermodynamic quantitites such as 
atomic compressibility, specific heat and spin susceptibility have 
power law dependences on the temperature $\sim T^{\beta_{\ell \ne 0}}$ in the BCS side, 
they have exponential dependences on the temperature and the minimum energy of 
quasiparticle excitations $\sim \exp[-\min \{ E_{\ell \ne 0} (\mathbf{k}) \}/T]$ in the BEC side.
Here, $\beta_{\ell \ne 0}$ is a real number which depends on particular $\ell$ state.
For $\ell = 1$, using Eqs.~(\ref{eqn:pwave.qp.bcs}) and~(\ref{eqn:pwave.qp.bec})
leads to $\sim T^{\beta_1}$ in the BCS side ($\mu_1 > 0$) and
$\sim \exp(-|\mu_1|/T)$ in the BEC side ($\mu_1 < 0$)
as shown in Fig.~\ref{fig:pwave.phase}.
Notice the change in qualitative behavior across $\mu_1 = 0$ (as well as other $\ell \ne 0$ states) 
at small but finite temperatures. This change occurs within the QCR and signals the existence
of a quantum phase transition ($T = 0$) for $\ell \ne 0$ superfluids.

\subsection{Thermodynamic Potential}
\label{sec:gse}

Now, we discuss the thermodynamic potential $\Omega_{\ell}$ at $T = 0$ in the 
asmyptotic BCS and BEC limits. In the weak coupling limit, we obtain 
\begin{equation}
\Omega_{\ell}^{\rm sp} = - \frac{2}{5} N_\ell \epsilon_{\rm F},
\end{equation}
which is identical to the full thermodynamic potential $\Omega_{\ell}$.
This indicates that $\Omega_{\ell}^{\rm fluct}$ is negligible in the
BCS limit.

However, in the strong coupling limit, we obtain
\begin{equation}
\Omega_{\ell}^{\rm sp} = - \frac{1}{2} N_\ell (2\mu_\ell - E_{{\rm b}, \ell}).
\end{equation}
Notice that, $\mu_{{\rm B},\ell} = 2\mu_\ell - E_{{\rm b},\ell} > 0$ 
is the Bosonic chemical potential and $N_{{\rm B},\ell} = N_{\ell}/2$ is the number of bosons.
To evaluate $\mu_{{\rm B},\ell}$, it is necessary to find the first nonvanishing correction 
for $2\mu_\ell - E_{{\rm b},\ell}$. In the specific case of $\ell = 0$, we obtain
$
\mu_{{\rm B}, 0} = 4\epsilon_{\rm F} k_{\rm F}a_0/(3\pi) = 4\pi a_{{\rm B},0}/M_{{\rm B},0}
$
for the chemical potential and 
$
\Omega_{\ell}^{\rm sp} = -\pi N_0^2a_0/(M{\cal V}) = - N_{{\rm B},0} \mu_{{\rm B},0}
$
for the thermodynamic potential of the pairs.
Here $a_{\rm B,0} = 2a_0$ and $M_{\rm B,0} = 2M$ is the scattering length and mass of the
corresponding bosons.
A better estimate for $a_{{\rm B},0} \approx 0.6a_0$ 
can be found in the literature~\cite{pieri-a,gora,kagan,gurarie-a}.
The main reason for this difference is that our theory does not include 
possible intermediate (virtual) scattering processes which renormalize $a_{{\rm B},0}$.
This is also the case when we analyze the collective modes in Sec.~\ref{sec:cm-sc} 
and the TDGL equation in Sec.~\ref{sec:tdgl.sc}.

Using $\mu_\ell = (\mu_{{\rm B},\ell} + E_{{\rm b},\ell})/2$ and 
the thermodynamic relation 
$\mu_\ell = (\partial E_\ell/\partial N_\ell)_{\cal V}$,
where $E_\ell$ is the ground state energy, we obtain 
\begin{equation}
\Omega_{\ell} = - \frac{1}{2} N_{{\rm B},\ell} \mu_{{\rm B},\ell}.
\end{equation}
Notice that this expression is identical to the thermodynamic potential of bosons
$
\Omega_{{\rm B},\ell} = E_{{\rm B},\ell} - N_{{\rm B},\ell}\mu_{{\rm B},\ell},
$
where $E_{{\rm B},\ell}$ is the ground state energy.
Therefore, the fermionic thermodynamic potential in the strong coupling limit
should lead to the thermodynamic potential of real bosons 
($\Omega_\ell \equiv \Omega_{{\rm B},\ell}$). Since
$
\Omega_{\ell} = \Omega_{\ell}^{\rm sp} + \Omega_{\ell}^{\rm fluct},
$
we conclude from this thermodynamic argument that
\begin{equation}
\Omega_{\ell}^{\rm fluct} = \frac{1}{2}N_{{\rm B},\ell} \mu_{{\rm B},\ell}
\end{equation}
in the strong coupling limit.
Therefore,
$
E_\ell - N_{\ell} E_{{\rm b},\ell}/2 \equiv E_{{\rm B},\ell},
$
or
$
E_\ell/N_\ell - \mu_\ell \equiv (E_{{\rm B},\ell}/N_{{\rm B},\ell} - \mu_{{\rm B},\ell})/2
$ 
which is consistent with quantum monte carlo calculations~\cite{giorgini}.

The fluctuation contribution to $\Omega_{\ell}^{\rm fluct}$ comes from the 
zero point energy of the collective excitations, which is discussed next.

\section{Gaussian Fluctuations}
\label{sec:gaussian.0}

The pole structure of $\mathbf{F}_\ell (\mathbf{q}, iv_j)$
determines the two-particle excitation spectrum of 
the superconducting state with $iv_j \to w + i0^+$, and
has to be taken into account to derive $\Omega_\ell^{\rm fluct}$.
The matrix elements of $\mathbf{F}_\ell (\mathbf{q}, iv_j)$ are
$\mathbf{F}_{\ell,m_\ell,m_\ell'} (\mathbf{q}, iv_j)$ for a given $\ell$.
We focus here only on the zero temperature limit and analyse
the collective phase modes. In this limit, we separate the diagonal matrix 
elements of $\mathbf{F}_{\ell,m_\ell,m_\ell'}^{-1}(q)$ into even and odd contributions 
with respect to $iv_j$
\begin{eqnarray}
(\mathbf{F}_{\ell,m_\ell,m_\ell'}^{-1})_{11}^E
&=&
\sum_{\mathbf{k}} \frac{(\xi_+ \xi_- + E_+ E_-) (E_+ + E_-)} {2E_+ E_- [ (iv_j)^2 - (E_+ + E_-)^2]} \nonumber \\
&& \Gamma_{\ell}^2(k) Y_{\ell,m_\ell}(\widehat{\mathbf{k}}) Y_{\ell,m_\ell'}^*(\widehat{\mathbf{k}})
- \frac{\delta_{m_\ell,m_\ell'}{\cal V}}{4\pi \lambda_\ell}, \\
(\mathbf{F}_{\ell,m_\ell,m_\ell'}^{-1})_{11}^O
&=&
- \sum_{\mathbf{k}} \frac{(\xi_+ \xi_- + E_+ E_-)(iv_j)} {2E_+ E_- [ (iv_j)^2 - (E_+ + E_-)^2 ]}  \nonumber \\
&& \Gamma_{\ell}^2(k) Y_{\ell,m_\ell}(\widehat{\mathbf{k}}) Y_{\ell,m_\ell'}^*(\widehat{\mathbf{k}}).
\end{eqnarray}
The off-diagonal term is even in $iv_j$, and it reduces to
\begin{eqnarray}
(\mathbf{F}_{\ell,m_\ell,m_\ell'}^{-1})_{12}
&=&
- \sum_{\mathbf{k}}\frac{\Delta_+ \Delta_- (E_+ + E_-)} {2E_+ E_- [(iv_j)^2 - (E_+ + E_-)^2]} \nonumber \\
&& \Gamma_{\ell}^2(k) Y_{\ell,m_\ell}(\widehat{\mathbf{k}}) Y_{\ell,m_\ell'}^*(\widehat{\mathbf{k}}).
\end{eqnarray}
Here the labels $\pm$ denote that the corresponding variables are functions of $\mathbf{k} \pm \mathbf{q}/2$.

In order to obtain the collective mode spectrum,
we express $\Lambda_{\ell,m_\ell}(q) = \tau_{\ell,m_\ell}(q) e^{i\vartheta_{\ell, m_\ell}(q)} = 
[ \rho_{\ell,m_\ell}(q) +i \theta_{\ell,m_\ell}(q) ] / \sqrt{2}$
where $\tau_{\ell,m_\ell}(q)$, $\vartheta_{\ell,m_\ell}(q)$, $\rho_{\ell,m_\ell}(q)$ 
and $\theta_{\ell,m_\ell}(q)$ are all real. Notice that the new fields 
$
\rho_{\ell,m_\ell} (q) = \tau_{\ell,m_\ell} (q) \cos [\vartheta_{\ell,m_\ell}(q)],
$ 
and 
$
\theta_{\ell,m_\ell} (q) = \tau_{\ell,m_\ell} (q) \sin [\vartheta_{\ell,m_\ell}(q)]
$ 
can be regarded essentially
as the amplitude and phase fields respectively, when $\vartheta_{\ell,m_\ell}(q)$ is small.
This change of basis can be described by the following unitary
transformation 
\begin{eqnarray}
\Lambda_{\ell,m_\ell} (q) = \frac{1}{\sqrt{2}}  \left( \begin{array}{cc} 
1 & i \\
1 &-i
\end{array}\right)   
\left[ \begin{array}{c} \rho_{\ell,m_\ell}(q) \\ \theta_{\ell,m_\ell}(q) \end{array} \right].
\end{eqnarray}
From now on, we take $\Delta_{\ell,m_\ell}$ as real without loss of generality.
The diagonal elements of the fluctuation matrix in the rotated basis are
$
(\mathbf{\widetilde{F}}_{\ell,m_\ell,m_\ell'}^{-1})_{11} =
(\mathbf{F}_{\ell,m_\ell,m_\ell'}^{-1})_{11}^E + (\mathbf{F}_{\ell,m_\ell,m_\ell'}^{-1})_{12},
$
and 
$
(\mathbf{\widetilde{F}}_{\ell,m_\ell,m_\ell'}^{-1})_{22} =
(\mathbf{F}_{\ell,m_\ell,m_\ell'}^{-1})_{11}^E - (\mathbf{F}_{\ell,m_\ell,m_\ell'}^{-1})_{12};
$
and the off-diagonal elements are
$
(\mathbf{\widetilde{F}}_{\ell,m_\ell,m_\ell'}^{-1})_{12} =
(\mathbf{\widetilde{F}}_{\ell,m_\ell,m_\ell'}^{-1})_{21}^* =
i(\mathbf{F}_{\ell,m_\ell,m_\ell'}^{-1})_{11}^O
$
with the $q$ dependence being implicit.

\subsection{Collective (Goldstone) Modes}
\label{sec:collective-modes}

The collective modes are determined by the poles of the propagator matrix 
$\mathbf{F}_\ell(q)$ for the pair fluctuation fields 
$\Lambda_{\ell,m_\ell}(q)$, which describe
the Gaussian deviations about the saddle point order parameter. 
The poles of $\mathbf{F}_\ell(q)$ are determined
by the condition $\det \mathbf{F}^{-1}_\ell (q) = 0$, which leads to 
$2(2\ell + 1)$ collective (amplitude and phase) modes, when the usual analytic continuation
$iv_j \rightarrow w + i0^+$ is performed.
Among them, there are $2\ell + 1$ amplitude modes which we do not discuss here.

The easiest way to get the phase collective modes is to integrate out the 
amplitude fields to obtain a phase-only effective action.
Notice that, for $\ell \ne 0$ channels at any temperature, and
for $\ell = 0$ channel at finite temperature, 
a well defined low frequency expansion is not possible for $\mu_\ell > 0$ 
due to Landau damping which causes the collective modes to decay into the two quasiparticle continuum.
A well defined expansion [collective mode dispersion $w$]
must satisfy the following condition
$
w \ll \min \{E_+ + E_- \}.
$
Thus, a zero temperature expansion is always possible when Landau damping is
subdominant (underdamped regime).
To obtain the long wavelength dispersions for the collective modes at $T = 0$, 
we expand the matrix elements of $\mathbf{\widetilde{F}}_{\ell,m_\ell,m_\ell'}^{-1}$ 
to second order in $|\mathbf{q}|$ and $w$ to get
\begin{eqnarray}
\label{eqn:matrix-elements-second}
(\mathbf{\widetilde{F}}_{\ell,m_\ell,m_\ell'}^{-1})_{11} &=& 
A_{\ell,m_\ell,m_\ell'} + \sum_{i,j} C_{\ell,m_\ell,m_\ell'}^{i,j} q_i q_j \nonumber \\
&-& D_{\ell,m_\ell,m_\ell'} w^2, \\
(\mathbf{\widetilde{F}}_{\ell,m_\ell,m_\ell'}^{-1})_{22}  
&=& P_{\ell,m_\ell,m_\ell'} + \sum_{i,j} Q_{\ell,m_\ell,m_\ell'}^{i,j} q_i q_j \nonumber \\
&-& R_{\ell,m_\ell,m_\ell'} w^2, \\
(\mathbf{\widetilde{F}}_{\ell,m_\ell,m_\ell'}^{-1})_{12} 
&=& iB_{\ell,m_\ell,m_\ell'} w.
\end{eqnarray}
The expressions for the expansion coefficients are given in App.~\ref{sec:appa}.

For $\ell = 0$, the coefficients $C_{0,0,0}^{i,j} = C_{0,0,0} \delta_{i,j}$ 
and $Q_{0,0,0}^{i,j} = Q_{0,0,0} \delta_{i,j}$ are diagonal and isotropic
in $(i,j)$, and $P_{0,0,0} = 0$ vanishes. Here, $\delta_{i,j}$ is the Kronecker delta.
Thus, the collective mode is the isotropic Goldstone mode with dispersion
\begin{eqnarray}
W_{0,0}(\mathbf{q}) &=& {\cal C}_{0,0} |\mathbf{q}|, \\
{\cal C}_{0,0} &=& \left( \frac{A_{0,0,0} Q_{0,0,0}}
{A_{0,0,0} R_{0,0,0} + B_{0,0,0}^2} \right)^\frac{1}{2},
\label{eqn:soundv.0}
\end{eqnarray}
where ${\cal C}_{0,0}$ is the speed of sound. 
Notice that the quasiparticle excitations are always fully gapped 
from weak to strong coupling, and thus, 
the Goldstone mode is not damped at $T = 0$ for all couplings.

For $\ell \ne 0$, the dispersion for collective modes is not easy to extract in general, 
and therefore, we consider the case when only one of the spherical harmonics
$Y_{\ell,m_\ell}(\widehat{\mathbf{k}})$ is dominant and characterizes the order parameter.
In this case, $P_{\ell,m_\ell,m_\ell}$ = 0 due to the order parameter equation, 
and the collective mode is the anisotropic Goldstone mode with dispersion
\begin{eqnarray}
W_{\ell \ne 0,m_\ell}(\mathbf{q}) &=& \left[ \sum_{i,j} ({\cal C}_{\ell,m_\ell}^{i,j})^2 q_i q_j \right]^\frac{1}{2}, \\
{\cal C}_{\ell \ne 0, m_\ell}^{i,j} &=& \left( \frac{A_{\ell,m_\ell,m_\ell} Q_{\ell,m_\ell,m_\ell}^{i,j}} 
{A_{\ell,m_\ell,m_\ell} R_{\ell,m_\ell,m_\ell} + B_{\ell,m_\ell,m_\ell}^2} \right)^\frac{1}{2}.
\label{eqn:soundv.l}
\end{eqnarray}
Notice that the speed of sound has a tensor structure and is anisotropic.
Furthermore, the quasiparticle excitations are gapless when $\mu_{\ell \ne 0} > 0$, 
and thus, the Goldstone mode is damped even at $T = 0$.
However, Landau damping is subdominant and the real 
part of the pole dominates for small momenta.
In addition, quasiparticle excitations are fully gapped
when $\mu_{\ell \ne 0} < 0$, and thus, the Goldstone mode is not damped.
Therefore, the pole contribution to $\Omega_{\ell \ne 0}^{\rm fluct}$
comes from the Goldstone mode for all couplings.
In addition, there is also a branch cut representing the continuum of two particle 
scattering states, but the contribution from the Goldstone mode dominates at sufficiently 
low temperatures.

It is also illustrative to analyze the eigenvectors of $\widetilde{\mathbf{F}}^{-1}_\ell(q)$
in the amplitude-phase representation corresponding to small
$W_{\ell,m_\ell}(\mathbf{q})$ mode
\begin{equation}
\left[ \begin{array}{c} \rho_{\ell,m_\ell}(q)  \\ \theta_{\ell,m_\ell}(q) \end{array} \right]
=
\left[ \begin{array}{c} -i \frac{B_{\ell,m_\ell,m_\ell}}{A_{\ell,m_\ell,m_\ell}}
W_{\ell,m_\ell}(\mathbf{q})  \\ 1 \end{array} \right].
\end{equation}
Notice that, when $B_{\ell,m_\ell,m_\ell} \to 0$ the amplitude and phase modes are not mixed.

Next, we discuss the dispersion of collective modes in the weak and strong coupling limits, 
where the expansion coefficients are analytically tractable for a fixed ($\ell,m_\ell$)
state.

\subsection{Weak coupling (BCS) regime}
\label{sec:cm-wc}

The $s$-wave $(\ell = 0, m_\ell = 0)$ weak coupling limit is characterized by
the criteria $\mu_0 > 0$ and $\mu_0 \approx \epsilon_{\rm F} \gg |\Delta_{0,0}|$.
The expansion of the matrix elements to order $|{\mathbf q}|^2$ and $w^2$ is
performed under the condition $[w,|\mathbf{q}|^2/(2M)] \ll |\Delta_{0,0}|$.
Analytic calculations are particularly simple in this case
since all integrals for the coefficients needed to calculate the collective mode
dispersions are peaked near the Fermi surface.
We first introduce a shell about the Fermi energy
$|\xi_0(\mathbf{k})| \le w_{\rm D}$ such that
$\epsilon_{\rm F} \gg w_{\rm D} \gg \Delta_0(\mathbf{k}_{\rm F})$, inside of
which one may ignore the 3D density of states factor $\sqrt{\epsilon/\epsilon_{\rm F}}$
and outside of which one may ignore $\Delta_0(\mathbf{k})$.
In addition, we make use of the nearly perfect particle-hole symmetry, which forces 
integrals to vanish when their integrands are odd under the transformation 
$\xi_0(\mathbf{k}) \rightarrow -\xi_0(\mathbf{k})$.
For instance, the coefficient that couple phase and amplitude
modes vanish ($B_{0,0,0} = 0$) in this limit. Thus, there
is no mixing between phase and amplitude fields in weak coupling,
as can be seen by inspection of the fluctuation matrix $\widetilde\mathbf{F}_0(q)$.

For $\ell = 0$, the zeroth order coefficient is
\begin{eqnarray}
A_{0,0,0} = \frac{N(\epsilon_{\rm F})}{4\pi} ,
\end{eqnarray}
and the second order coefficients are
\begin{eqnarray}
C_{0,0,0}^{i,j} &=& \frac{Q_{0,0,0}^{i,j}}{3} = 
\frac{N(\epsilon_{\rm F}) v_{\rm F}^2}{36|\Delta_{0,0}|^2} \delta_{i,j}, \\
D_{0,0,0} &=& \frac{R_{0,0,0}}{3} = 
\frac{N(\epsilon_{\rm F})}{12|\Delta_{0,0}|^2}.
\end{eqnarray}
Here, $v_{\rm F} = k_{\rm F}/M$ is the Fermi velocity and 
$N(\epsilon_{\rm F}) = M{\cal V} k_{\rm F}/(2\pi^2)$ is the 
density of states per spin at the Fermi energy.

In weak coupling, since
$
B_{\ell,m_\ell,m_\ell}^2 \ll A_{\ell,m_\ell,m_\ell}R_{\ell,m_\ell,m_\ell},
$ 
the sound velocity is simplified to
$
C_{\ell,m_\ell}^{i,j} \approx [Q_{\ell,m_\ell,m_\ell}^{i,j}/ A_{\ell,m_\ell,m_\ell}]
$
for any $\ell$.
Using the coefficients above in Eq.~(\ref{eqn:soundv.0}), for $\ell = 0$, we obtain
\begin{eqnarray}
{\cal C}_{0,0} = \frac{v_{\rm F}}{\sqrt{3}}
\end{eqnarray}
which is the well known Anderson-Bogoliubov relation.
For $\ell \ne 0$, the expansion coefficients require more detailed and lengthy analysis,
and therefore, we do not discuss here.
On the other hand, the expansion coefficients can be calculated for any $\ell$ 
in the strong coupling BEC regime, which is discussed next.

\subsection{Strong coupling (BEC) regime}
\label{sec:cm-sc}

The strong coupling limit is characterized by
the criteria $\mu_\ell < 0$, $|\mu_\ell| \ll \epsilon_0 = k_0^2/(2M)$ 
and $|\xi_\ell(\mathbf{k})| \gg |\Delta_{\ell}(\mathbf{k})|$.
The expansion of the matrix elements to order $|{\mathbf q}|^2$ and $w^2$ is
performed under the condition $[w, |\mathbf{q}|^2/(2M)] \ll |\mu_\ell|$.
The situation encountered here is very different from the weak coupling 
limit, because one can no longer invoke particle-hole symmetry to simplify the
calculation of many of the coefficients appearing in the fluctuation matrix
$\widetilde \mathbf{F}_\ell (q)$.
In particular, the coefficient $B_{\ell,m_\ell,m_\ell'} \ne 0$
indicates that the amplitude and phase fields are mixed.
Furthermore, $P_{\ell,m_\ell,m_\ell'} = 0$ , since this coefficient reduces to
the order parameter equation in this limit.

For $\ell = 0$, the zeroth order coefficient is
\begin{eqnarray}
A_{0,0,0} &=& \frac{\kappa |\Delta_{0,0}|^2}{8\pi|\mu_0|},
\end{eqnarray}
the first order coefficient is
\begin{eqnarray}
B_{0,0,0} = \kappa,
\end{eqnarray}
and the second order coefficients are
\begin{eqnarray}
C_{0,0,0}^{i,j} &=& Q_{0,0,0}^{i,j} = \frac{\kappa}{4M} \delta_{i,j}, \\
D_{0,0,0} &=& R_{0,0,0} = \frac{\kappa}{8|\mu_0|},
\end{eqnarray}
where
$
\kappa = N(\epsilon_{\rm F}) / (32\sqrt{|\mu_0|\epsilon_{F}}).
$

Using the expressions above in Eq.~(\ref{eqn:soundv.0}), we obtain the sound velocity
\begin{eqnarray}
{\cal C}_{0,0} = \left( \frac{|\Delta_{0,0}|}{32M|\mu_0|\pi} \right)^{\frac{1}{2}}
= v_{\rm F} \sqrt{\frac{k_{\rm F}a_0}{3\pi}}.
\label{eqn:sound.0}
\end{eqnarray}
Notice that the sound velocity is very small and its smallness is controlled
by the scattering length $a_0$.
Furthermore, in the theory of weakly interacting dilute Bose gas, 
the sound velocity is given by 
$
{\cal C}_{{\rm B},0} = 4\pi a_{{\rm B},0} n_{{\rm B},0} / M_{{\rm B},0}^2.
$
Making the identification that the density of 
pairs is $n_{{\rm B},0} = n_0/2$, the mass of the pairs is 
$M_{{\rm B},0} = 2M$ and that the Bose scattering length
is $a_{{\rm B},0} = 2a_0$, it follows that Eq.~(\ref{eqn:sound.0}) 
is identical to the Bogoliubov result ${\cal C}_{{\rm B},0}$.
Therefore, our result for the fermionic system represents in fact a weakly interacting Bose gas
in the strong coupling limit.
A better estimate for $a_{{\rm B},0} \approx 0.6a_0$
can be found in the literature~\cite{pieri-a,gora,kagan,gurarie-a}.
This is also the case when we construct the TDGL equation in Sec.~\ref{sec:tdgl.sc}.

For $\ell \ne 0$, the zeroth order coefficient is
\begin{eqnarray}
A_{\ell \ne 0,m_\ell,m_\ell} &=& \frac{15\widehat{\phi}_\ell \widetilde{\kappa}}{2\epsilon_0\sqrt{\pi}} 
|\Delta_{\ell,m_\ell}|^2 \gamma_{\ell,\{m_\ell\}},
\end{eqnarray}
the first order coefficient is
\begin{eqnarray}
B_{\ell \ne 0,m_\ell,m_\ell'} = \frac{\phi_\ell \widetilde{\kappa}}{\sqrt{\pi}} \delta_{m_\ell,m_\ell'},
\end{eqnarray}
and the second order coefficients are
\begin{eqnarray}
C_{\ell \ne 0,m_\ell,m_\ell'}^{i,i} &=& Q_{\ell \ne 0,m_\ell,m_\ell'}^{i,i} = 
\frac{\phi_\ell \widetilde{\kappa}}{4M\sqrt{\pi}} \delta_{m_\ell,m_\ell'}, \\
D_{1,m_\ell,m_\ell'} &=& R_{1,m_\ell,m_\ell'} = \frac{3\widetilde{\kappa}}{8\sqrt{\epsilon_0|\mu_1|}} \delta_{m_\ell,m_\ell'}, \\
D_{\ell > 1,m_\ell,m_\ell'} &=& R_{\ell > 1,m_\ell,m_\ell'} = 
\frac{3 \bar{\phi}_\ell \widetilde{\kappa}}{4 \sqrt{\pi}\epsilon_0} \delta_{m_\ell,m_\ell'},
\end{eqnarray}
where
$
\widetilde{\kappa} = N(\epsilon_{\rm F}) / (32\sqrt{\epsilon_0 \epsilon_{F}}),
$
$
\phi_{\ell} = \Gamma(\ell - 1/2) / \Gamma(\ell + 1),
$
$
\bar{\phi}_{\ell} = \Gamma(\ell - 3/2) / \Gamma(\ell + 1)
$
and
$
\widehat{\phi}_{\ell} = \Gamma(2\ell - 3/2) / \Gamma(2\ell + 2).
$
Here $\Gamma(x)$ is the Gamma function, and 
$\gamma_{\ell,{m_\ell}}$ is an angular averaged quantity defined in App.~\ref{sec:appb}.

In strong coupling, since
$
B_{\ell,m_\ell,m_\ell}^2 \gg A_{\ell,m_\ell,m_\ell}R_{\ell,m_\ell,m_\ell},
$ 
the sound velocity is simplified to 
$
C_{\ell,m_\ell}^{i,j} \approx [A_{\ell,m_\ell,m_\ell} Q_{\ell,m_\ell,m_\ell}^{i,j}/ B_{\ell,m_\ell,m_\ell}^2]^{1/2}
$
for any $\ell$.
Using the expressions above in Eq.~(\ref{eqn:soundv.l}), for $\ell \ne 0$, we obtain
\begin{eqnarray}
{\cal C}_{\ell \ne 0,m_\ell}^{i,i}
&=& \left(\frac{15\gamma_{\ell,\{m_\ell\}} |\Delta_{\ell ,m_\ell}|^2 \widehat{\phi}_\ell}
{8M\phi_\ell \epsilon_0}\right)^\frac{1}{2} \\
&=& v_{\rm F} \left( \frac{20 \gamma_{\ell,\{m_\ell\}}\sqrt{\pi} 
\widehat{\phi}_\ell}{\phi^2_\ell} \frac{k_{\rm F}}{k_0} \right)^\frac{1}{2}.
\end{eqnarray}
Therefore, the sound velocity is also very small and its smallness is controlled 
by the interaction range $k_0$ through the diluteness condition
i.e. $(k_0/k_F)^3 \gg 1$, for $\ell \ne 0$.
Notice that, the sound velocity is independent of the scattering
parameter for $\ell \ne 0$.

Now, we turn our attention to a numerical analysis of the phase collective modes 
during the evolution from weak coupling BCS to strong coupling BEC limits.

\subsection{Evolution from BCS to BEC regime}
\label{sec:cm-bcsbec}

We focus only on $s$-wave ($\ell = 0, m_\ell = 0$) and $p$-wave ($\ell = 1, m_\ell = 0$)
cases, since they may be the most relevant to current experiments involving ultracold atoms.

\begin{figure} [htb]
\centerline{\scalebox{0.67}{\includegraphics{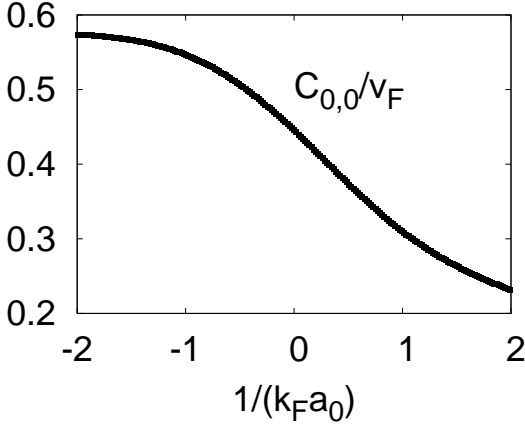}}}
\caption{\label{fig:swave.collective} Plots of reduced 
Goldstone (sound) velocity $({\cal C}_{0,0})_r = {\cal C}_{0,0}/v_{\rm F}$
versus interaction strength $1/(k_{\rm F} a_0)$ for $k_0 \approx 200k_{\rm F}$.
}
\end{figure}

In Fig.~\ref{fig:swave.collective}, we show the evolution of ${\cal C}_{0,0}$ 
as a function of $1/(k_{\rm F}a_0)$ for $s$-wave case. 
The weak coupling Anderson-Bogoliubov velocity 
$
{\cal C}_{0,0} = v_{\rm F}/\sqrt{3}
$ 
evolves continuously to the strong coupling Bogoliubov velocity 
$
{\cal C}_{0,0} = v_{\rm F}\sqrt{k_{\rm F}a_0/(3\pi)}.
$
Notice that the sound velocity is a monotonically decreasing function of 
$1/(k_{\rm F}a_0)$, and the evolution across $\mu_0 = 0$ is a crossover.

\begin{figure} [htb]
\centerline{\scalebox{0.67}{\includegraphics{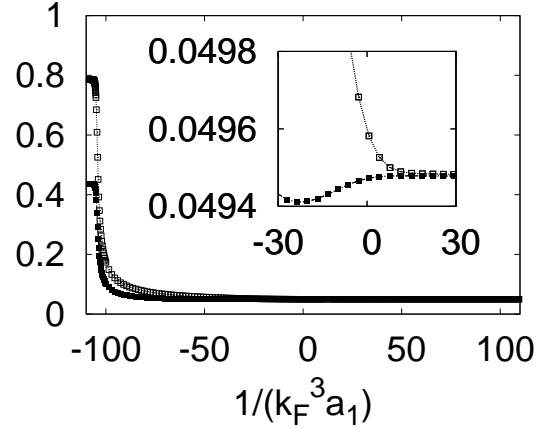}}}
\caption{\label{fig:pwave.collective} Plots of reduced Goldstone (sound) velocity
$({\cal C}_{1,0}^{x,x})_r = {\cal C}_{1,0}^{x,x}/v_{\rm F}$ (solid squares) and
$({\cal C}_{1,0}^{z,z})_r = {\cal C}_{1,0}^{z,z}/v_{\rm F}$ (hollow squares)
versus interaction strength $1/(k_{\rm F}^3 a_1)$ for $k_0 \approx 200k_{\rm F}$. 
The inset zooms into the unitarity region.
}
\end{figure}

In Fig.~\ref{fig:pwave.collective}, we show the evolution of ${\cal C}_{1,0}^{i,j}$ 
as a function of $1/(k_{\rm F}^3 a_1)$ for $p$-wave case. 
Notice that ${\cal C}_{1,0}^{i,i}$ is strongly anisotropic in weak coupling, since
${\cal C}_{1,0}^{x,x} = {\cal C}_{1,0}^{y,y} \approx 0.44 v_{\rm F}$
and ${\cal C}_{1,0}^{z,z} = \sqrt{3} {\cal C}_{1,0}^{x,x} \approx 0.79 v_{\rm F}$, 
thus reflecting the order parameter symmetry.
In addition, ${\cal C}_{1,0}^{i,i}$ is isotropic in strong coupling, since 
$
{\cal C}_{1,0}^{i,i} = v_{\rm F} \sqrt{3k_{\rm F}/(2\pi k_0)} \approx 0.049 v_{\rm F}
$
for $k_0 \approx 200 k_{\rm F}$, thus revealing the secondary role of
the order parameter symmetry in this limit.
The anisotropy is very small in the intermediate regime beyond $\mu_1 < 0$.
Notice also that, ${\cal C}_{1,0}^{z,z}$ is a monotonically decreasing function of
$1/(k_{\rm F}^3 a_1)$ in BCS side until $\mu_1 = 0$, where it saturates.
However, ${\cal C}_{1,0}^{x,x} = {\cal C}_{1,0}^{y,y}$ is a nonmonotonic function
of $1/(k_{\rm F}a_1)^3$, and it also saturates beyond $\mu_1 = 0$.
Therefore, the behavior of ${\cal C}_{1,0}^{i,i}$ reflects the disapperance of nodes
of the quasiparticle energy $E_1(\mathbf{k})$ as $\mu_1$ changes sign.

These collective excitations may contribute significantly to the 
thermodynamic potential, which is discussed next.

\subsection{Corrections to $\Omega_\ell^{\rm sp}$ due to collective modes}
\label{sec:cm-corr}

In this section, we analyze corrections to the saddle point thermodynamic
potential $\Omega_\ell^{\rm sp}$ due to low energy collective excitations.
The evaluation of bosonic Matsubara frequency sums in Eq.~(\ref{eqn:flucttp}) leads to
$
\Omega_\ell^{\rm fluct} \to \Omega_\ell^{\rm coll},
$
where
\begin{equation}
\Omega_\ell^{\rm coll} = {\sum_{\mathbf{q}}}^\prime \big\lbrace W_\ell(\mathbf{q}) + \frac{2}{\beta}
\ln \left[ 1 - \exp(-\beta W_\ell(\mathbf{q})) \right] \big\rbrace
\label{eqn:omega.coll}
\end{equation}
is the collective mode contribution to the thermodynamic potential.
The prime on the summation indicates that a momentum cutoff is required
since a long wavelength and low frequency approximation is used to derive the
collective mode dispersion.
Notice that the first term in Eq.~(\ref{eqn:omega.coll}) contributes to the ground state energy
of the interacting Fermi system. This contribution is necessary to
recover the ground state energy of the effective Bose system in
the strong coupling limit as discussed in Sec~\ref{sec:gse}.

The corrections to the saddle point number equation 
$
N_\ell^{\rm coll} = - \partial \Omega_\ell^{\rm coll} / \partial \mu_\ell
$ 
are due to the zero point motion ($N_\ell^{\rm zp})$ and 
thermal excitation ($N_\ell^{\rm te}$) of the collective modes
\begin{eqnarray}
N_\ell^{\rm zp} &=& - \frac{\partial} {\partial \mu_\ell} 
{\sum_{\mathbf{q}}}^\prime W_{\ell}(\mathbf{q}) , \\
N_\ell^{\rm te} &=& - {\sum_{\mathbf{q}}}^\prime \frac{\partial W_{\ell}(\mathbf{q})} {\partial \mu_\ell} 
n_{\rm B}[W_{\ell}(\mathbf{q})].
\label{eqn:n.coll}
\end{eqnarray}
Here $n_{\rm B}(x) = 1/[\exp(\beta x) -1]$ is the Bose distribution.
For $\ell = 0$, the last equation can be solved to obtain
$
N_0^{\rm te} = -6 (\partial {\cal C}_{0,0} / \partial \mu_0) \zeta(4) T^4 /(\pi^2 {\cal C}_{0,0}^2),
$
which vanishes at $T = 0$.
Here $\zeta(x)$ is the Zeta function.
Similarly, $N_{\ell \ne 0}^{\rm te}$ has a power law dependence on $T$, 
and therefore, vanishes at $T = 0$ since the collective modes are not excited.
$N_\ell^{\rm zp}$ gives small contributions to the number equation in
weak and strong couplings, but may lead to significant 
contributions in the intermediate regime for all $\ell$.
The impact of $N_\ell^{\rm zp}$ on the order parameter and chemical potential in the
intermediate regime may require a careful analysis of the full fluctuation contributions~\cite{pieri}.

Until now, we discussed the evolution of superfluidity from the BCS to the BEC regime 
at zero temperature. In the rest of the manuscript, we analyze the evolution of
superfluidity from the BCS to the BEC limit at finite temperatures.

\section{Bcs to Bec evolution near $T = T_{{\rm c},\ell}$}
\label{sec:normal-state}
In this section, we concentrate on physical properties near critical
temperatures $T = T_{{\rm c},\ell}$. To calculate $T_{{\rm c}, \ell}$, the self-consistency
(order parameter and number) equations have to be solved simultaneously.
At $T = T_{{\rm c},\ell}$, then $\Delta_{\ell,m_\ell} = 0$, and
the saddle point order parameter equation Eq.~(\ref{eqn:opeqn}) reduces to
\begin{equation}
\frac{M{\cal V}}{4\pi k_0^{2\ell} a_\ell} = \sum_{\mathbf{k}}\Gamma_\ell^2(k)
\left[ \frac{1}{2\epsilon(\mathbf{k})}
- \frac{\tanh[\xi_\ell(\mathbf{k})/(2T_{{\rm c},\ell})]} {2\xi_\ell(\mathbf{k})}
\right].
\label{eqn:tceqn}
\end{equation}
This expression is independent of $m_\ell$ since the interaction 
amplitude $\lambda_\ell$ depends only on $\ell$.
Similarly, the saddle point number equation reduces to
\begin{equation}
N_\ell^{\rm{sp}} = \sum_{\mathbf{k}, s} n_{\rm F}[\xi_\ell(\mathbf{k})],
\label{eqn:tcnsp}
\end{equation}
where $n_{\rm F}(x) = 1/[\exp(\beta x) + 1]$ is the Fermi distribution.
Notice that the summation over spins ($s$) is not present in the SHS case.
It is important to emphasize that the inclusion of $N_\ell^{\rm{fluct}}$ 
around $T_\ell = T_{{\rm c},\ell}$ is essential to produce the qualitatively 
correct physics with increasing coupling, as discussed next.

\subsection{Gaussian Fluctuations}
\label{sec:gaussian.Tc}

To evaluate the gaussian contribution to the thermodynamic potential,
we sum over the fermionic Matsubara frequencies in Eq.~(\ref{eqn:flucttp}), 
and obtain the action
\begin{equation}
S_\ell^{\rm{fluct}} = \frac{\beta}{2} \sum_{q,m_\ell,m'_\ell} \Lambda_{\ell,m_\ell}^\dagger(q) 
L^{-1}_{\ell,m_\ell,m'_\ell}(q) \Lambda_{\ell,m'_\ell}(q),
\end{equation}
where $L^{-1}_{\ell,m_\ell,m'_\ell}(q) = (\mathbf{F}^{-1}_{\ell,m_\ell,m'_\ell})_{11}$
is the element of the fluctuation propagator given by
\begin{eqnarray}
L^{-1}_{\ell,m_\ell,m'_\ell}(q) &=& \frac{\delta_{m_\ell,m_\ell'}}{4\pi {\cal V}^{-1} \lambda_\ell}  - 
\sum_{\mathbf{k}} 
\frac{1 - n_{\rm F}(\xi_+) - n_{\rm F}(\xi_-)} {\xi_+ + \xi_- - iv_j} \nonumber \\
& & \Gamma_\ell^2(k) Y_{\ell,m_\ell}(\widehat{\mathbf{k}}) Y_{\ell,m'_\ell}^*(\widehat{\mathbf{k}}).
\label{eqn:L-1}
\end{eqnarray}
This is the generalization of the $\ell = 0$ case to $\ell \ne 0$,
where $\xi_\pm = \xi_\ell(\mathbf{k} \pm \mathbf{q}/2)$.
From $S_\ell^{\rm fluct}$, we obtain the thermodynamic potential
$\Omega_\ell^{\rm{gauss}} = \Omega_\ell^{\rm{sp}} + \Omega_\ell^{\rm{fluct}}$,
where $\Omega_\ell^{\rm sp}$ is the saddle point contribution 
with $\Delta_\ell(\mathbf{k}) = 0$, and
\begin{equation}
\Omega_\ell^{\rm fluct} = - \frac{1}{\beta}\sum_{q} \ln \det [L_{\ell}(q)/\beta]
\label{eqn:omega1}
\end{equation}
is the fluctuation contribution.

We evaluate the bosonic Matsubara frequency ($iv_j$) sums by using contour integration,
and determine the branch cut and pole terms.
We use the phase shift
\begin{equation}
\varphi_{\ell}^{\rm fluct} (\mathbf{q},w) = 
{\rm Arg} [\det L_{\ell}(\mathbf{q}, iv_j \to w + i0^+)]
\end{equation}
to replace $\det L_{\ell}(q)$ in Eq.~(\ref{eqn:omega1}), leading to
\begin{equation}
\Omega_\ell^{\rm fluct} = -\sum_{\mathbf{q}} 
\int_{-\infty}^{\infty} \frac{dw}{\pi} n_{\rm B}(w) 
\widetilde{\varphi}_{\ell}^{\rm fluct} (\mathbf{q},w),
\end{equation}
where
$
\widetilde{\varphi}_{\ell}^{\rm fluct} (\mathbf{q},w) 
= \varphi_{\ell}^{\rm fluct} (\mathbf{q},w) 
- \varphi_{\ell}^{\rm fluct} (\mathbf{q},0)
$
and $n_{\rm B}(x) = 1/\left[\exp(\beta x) - 1\right]$ is the Bose distribution.
Notice that, this equation is the generalization of 
the $s$-wave ($\ell = 0$) case~\cite{nsr,carlos} for $\ell \ne 0$.
Furthermore, the phase shift can be written as
$
\widetilde{\varphi}_{\ell}^{\rm fluct} (\mathbf{q},w) = 
\widetilde{\varphi}_{\ell}^{\rm sc} (\mathbf{q},w)
+ \widetilde{\varphi}_{\ell}^{\rm bs} (\mathbf{q},w),
$
where
\begin{eqnarray}
\widetilde{\varphi}_{\ell}^{\rm sc} (\mathbf{q},w) 
&=& \widetilde{\varphi}_{\ell}(\mathbf{q},w) \Theta(w - w_{\mathbf{q}}^*),
\end{eqnarray}
is the branch cut (scattering) and 
$\widetilde{\varphi}_{\ell}^{\rm bs} (\mathbf{q},w)$ 
is the pole (bound state) contribution. 
Here, $\Theta(x)$ is the Heaviside function, $w_{\mathbf{q}}^* = w_{\mathbf{q}} - 2 \mu_\ell$
with $w_{\mathbf{q}} = |\mathbf{q}|^2/(4M)$ is the branch frequency
and $\mu_\ell$ is the fermionic chemical potential.

The branch cut (scattering) contribution to the thermodynamic potential becomes
\begin{equation}
\Omega_\ell^{\rm{sc}} = -\sum_{\mathbf{q}} 
\int_{-\infty}^{\infty} \frac{dw}{\pi} n_{\rm B}(w) 
\widetilde{\varphi}_{\ell}^{\rm sc} (\mathbf{q},w).
\end{equation}
For each $\mathbf{q}$, the integrand is nonvanishing only
for $w > w_\mathbf{q}^*$ since
$\widetilde{\varphi}_{\ell}^{\rm sc} (\mathbf{q},w) = 0$ 
otherwise.
Thus, the branch cut (scattering) contribution to the number equation 
$N_\ell^{\rm{sc}} = -\partial \Omega_\ell^{\rm{sc}} / \partial \mu_\ell$ is given by
\begin{eqnarray}
N_\ell^{\rm{sc}} = \sum_{\mathbf{q}} \int_{0}^{\infty} \frac{dw}{\pi}
&&\left[\frac{\partial n_{\rm B}(\widetilde{w})}{\partial \mu_\ell} - n_{\rm B}(\widetilde{w})
\frac{\partial}{\partial \mu_\ell}\right] \nonumber \\
&& \widetilde{\varphi}_{\ell}(\mathbf{q}, \widetilde{w}),
\label{eqn:sc}
\end{eqnarray}
where $\widetilde{w} = w + w_{\mathbf{q}}^*$.

When $a_\ell < 0$, there are no bound states above $T_{\rm c,\ell}$ 
and $N_\ell^{\rm{sc}}$ represents the correction due to scattering states.
However, when $a_\ell > 0$, there are bound states 
represented by poles at $w < w_\mathbf{q}^*$.
The pole (bound state) contribution to the number equation is
\begin{equation}
N_\ell^{\rm{bs}} = -\sum_{\mathbf{q}} n_{\rm B}[{\cal W}_\ell (\mathbf{q})]
\eta_{\ell}[\mathbf{q},{\cal W}_\ell(\mathbf{q})],
\end{equation}
where ${\cal W}_\ell(\mathbf{q})$ corresponds to the poles of $L_{\ell}^{-1} (q)$ and
\begin{equation}
\eta_{\ell}[\mathbf{q},{\cal W}_\ell(\mathbf{q})] =
{\rm Res} \big\lbrace \frac{\partial \det L^{-1}_{\ell}[\mathbf{q},{\cal W}_\ell(\mathbf{q})]/\partial \mu_\ell}
{\det L^{-1}_{\ell}[\mathbf{q},{\cal W}_\ell(\mathbf{q})]} \big \rbrace
\label{eqn:residue}
\end{equation}
is the residue.
Heavy numerical calculations are necessary to find the poles
as a function of $\mathbf{q}$ for all couplings.
However, in sufficiently strong coupling, when $n_{\rm F}(\xi_\pm) \ll 1$ in Eq.~(\ref{eqn:L-1}),
the pole (bound state) contribution can be evaluated analytically by eliminating 
$\lambda_\ell$ in favor of the two body bound state energy $\widetilde{E}_{{\rm b},\ell}$ in vacuum.
Notice that, $\widetilde{E}_{{\rm b},\ell}$ is related to the $E_{{\rm b},\ell}$ obtained from 
the T-matrix approach, where multiple scattering events are included. 
However, they become identical in the dilute limit. 

A relation between $\lambda_\ell$ and $\widetilde{E}_{{\rm b},\ell}$ can be obtained by solving
the Schroedinger equation for two fermions interacting via a pairing potential $V(r)$. 
After Fourier transforming from real to momentum space,
the Schroedinger equation for the pair wave function $\psi(\mathbf{k})$ is
\begin{equation}
2\epsilon(\mathbf{k})\psi(\mathbf{k}) + \frac{1}{{\cal V}}
\sum_{\mathbf{k'}}V(\mathbf{k},\mathbf{k'})\psi(\mathbf{k'}) = \widetilde{E}_{\rm b}\psi(\mathbf{k}).
\end{equation}
Using the Fourier expansion of $V(\mathbf{k}, \mathbf{k'})$ given in Eq.~(\ref{eqn:Vsh}) 
and choosing only the $\ell$th angular momentum channel, we obtain
\begin{equation}
\frac{1}{\lambda_\ell} = \frac{1}{{\cal V}} \sum_{\mathbf{k}} 
\frac{\Gamma_\ell^2(k)}{2\epsilon(\mathbf{k}) - \widetilde{E}_{{\rm b},\ell}}.
\label{eqn:binding}
\end{equation}
This expression relates $\widetilde{E}_{{\rm b},\ell} < 0$ to $\lambda_\ell$
in order to express Eq.~(\ref{eqn:bs}) in terms of binding energy $\widetilde{E}_{{\rm b},\ell} < 0$.
Notice that, this equation is similar to the order parameter equation
in the strong coupling limit ($\mu_\ell < 0$ and $|\mu_\ell| \gg T_{c,\ell}$), where
\begin{equation}
\frac{1}{\lambda_\ell} = \frac{1}{{\cal V}} \sum_{\mathbf{k}} 
\frac{\Gamma_\ell^2(k)}{2\epsilon(\mathbf{k}) - 2\mu_\ell}.
\label{eqn:opsc}
\end{equation}
Therefore, $\mu_\ell \to \widetilde{E}_{{\rm b},\ell}/2$ as the coupling increases.

Substitution of Eq.~(\ref{eqn:binding}) in Eq.~(\ref{eqn:residue}) yields the pole contribution which
is given by ${\cal W}_{\ell}(\mathbf{q}) = w_\mathbf{q} + \widetilde{E}_{{\rm b},\ell} - 2\mu_\ell$,
and the residue at this pole is $\eta_{\ell}[\mathbf{q},{\cal W}_\ell(\mathbf{q})] = - 2 \sum_{m_\ell}$.
Therefore, the bound state contribution to the phase shift in the sufficiently 
strong coupling limit is given by
\begin{eqnarray}
\widetilde{\varphi}_{\ell}^{\rm bs} (\mathbf{q},w) 
&=& \pi \Theta(w - w_{\mathbf{q}} + \mu_{{\rm B},\ell}),
\end{eqnarray}
which leads to the bound state number equation
\begin{equation}
N_\ell^{\rm{bs}} = 2\sum_{\mathbf{q},m_\ell} n_{\rm B}[w_\mathbf{q} - \mu_{{\rm B},\ell}],
\label{eqn:bs}
\end{equation}
where $\mu_{{\rm B},\ell} = 2\mu_\ell - \widetilde{E}_{{\rm b},\ell} \le 0$ 
is the chemical potential of the bosonic molecules.
Notice that, Eq.~(\ref{eqn:bs}) is only valid for interaction strengths where $\mu_{{\rm B},\ell} \le 0$.
Thus, this expression can not be used over a region of coupling strengths
where $\mu_{{\rm B},\ell}$ is positive.

\subsection{Critical Temperature and Chemical Potential}
\label{sec:critical-temperature}

To obtain the evolution from BCS to BEC, the number
\begin{equation}
N_\ell \approx N_\ell^{\rm gauss} = N_\ell^{\rm{sp}} + N_\ell^{\rm{sc}} + N_\ell^{\rm{bs}}
\label{eqn:numberTc}
\end{equation}
and order parameter [Eq.~(\ref{eqn:tceqn})] equations have to be solved self-consistently
for $T_{{\rm c},\ell}$ and $\mu_\ell$. 
First, we analyze the number of unbound, scattering and bound fermions as a 
function of the scattering parameter for the $s$-wave ($\ell = 0$) and $p$-wave ($\ell = 1$) cases.

\begin{figure} [htb]
\centerline{\scalebox{0.67}{\includegraphics{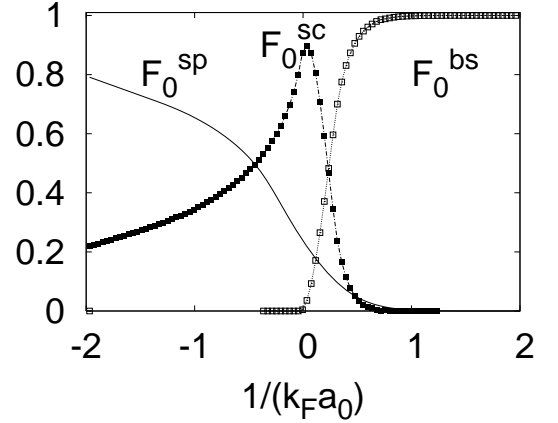}}}
\caption{\label{fig:swave.tc.n} Fractions of unbound $F_0^{\rm sp} = N_0^{\rm sp}/N_0$, 
scattering $F_0^{\rm sc} = N_0^{\rm sc}/ N_0$, 
bound $F_0^{\rm bs} = N_0^{\rm bs} / N_0$ fermions at $T = T_{{\rm c}, 0}$
versus $1/(k_{\rm F} a_0)$ for $k_0 \approx 200k_{\rm F}$. 
}
\end{figure}

In Fig.~{\ref{fig:swave.tc.n}}, we plot different contributions
to the number equation as a function of $1/(k_{\rm F} a_0)$ for the
$s$-wave $(\ell = 0, m_\ell = 0)$ case.
Notice that, $N_0^{\rm sp}$ ($N_0^{\rm bs}$) dominates in weak (strong) coupling,
while $N_0^{\rm sc}$ is the highest for intermediate couplings.
Thus, all fermions are unbound in the strictly BCS limit (not shown in the figure), 
while all fermions are bound in the strictly BEC limit.

\begin{figure} [htb]
\centerline{\scalebox{0.67}{\includegraphics{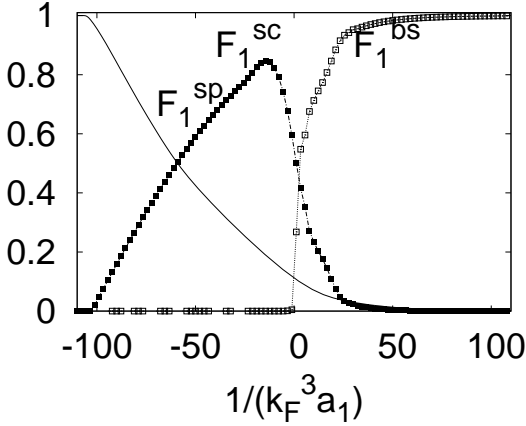}}}
\caption{\label{fig:pwave.tc.n} Fractions of unbound $F_1^{\rm sp} = N_1^{\rm sp}/N_1$, 
scattering $F_1^{\rm sc} = N_1^{\rm sc}/ N_1$, 
bound $F_1^{\rm bs} = N_1^{\rm bs} / N_1$ fermions at $T = T_{{\rm c}, 1}$
versus $1/(k_{\rm F}^3 a_1)$ for $k_0 \approx 200k_{\rm F}$. 
}
\end{figure}

In Fig.~{\ref{fig:pwave.tc.n}}, we present plots of different contributions 
to the number equation as a function of $1/(k_{\rm F}^3 a_1)$ for the
$p$-wave $(\ell = 1, m_\ell = 0)$ case.
Notice also that, $N_1^{\rm sp}$ ($N_1^{\rm bs}$) dominates in weak (strong) coupling,
while $N_1^{\rm sc}$ is the highest for intermediate couplings.
Thus, again all fermions are unbound in the strictly BCS limit, 
while all fermions are bound in the strictly BEC limit.

Therefore, the total fluctuation contribution
$N_\ell^{\rm sc} + N_\ell^{\rm bs}$ is negligible in weak coupling 
and $N_\ell^{\rm sp}$ is sufficient. However, the inclusion of fluctuations 
is necessary for strong coupling to recover the physics of BEC.
However, in the vicinity of the unitary limit [$1/(k_{\rm F}^{2\ell + 1} a_\ell) \to 0$], 
our results are not strictly applicable and should be regarded as qualitative.

Next, we discuss the chemical potential and critical temperature.
In weak coupling, we introduce a shell about the Fermi energy
$|\xi_\ell(\mathbf{k})| \le w_{\rm D}$, such that
$\mu_\ell \approx \epsilon_{\rm F} \gg w_{\rm D} \gg T_{{\rm c},\ell}$. 
Then, in Eq.~(\ref{eqn:tceqn}), we set 
$\tanh[|\xi_\ell(\mathbf{k})|/(2T_{{\rm c},\ell})] = 1$ 
outside the shell and treat the integration within the shell as usual in the BCS theory.
In strong coupling, we use that $\min [\xi_\ell(\mathbf{k})] = |\mu_\ell| \gg T_{{\rm c},\ell}$ 
and set $\tanh[\xi_\ell(\mathbf{k})/(2T_{{\rm c},\ell})] = 1$.
Therefore, in strictly weak and strong coupling, 
the self-consistency equations are decoupled, and play  
reversed roles. In weak (strong) coupling the order parameter 
equation determines $T_{{\rm c},\ell}$ ($\mu_\ell$)
and the number equation determines $\mu_\ell$ ($T_{{\rm c},\ell}$).

In weak coupling, the number equation $N_\ell \approx N_\ell^{\rm{sp}}$ leads to
\begin{equation}
\mu_\ell \approx \epsilon_{\rm F}
\end{equation}
for any $\ell$.
In strong coupling, the order parameter equation leads to
\begin{eqnarray}
\mu_0 &=& - \frac{1}{2Ma_0^2}, 
\label{eqn:mu0.0} \\
\mu_{\ell \ne 0} &=& - \frac{\sqrt{\pi}}{M k_0^{2\ell -1} a_\ell \phi_\ell},
\label{eqn:muTc.0}
\end{eqnarray}
where 
$
\phi_\ell = \Gamma(\ell - 1/2) / \Gamma(\ell + 1)
$
and $\Gamma(x)$ is the Gamma function.
This calculation requires $a_0k_0 > 1$ for $\ell = 0$,
and $k_0^{2\ell + 1} a_\ell \phi_\ell > (\ell + 1) \sqrt{\pi}$ for $\ell \ne 0$
for the order parameter equation to have a solution with $\mu_\ell < 0$.
Furthermore, we assume $|\mu_\ell| \ll \epsilon_0 = k_0^2/(2M)$ to obtain 
Eqs.~(\ref{eqn:mu0.0}) and~(\ref{eqn:muTc.0}).
Notice that, $\mu_\ell = E_{{\rm b},\ell}/2$ in this limit.

On the other hand, the solution of the order parameter equation 
in weak coupling is
\begin{eqnarray}
T_{{\rm c},0} &=& \frac{8}{\pi} \epsilon_{\rm F} \exp\left[\gamma - 2  
+ \frac{\pi}{2} \frac{k_{\rm F}}{k_0} - \frac{\pi}{2k_{\rm F}|a_0|}\right], \\
T_{{\rm c},\ell} &\sim& \epsilon_{\rm F} \exp\left[t_\ell \left(\frac{k_0}{k_{\rm F}}\right)^{2\ell -1} 
- \frac{\pi}{2k_{\rm F}^{2\ell+1}|a_\ell|}\right],
\label{eqn:tc.weak}
\end{eqnarray}
where $\gamma \approx 0.577$ is the Euler's constant,
$t_1 = \pi /4$ and $t_{\ell > 1} = \pi 2^{\ell + 1} (2\ell - 3)!! / \ell!$.
These expressions are valid only when the exponential terms are small.
In strong coupling, the number equation $N_\ell \approx N_\ell^{\rm bs}$ leads to
\begin{equation}
T_{{\rm c},\ell}^{\rm{THS}} = \frac{2\pi}{M_{{\rm B},\ell}} \left[\frac{n_\ell}{\sum_{m_\ell}\zeta(3/2)}\right]^{\frac{2}{3}} 
= \frac{0.218}{(\sum_{m_\ell})^{\frac{2}{3}}}\epsilon_{\rm F},
\end{equation}
where $M_{B,\ell} = 2M$ is the mass of the bosonic molecules.
Here, $n_\ell = k_{\rm F}^3/(3\pi^2)$ is the density of fermions.
For THS Fermi gases, we conclude that the BEC critical temperature of $s$-wave
superfluids is the highest, and this temperature is reduced
for higher angular momentum states.
However, for SHS Fermi gases
\begin{equation}
T_{{\rm c},\ell \ne 0}^{\rm{SHS}} = \frac{2\pi}{M_{{\rm B},\ell}} \left[\frac{n_\ell}{\sum_{m_\ell}\zeta(3/2)}\right]^{\frac{2}{3}} 
= \frac{0.137\epsilon_{\rm F}}{(\sum_{m_\ell})^\frac{2}{3}}.
\end{equation}
where $n_\ell = k_{\rm F}^3/(6\pi^2)$ and $\zeta(x)$ is the Zeta function.
Here, the summation over $m_\ell$ is over degenerate spherical harmonics 
involved in the order parameter of the system, and can be at most
$\sum_{m_\ell} = 2\ell + 1$. 
For SHS states, we conclude that the BEC critical temperature of $p$-wave
superfluids is the highest, and this temperature is reduced
for higher angular momentum states.

For completeness, it is also possible to relate $a_\ell$ and
$T_{{\rm c},\ell}$ when chemical potential vanishes.
When $\mu_\ell = 0$, the solution of number equation Eq.~(\ref{eqn:numberTc})
is highly nontrivial and it is difficult to find the value 
of the scattering parameter $a_\ell^*$ at $\mu_\ell = 0$.
However, the critical temperature in terms of $a_\ell^*$ can be found analytically 
from Eq.~(\ref{eqn:tceqn}) as
\begin{equation}
\left( \frac{T_{{\rm c},\ell}}{\epsilon_{\rm F}}\right)^{\ell + \frac{1}{2}} = 
\frac{\pi /(k_{\rm F}^{2\ell + 1} a_\ell^*)}{(2-2^{-\ell + \frac{3}{2}}) 
\Gamma(\ell + \frac{1}{2}) \zeta(\ell + \frac{1}{2})}.
\label{eqn:tcc}
\end{equation}
to order $T_{{\rm c},\ell}/\epsilon_0$, where $\epsilon_0 = k_0^2/(2M) \gg T_{{\rm c},\ell}$.
Notice that, this relation depends on $k_0$ only through $a_\ell^*$.

On the other hand, if temporal fluctuations are neglected, the solution for $T_{0,\ell}$ 
from the saddle point self-consistency equations is
$
|\widetilde{E}_{{\rm b},\ell}| = 2T_{0,\ell} \ln\left[3\sqrt{\pi}(T_{0,\ell}/\epsilon_{\rm F})^{3/2}/4\right]
$
and $\mu_\ell = \widetilde{E}_{{\rm b},\ell}/2$ which leads to
\begin{equation}
T_{0,\ell} \sim \frac{|\widetilde{E}_{{\rm b},\ell}|} 
{2\ln\left(|\widetilde{E}_{{\rm b},\ell}|/\epsilon_{\rm F}\right)^{\frac{3}{2}}}.
\label{eqn:t0}
\end{equation}
up to logarithmic accuracy. Therefore, $T_{0,\ell}$ grows without bound as the
coupling increases. Within this calculation, the normal state for $T > T_{0,\ell}$ 
is described by unbound and nondegenerate fermions since $\Delta_\ell(\mathbf{k}) = 0$ and
$|\mu_\ell| / T_{0,\ell} \sim \ln(|\widetilde{E}_{{\rm b},\ell}|/\epsilon_{\rm F})^{3/2} \gg 1$.
Notice that the saddle point approximation becomes progressively worse 
with increasing coupling, since the formation of bound states is neglected.

We emphasize that, there is no phase transition across $T_{0,\ell}$ in strong coupling.
However, this temperature is related to the pair breaking or dissociation energy scale.
To see this connection, we consider the chemical equilibrium between nondegenerate unbound
fermions (f) and bound pairs (b) such that $b \leftrightarrow f \uparrow + f \downarrow$
for THS singlet states and $b \leftrightarrow f \uparrow + f \uparrow$
for SHS triplet states.

Notice that $T_{0,\ell}$ is sufficiently high that the chemical potential of the bosons 
and the fermions satisfy $|\mu_{\rm b}| \gg T$ and $|\mu_{\rm f}| \gg T$ at the temperature
$T$ of interest. Thus, both the unbound fermions (f) and molecules (b) can be
treated as classical ideal gases. The equilibrium condition $\mu_{\rm b} = 2\mu_{\rm f}$
for these non-degenerate gases may be written as
$
T \ln \big\lbrace n_{\rm b} [2\pi/(M_{\rm b} T)]^{3/2} \big\rbrace - \widetilde{E}_{{\rm b},\ell} 
= 2T \ln \big\lbrace n_{\rm f} [2\pi/(M_{\rm f} T)]^{3/2} \big\rbrace,
$
where $n_{\rm b}$ ($n_{\rm f}$) is the boson (fermion) density, $M_{\rm b}$ ($M_{\rm f}$) is the boson
(fermion) mass, and $\widetilde{E}_{{\rm b},\ell}$ is the binding energy of a bosonic molecule.
The dissociation temperature above which some fraction of the bound pairs (molecules)
are dissociated, is then found to be 
\begin{equation}
T_{\rm dissoc, \ell} \approx \frac{|\widetilde{E}_{{\rm b},\ell}|}
{\ln \left(|\widetilde{E}_{{\rm b},\ell}|/\epsilon_{\rm F}\right)^{\frac{3}{2}}},
\label{eqn:tdissoc}
\end{equation}
where we dropped a few constants of order unity. 
Therefore, the logarithmic term is an entropic contribution which favors broken
pairs and leads to a dissociation temperature considerably lower than the
absolute value of binding energy $|\widetilde{E}_{{\rm b},\ell}|$.
The analysis above gives insight into the logarithmic factor appearing in 
Eq.~(\ref{eqn:t0}) since $T_{0,\ell} \sim T_{{\rm dissoc}, \ell}/2$.
Thus, $T_{0,\ell}$ is essentially the pair dissociation temperature of bound pairs 
(molecules), while $T_{{\rm c},\ell}$ is the phase coherence temperature corresponding
to BEC of bound pairs (bosonic molecules).

\begin{figure} [htb]
\centerline{\scalebox{0.67}{\includegraphics{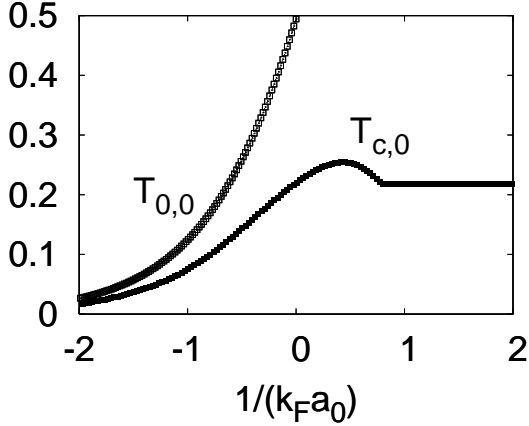}}}
\caption{\label{fig:swave.tc} Plot of 
reduced critical temperature $T_{\rm r} = T_{{\rm c},0}/\epsilon_{\rm F}$ 
versus interaction strength $1/(k_{\rm F} a_0)$ at $T = T_{{\rm c},0}$ for $k_0 \approx 200k_{\rm F}$.
}
\end{figure}

In Fig.~{\ref{fig:swave.tc}}, we show $T_{{\rm c},0}$ for the $s$-wave ($\ell = 0, m_\ell = 0$) case.
Notice that $T_{{\rm c},0}$ grows from an exponential dependence in weak coupling to
a constant in strong coupling with increasing interaction.
Furthermore, the mean field $T_{0,0}$ and gaussian $T_{{\rm c},0}$ are similar
only in weak coupling, while $T_{0,0}$ increases without bound as 
$
T_{0,0} \sim 1/[(M a_0^2) |\ln(k_{\rm F} a_0)|]
$ 
in strong coupling. When $\mu_0 = 0$, we also obtain analytically
$
T_{{\rm c},0}/\epsilon_{\rm F} \approx 2.15/(k_{\rm F}a_0^*)^2
$
from Eq.~(\ref{eqn:tcc}).
The hump in $T_{{\rm c},0}$ around $1/(k_{\rm F} a_0) \approx 0.5$ is similar to the
those in Ref.~\cite{carlos}, and might be an artifact of the approximations used here. 
Thus, a more detailed self-consistent numerical analysis is needed 
to determine if this hump is real.

\begin{figure} [htb]
\centerline{\scalebox{0.67}{\includegraphics{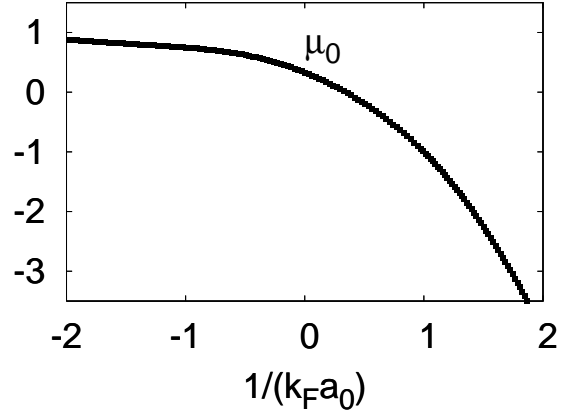}}}
\caption{\label{fig:swave.mutc} Plot
reduced chemical potential $\mu_{\rm r} = \mu_0/\epsilon_{\rm F}$ (inset)
versus interaction strength $1/(k_{\rm F} a_0)$ at $T = T_{{\rm c},0}$ for $k_0 \approx 200k_{\rm F}$.
}
\end{figure}

In Fig.~{\ref{fig:swave.mutc}}, we show $\mu_0$ for the $s$-wave case, where it 
changes from $\epsilon_{\rm F}$ in weak coupling to $E_{{\rm b},0}/2 = - 1/(2M a_0^2)$ 
in strong coupling. Notice that, $\mu_0$ at $T_{{\rm c},0}$ is qualitatively
similar to $\mu_0$ at $T = 0$, however, it is reduced at $T_{{\rm c},0}$
in weak coupling. Furthermore, $\mu_0$ changes sign at $1/(k_{\rm F}a_0) \approx 0.32$.

\begin{figure} [htb]
\centerline{\scalebox{0.67}{\includegraphics{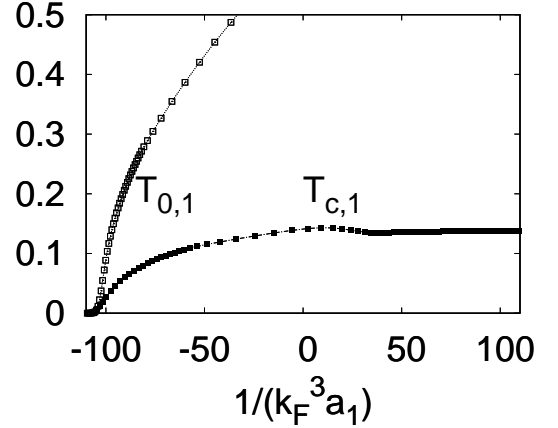}}}
\caption{\label{fig:pwave.tc} Plot of 
reduced critical temperature $T_{\rm r} = T_{c,1}/\epsilon_{\rm F}$
versus interaction strength $1/(k_{\rm F}^3 a_1)$ at $T = T_{c,1}$ for $k_0 \approx 200k_{\rm F}$.
}
\end{figure}

In Fig.~\ref{fig:pwave.tc}, we show $T_{{\rm c},1}$ for the $p$-wave ($\ell = 1, m_\ell = 0$) case.
$T_{{\rm c},1}$ grows from an exponential dependence in weak coupling to 
a constant in strong coupling with increasing interaction.
For completeness, we present the limiting expressions
\begin{eqnarray}
T_{{\rm c},1} &=& \frac{8}{\pi} \epsilon_{\rm F} \exp \left[\gamma - \frac{8}{3} + \frac{\pi k_0}{4k_{\rm F}} 
- \frac{\pi}{2k_{\rm F}^3|a_1|} \right], \\
T_{{\rm c},1} &=& \frac{2\pi}{M_{{\rm B}, 1}} \left[\frac{n_1}{\zeta(3/2)}\right]^\frac{2}{3} = 0.137\epsilon_{\rm F},
\end{eqnarray}
in the weak and strong coupling limits, respectively.
Furthermore, the mean field $T_{0,1}$ and gaussian $T_{{\rm c},1}$ are similar
only in weak coupling, while $T_{0,1}$ increases without bound as 
$T_{0,1} \sim 1/[(M k_0 a_1) |\ln(k_{\rm F}^2 k_0 a_1)|]$ in strong coupling.
When $\mu_1 = 0$, we also obtain analytically
$
T_{{\rm c},1}/\epsilon_{\rm F} \approx 1.75/(k_{\rm F}^3 a_1^*)^{2/3}
$
from Eq.~(\ref{eqn:tcc}).
The hump in the intermediate regime is similar to the 
one found in fermion-boson model~\cite{ohashi}.
But to determine if this hump is real, it may be necessary 
to develop a fully self-consistent numerical calculation.

\begin{figure} [htb]
\centerline{\scalebox{0.67}{\includegraphics{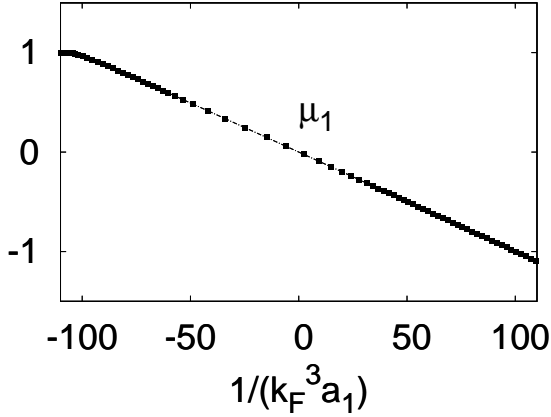}}}
\caption{\label{fig:pwave.mutc} Plot of 
reduced chemical potential $\mu_{\rm r} = \mu_1/\epsilon_{\rm F}$ (inset)
versus interaction strength $1/(k_{\rm F}^3 a_1)$ at $T = T_{c,1}$ for $k_0 \approx 200k_{\rm F}$.
}
\end{figure}

In Fig.~{\ref{fig:pwave.mutc}}, we show $\mu_1$ for the $p$-wave case, where it 
changes from $\epsilon_{\rm F}$ in weak coupling to 
$E_{{\rm b},1}/2 = - 1/(M k_0 a_1)$ in strong coupling. 
Notice that, $\mu_1$ at $T_{{\rm c},1}$ is both qualitatively and quantitatively
similar to $\mu_1$ at $T = 0$.
Furthermore, $\mu_1$ changes sign at $1/(k_{\rm F}^3 a_1) \approx 0.02$.

For any given $\ell$, mean field and gaussian 
theories lead to similar results for $T_{{\rm c},\ell}$ and $T_{0,\ell}$ in the BCS regime, while
they are very different in the BEC side.
In the latter case, $T_{0,\ell}$ increases without bound, however, gaussian theory results in
a constant critical temperature which coincides with the BEC temperature of bosons. 
Notice that the pseudogap region $T_{{\rm c},\ell} < T < T_{0,\ell}$ 
for $\ell = 0$ state is much larger than $\ell \ne 0$ states since
$T_{0,{\ell \ne 0}}$ grows faster than $T_{0, \ell \ne 0}$.
Furthermore, similar humps in $T_{{\rm c},\ell}$ around $1/(k_{\rm F}^{2\ell + 1} a_\ell) = 0$ 
are expected for any $\ell$ as shown for the $s$-wave and $p$-wave cases, 
however, whether these humps are physical or not may require 
a fully self-consistent numerical approach.

As shown in this section, the frequency (temporal) dependence of fluctuations
about the saddle point is crucial to describe adequately the bosonic degrees of freedom
that emerge with increasing coupling. 
In the next section, we derive the TDGL functional near $T_{{\rm c},\ell}$ 
to emphasize further the importance of these fluctuations.

\section{TDGL Functional near $T_{c,\ell}$}
\label{sec:tdgl}

Our basic motivation here is to investigate the low frequency and long wavelength 
behavior of the order parameter near $T_{{\rm c},\ell}$.
To study the evolution of the time-dependent Ginzburg-Landau (TDGL) 
functional near $T_{{\rm c},\ell}$, we need to
expand the effective action $S_\ell^{\rm{eff}}$ in Eq.~(\ref{eqn:s_eff})
around $\Delta_{\ell,m_\ell} = 0$ leading to
\begin{eqnarray}
S_\ell^{\rm{eff}} &=& S_\ell^{\rm{sp}} + S_\ell^{\rm{gauss}} \nonumber \\
&+& \frac{\beta}{2}\sum_{\{q_n,m_{\ell_n}\}} b_{\ell,\{m_{\ell_n}\}}(\{q_n\}) 
\Lambda^\dagger_{\ell,m_{\ell_1}}(q_1) \Lambda_{\ell,m_{\ell_2}}(q_2) \nonumber \\
&& \hskip 1cm \Lambda^\dagger_{\ell,m_{\ell_3}}(q_3) \Lambda_{\ell,m_{\ell_4}}(q_1-q_2+q_3).
\end{eqnarray}
Here, $\Lambda_{\ell,m_\ell}(q)$ is the pairing fluctuation field.

We first consider the static part of $L^{-1}_{\ell,m_\ell,m'_\ell}(q)$, 
and expand it in powers of $q_i$ to get
\begin{equation}
L^{-1}_{\ell,m_\ell,m'_\ell}(\mathbf{q},0) = a_{\ell,m_\ell,m_\ell'} 
+ \sum_{i,j} c_{\ell,m_\ell,m'_\ell}^{i,j} \frac{q_i q_j}{2M} + ... .
\end{equation}
Next, we consider the time-dependence of the TDGL equation, 
where it is necessary to expand
$L^{-1}_{\ell,m_\ell,m'_\ell}(0, iv_j) - L^{-1}_{\ell,m_\ell,m'_\ell}(0,0)$
in powers of $w$ after analytic continuation $iv_j \to w + i0^+$.

In the $x = (\mathbf{x},t)$ representation, the calculation above leads to the TDGL equation
\begin{eqnarray}
\sum_{m_{\ell_2}}
\left[ a_{\ell,m_{\ell_1},m_{\ell_2}} \right.
&-& \sum_{i,j} c_{\ell,m_{\ell_1},m_{\ell_2}}^{i,j} \frac{\nabla_i \nabla_j}{2M} \nonumber \\
&+& \sum_{m_{\ell_3},m_{\ell_4}}b_{\ell,\{m_{\ell_n}\}}(0)
\Lambda^\dagger_{\ell,m_{\ell_3}}(x)
\Lambda_{\ell,m_{\ell_4}}(x) \nonumber \\
&-& \left.i d_{\ell,m_{\ell_1},m_{\ell_2}} \frac{\partial}{\partial t} 
\right]\Lambda_{\ell,m_{\ell_2}}(x) = 0,
\end{eqnarray}
which is the generalization of the TDGL equation to higher momentum channels
of THS singlet and SHS triplet states.
Notice that, for THS triplet states, there may be extra gradient
mixing textures and fourth order terms in the expansion~\cite{leggett-review}, which are not discussed here.
All static and dynamic expansion coefficients are presented in Appendix~\ref{sec:appb}.
The condition $\det a_{\ell} = 0$ with matrix elements $a_{\ell, m_{\ell_1}, m_{\ell_2}}$ 
is the Thouless criterion, which leads to the order parameter equation.
The coefficient $c_{\ell,m_{\ell_1},m_{\ell_2}}^{i,j}$ reflects a major difference
between $\ell = 0$ and $\ell \ne 0$ cases.
While $c_{0,0,0}^{i,j} = c_{0,0,0} \delta_{i,j}$ is isotropic in space, 
$c_{\ell \ne 0, m_{\ell_1}, m_{\ell_2}}^{i,j}$ is anisotropic, thus characterizing 
the anisotropy of the order parameter.
The coefficient $b_{\ell,\{m_{\ell_n}\}}(0)$ is positive and guarantees the stability
of the theory.
The coefficient $d_{\ell, m_{\ell_1}, m_{\ell_2}}$ is a complex number.
Its imaginary part reflects the decay of Cooper pairs into the two-particle continuum for $\mu_\ell > 0$.
However, for $\mu_\ell < 0$, imaginary part of $d_{\ell, m_{\ell_1}, m_{\ell_2}}$
vanishes and the behavior of the order parameter is
propagating reflecting the presence of stable bound states.

Next, we present the asymptotic forms of
$a_{\ell,m_{\ell_1},m_{\ell_2}}$; $b_{\ell,\{m_{\ell_n}\}}(0)$; $c_{\ell,m_{\ell_1},m_{\ell_2}}^{i,j}$ 
and $d_{\ell,m_{\ell_1},m_{\ell_2}}$ which are used to recover the usual Ginzburg-Landau (GL) equation  
for BCS superfluids in weak coupling and the Gross-Pitaevskii (GP) equation
for a weakly interacting dilute Bose gas in strong coupling.

\subsection{Weak coupling (BCS) regime}
\label{sec:tdgl.wc}

The weak coupling BCS regime is characterized by $\mu_\ell > 0$ and 
$\mu_\ell \approx \epsilon_{\rm F} \gg T_{{\rm c},\ell}$.
For any given $\ell$, we find the following values for the coefficients
\begin{eqnarray}
a_{\ell,m_{\ell_1},m_{\ell_2}} &=& \kappa^{\rm w}_\ell \ln \left( {\frac{T}{T_{{\rm c},\ell}}}\right) 
\delta_{m_{\ell_1},m_{\ell_2}}, \\
b_{\ell,\{m_{\ell_n}\}}(0) &=& 7\gamma_{\ell,\{m_{\ell_n}\}} \frac{\kappa^{\rm w}_\ell\zeta(3)}{8T_{{\rm c},\ell}^2} 
\left(\frac{\epsilon_{\rm F}}{\epsilon_0}\right)^\ell, \\
c_{\ell,m_{\ell_1},m_{\ell_2}}^{i,j} &=& 7\alpha_{\ell,m_{\ell_1},m_{\ell_2}}^{i,j} 
\frac{\kappa^{\rm w}_\ell \epsilon_{\rm F} \zeta(3)}{4\pi^2 T_{{\rm c},\ell}^2}, \\
d_{\ell,m_{\ell_1},m_{\ell_2}} &=& \kappa^{\rm w}_\ell 
\left(\frac{1}{4\epsilon_{\rm F}} + i\frac{\pi}{8T_{{\rm c},\ell}}\right)
\delta_{m_{\ell_1},m_{\ell_2}},
\end{eqnarray}
where 
$
\kappa^{\rm w}_\ell = N(\epsilon_{\rm F}) (\epsilon_{\rm F}/\epsilon_0)^\ell /(4\pi)
$
with
$
N(\epsilon_{\rm F}) = M{\cal V}k_{\rm F}/(2\pi^2)
$
is the density of states per spin at the Fermi energy.
Here $\delta_{m_{\ell_1},m_{\ell_2}}$ is the Kronecker delta, and 
$
\alpha_{\ell,m_{\ell_1},m_{\ell_2}}^{i,j}
$
and
$
\gamma_{\ell,\{m_\ell\}}
$ 
are angular averaged quantities defined in App.~\ref{sec:appb}.
Notice that the critical transition temperature is determined by 
$\det a_{\ell} = 0$.

In the particular case, where only one of the spherical harmonics 
$Y_{\ell,m_\ell}(\widehat{\mathbf{k}})$ is dominant and
characterizes the order parameter, we can rescale the pairing field as
\begin{equation}
\Psi^{\rm w}_{\ell,m_\ell}(x) = 
\sqrt{\frac{b_{\ell,\{m_\ell\}}(0)}{\kappa^{\rm w}_\ell}} \Lambda_{\ell,m_\ell}(x)
\end{equation}
to obtain the conventional TDGL equation
\begin{equation}
\left[ - \varepsilon_{\ell} + |\Psi^{\rm w}_{\ell,m_\ell}|^2 - \sum_{i} (\xi^{\rm{GL}}_{\ell,m_\ell})_i^2 \nabla_i^2 
+ \tau^{\rm{GL}}_{\ell,m_\ell} \frac{\partial}{\partial t} \right] \Psi^{\rm w}_{\ell,m_\ell} = 0.
\end{equation}
Here, 
$
\varepsilon_{\ell} = (T_{{\rm c},\ell} - T)/T_{{\rm c},\ell}
$
with $|\varepsilon_\ell| \ll 1$,
$
(\xi_{\ell, m_\ell})_i^2 (T)
= c_{\ell, m_\ell, m_\ell}^{i,i} /[2 M a_{\ell, m_\ell, m_\ell}]
= (\xi^{\rm{GL}}_{\ell,m_\ell})_i^2 / \varepsilon_\ell
$
is the characteristic GL length and
$
\tau_{\ell,m_\ell} 
= -i d_{\ell, m_\ell, m_\ell} / a_{\ell, m_\ell, m_\ell}
= \tau^{\rm{GL}}_{\ell,m_\ell}/\varepsilon_\ell
$
is the characteristic GL time.

In this limit, the GL coherence length is given by
$
k_{\rm F}(\xi^{\rm{GL}}_{\ell,m_\ell})_i 
= [7\alpha_{\ell,m_\ell,m_\ell}^{i,i} \zeta(3)/(4\pi^2)]^{1/2} (\epsilon_{\rm F} / T_{{\rm c},\ell}),
$
which makes $(\xi^{\rm{GL}}_{\ell,m_\ell})_i$ much larger than the interparticle spacing $k_{\rm F}^{-1}$.
There is a major difference between $\ell = 0$ and $\ell \ne 0$ pairings regarding 
$(\xi^{\rm{GL}}_{\ell,m_\ell})_i$.
While $c_{0,0,0}^{i,j} = c_{0,0,0}\delta_{i,j}$ is isotropic,
$
c_{\ell \ne 0,m_{\ell_1},m_{\ell_2}}^{i,j} = c_{\ell,m_{\ell_1},m_{\ell_2}}^{i,i} \delta_{i,j}
$
is in general anisotropic in space (see App.~\ref{sec:appb}). 
Thus, $(\xi^{\rm{GL}}_{0,0})_i$ is isotropic and $(\xi^{\rm{GL}}_{\ell \ne 0, m_\ell})_i$ is not.

Furthermore, 
$
\tau^{\rm{GL}}_{\ell,m_\ell} = -i/(4\epsilon_{\rm F}) + \pi/(8 T_{{\rm c},\ell})
$
showing that the dynamics of $\Psi^{\rm w}_{\ell,m_\ell}(x)$ is overdamped reflecting
the continuum of fermionic excitations into which a pair can decay.
In addition, there is a small propagating term since there is no perfect particle-hole symmetry.
As the coupling grows, the coefficient of the propagating term
increases while that of the damping term vanishes for $\mu_\ell \le 0$.
Thus, the mode is propagating in strong coupling reflecting the
stability of the bound states against the two particle continuum.

\subsection{Strong coupling (BEC) regime}
\label{sec:tdgl.sc}

The strong coupling BEC regime is characterized by $\mu_\ell < 0$ and 
$\epsilon_0 = k_0^2/(2M) \gg |\mu_\ell| \gg T_{{\rm c},\ell}$.
For $\ell = 0$, we find the following coefficients
\begin{eqnarray}
a_{0,0,0} &=& 2\kappa^{\rm s}_0 \left(2|\mu_0| - |\widetilde{E}_{{\rm b},0}| \right), \\ 
b_{0,\{0\}}(0) &=& \frac{\kappa^{\rm s}_0}{8\pi|\mu_0|}, \\
c_{0,0,0}^{i,j} &=& \kappa^{\rm s}_0 \delta_{i,j}, \\
d_{0,0,0} &=& 2\kappa^{\rm s}_0,
\end{eqnarray}
where 
$
\kappa^{\rm s}_0 = N(\epsilon_{\rm F}) / (64\sqrt{\epsilon_{\rm F}|\mu_0|}).
$
Similarly, for $\ell \ne 0$, we obtain
\begin{eqnarray}
a_{\ell \ne 0, m_{\ell_1}, m_{\ell_2}} &=& 2\kappa^{\rm s}_\ell\phi_\ell 
\left(2|\mu_\ell| - |\widetilde{E}_{{\rm b},\ell}|\right) \delta_{m_{\ell_1},m_{\ell_2}}, \\
b_{\ell \ne 0, \{m_{\ell_n}\}}(0) &=& 15\gamma_{\ell,\{m_{\ell_n}\}} \frac{\kappa^{\rm s}_\ell \widehat{\phi}_\ell}{2\epsilon_0}, \\
c_{\ell \ne 0, m_{\ell_1}, m_{\ell_2}}^{i,j} &=& \kappa^{\rm s}_\ell\phi_\ell \delta_{m_{\ell_1},m_{\ell_2}} \delta_{i,j}, \\
d_{\ell \ne 0, m_{\ell_1}, m_{\ell_2}} &=& 2\kappa^{\rm s}_\ell\phi_\ell \delta_{m_{\ell_1},m_{\ell_2}},
\end{eqnarray}
where 
$
\kappa^{\rm s}_{\ell \ne 0} = N(\epsilon_{\rm F}) / (64\sqrt{\pi\epsilon_{\rm F}\epsilon_0}).
$
Here 
$
\phi_\ell = \Gamma(\ell - 1/2) / \Gamma(\ell + 1)
$
and
$
\widehat{\phi}_\ell = \Gamma(2\ell - 3/2) / \Gamma(2\ell + 2),
$
where $\Gamma(x)$ is the Gamma function.
Notice that, $c_{\ell \ne 0,m_{\ell_1},m_{\ell_2}}^{i,j}$ is isotropic in space for any $\ell$.
Thus, the anisotropy of the order parameter plays a secondary role in the TDGL theory in this limit.

In the particular case, where only one of the spherical harmonics 
$Y_{\ell,m_\ell}(\widehat{\mathbf{k}})$ is dominant and
characterizes the order parameter, we can rescale the pairing field as
\begin{equation}
\Psi^{\rm s}_{\ell,m_\ell}(x) = \sqrt{d_{\ell,m_\ell,m_\ell}}\Lambda_{\ell,m_\ell}(x),
\end{equation}
to obtain the conventional Gross-Pitaevskii (GP) equation
\begin{equation}
\left[ \mu_{{\rm B},\ell} + U_{\ell,m_\ell}|\Psi^{\rm s}_{\ell,m_\ell}|^2 - 
\frac{\nabla^2}{2M_{{\rm B},\ell}} - i\frac{\partial}{\partial t} \right] \Psi^{\rm s}_{\ell,m_\ell} = 0
\end{equation}
for a dilute gas of bosons. Here, 
$
\mu_{{\rm B},\ell} = - a_{\ell,m_\ell,m_\ell}/d_{\ell,m_\ell,m_\ell} = 2\mu_\ell - \widetilde{E}_{{\rm b},\ell}
$
is the chemical potential, 
$
M_{{\rm B},\ell} = M d_{\ell,m_\ell,m_\ell}/c_{\ell,m_\ell,m_\ell}^{i,i} = 2M
$
is the mass, and
$
U_{\ell,m_\ell} = b_{\ell,\{m_\ell\}}(0)/d_{\ell,m_\ell,m_\ell}^2
$
is the repulsive interactions of the bosons.
We obtain,
$
U_{0,0} = 4\pi a_0/M
$ 
and
$
U_{\ell \ne 0,m_\ell} = 240\pi^2\sqrt{\pi}\widehat{\phi}_\ell \gamma_{\ell,\{m_\ell\}}/(M \phi_{\ell}^2 k_0)
$
for $\ell = 0$ and $\ell \ne 0$, respectively.
Notice that the mass of the composite bosons is independent of the
anisotropy and symmetry of the order parameter for any given $\ell$.
However, this is not the case for the repulsive interactions between bosons,
which explicitly depends on $\ell$.

For $\ell = 0$, $U_{0,0} = 4\pi a_{{\rm B},0}/M_{{\rm B},0}$ is directly 
proportional to the fermion (boson) scattering length $a_0$ ($a_{{\rm B}, 0}$), 
where $a_{{\rm B}, 0} = 2a_0$ is the boson-boson scattering lenth.
A better estimate for $a_{{\rm B},0} \approx 0.6a_0$
can be found in the literature~\cite{pieri-a,gora,kagan,gurarie-a}.
While for $\ell \ne 0$, $U_{\ell, m_\ell}$ is a constant (independent of the scattering 
parameter $a_\ell$) depending only on the interaction range $k_0$ 
and the particular $(\ell, m_\ell)$ state.
For a finite range potential, $n_{{\rm B},\ell} U_{\ell,m_\ell}$ is small
compared to $\epsilon_{\rm F}$, where $n_{{\rm B},\ell} = n_\ell/2$ is the density of bosons.
In the $\ell = 0$ case $n_{{\rm B},0} U_{0,0}/\epsilon_{\rm F} =  4k_{\rm F} a_0/(3\pi)$ 
is much smaller than unity.
For $\ell \ne 0$ and even, $n_{{\rm B},\ell} U_{\ell,m_\ell}/\epsilon_{\rm F} = 
80\sqrt{\pi} \widehat{\phi}_\ell \gamma_{\ell,\{m_\ell\}}/\phi_{\ell}^2 (k_{\rm F}/k_0)$.
In the case of SHS states where $\ell \ne 0$ and odd, 
$n_{{\rm B},\ell} U_{\ell,m_\ell}/\epsilon_{\rm F} = 
40\sqrt{\pi}\widehat{\phi}_\ell\gamma_{\ell,\{m_\ell\}}/\phi_{\ell}^2 (k_{\rm F}/k_0)$.
The results for higher angular momentum channels reflect
the diluteness condition $(k_{\rm F}/k_0)^3 \ll 1$.

To calculate $(\xi^{\rm{GL}}_{\ell,m_\ell})_i$ in the strong coupling limit,
we need to know $\partial \mu_\ell / \partial T$ evaluated at $T_{{\rm c},\ell}$ (see below).
The temperature dependence of $\mu_\ell$ in the vicinity of $T_{{\rm c},\ell}$ 
can be obtained by noticing that
$
\mu_{{\rm B},\ell} = \widetilde{n}(T) U_{\ell,m_\ell},
$ 
where
$
\widetilde{n}(T) = n_{\rm B,\ell} \left[ 1 - (T/T_{{\rm c},\ell})^{3/2} \right].
$
This leads to
$
k_{\rm F}(\xi^{\rm{GL}}_{\ell,m_\ell})_i = [\pi^2/(2M k_{\rm F} U_{\ell,m_\ell})]^{1/2}
$ 
in the BEC regime.
Using the asymptotic values of $U_{\ell, m_\ell}$, we obtain
$
k_{\rm F} (\xi^{\rm{GL}}_{0,0})_i = \left[ \pi /(8 k_{\rm F} a_0)\right]^{1/2}
$ 
for $\ell = 0$ and
$
k_{\rm F} (\xi^{\rm{GL}}_{\ell \ne 0,m_\ell})_i = \left[\phi_\ell^2/
(480\sqrt{\pi}\gamma_{\ell,\{m_\ell\}} \widehat{\phi}_\ell)\right]^{1/2}(k_0/k_{\rm F})^{1/2}
$
for $\ell \ne 0$.
Therefore, $(\xi^{\rm{GL}}_{\ell,m_\ell})_i$ is also much larger 
than the interparticle spacing $k_{\rm F}^{-1}$ in this limit, since
$k_{\rm F}a_0 \to 0$ for $\ell = 0$ and $k_0 \gg k_{\rm F}$ for any $\ell$.

\subsection{Ginzburg-Landau Coherence length versus average Cooper pair size}
\label{sec:tdgl.cl}

In the particular case, where only one of the spherical harmonics 
$Y_{\ell, m_\ell}(\widehat{\mathbf{k}})$ is dominant and characterizes the order parameter, 
we can define the GL coherence length as
$
(\xi_{\ell,m_\ell})_i^2(T) = c_{\ell,m_\ell,m_\ell}^{i,i}/(2M a_{\ell,m_\ell,m_\ell}). 
$ 
An expansion of the parameters $a_{\ell,m_\ell,m_\ell}$ and $c_{\ell,m_\ell,m_\ell}^{i,i}$ 
in the vicinity of $T_{{\rm c},\ell}$ leads to
\begin{equation}
(\xi_{\ell,m_\ell})_i^2(T) \approx (\xi^{\rm{GL}}_{\ell,m_\ell})_i^2 \frac{T_{{\rm c},\ell}} {T_{{\rm c},\ell} - T},
\end{equation}
where the prefactor is the GL coherence length and given by
\begin{equation}
(\xi^{\rm{GL}}_{\ell,m_\ell})_i^2 = \frac{c_{\ell,m_\ell,m_\ell}^{i,i}} {2M T_{{\rm c},\ell}} 
\left[\frac{\partial a_{\ell,m_\ell,m_\ell}} {\partial T}\right]_{T = T_{{\rm c},\ell}}^{-1}.
\end{equation}
The slope of the coefficient $a_{\ell,m_\ell,m_\ell}$ with respect to $T$ is given by
\begin{eqnarray}
\frac{\partial a_{\ell,m_\ell,m_\ell}} {\partial T} &=& 
\sum_{\mathbf{k}} \left[ \frac{{\cal Y}_\ell(\mathbf{k})}{2T^2} + 
\frac{\partial \mu_\ell}{\partial T} \left( \frac{{\cal Y}_\ell(\mathbf{k})}{2T\xi_\ell(\mathbf{k})} 
- \frac{{\cal X}_\ell(\mathbf{k})}{\xi_\ell^2(\mathbf{k})} \right)
\right] \nonumber \\
&& \hskip 1cm \frac{\Gamma_\ell^2(k)}{8\pi}.
\end{eqnarray}
Here ${\cal X}_\ell(\mathbf{k})$ and ${\cal Y}_\ell(\mathbf{k})$ are defined in App.~\ref{sec:appb}.
Notice that, while $\partial \mu_\ell/\partial T$ vanishes at $T_{c,\ell}$ in weak coupling, 
it plays an important role in strong coupling.
Furthermore, while $(\xi_{\ell,m_\ell}^{\rm GL})_i$ representing the phase
coherence length is large compared to interparticle 
spacing in both BCS and BEC limits, it should have a minimum near $\mu_\ell \approx 0$.

The prefactor $(\xi^{\rm{GL}}_{\ell,m_\ell})_i$ of the GL coherence length must be
compared with the average Cooper pair size $\xi^{\rm{pair}}_{\ell}$ defined by
\begin{eqnarray}
(\xi^{\rm{pair}}_{\ell})^2 
&=& 
\frac{\langle {\cal Z}_{\ell}(\mathbf{k})| 
r^2 |{\cal Z}_{\ell}(\mathbf{k}) \rangle} 
{\langle {\cal Z}_{\ell}(\mathbf{k}) | {\cal Z}_{\ell}(\mathbf{k}) \rangle} \nonumber \\
&=&
- \frac{\langle {\cal Z}_{\ell}(\mathbf{k})| 
\nabla_{\mathbf{k}}^2 |{\cal Z}_{\ell}(\mathbf{k}) \rangle} 
{\langle {\cal Z}_{\ell}(\mathbf{k}) | {\cal Z}_{\ell}(\mathbf{k}) \rangle},
\end{eqnarray}
where 
$
{\cal Z}_{\ell}(\mathbf{k}) = \Delta_{\ell}(\mathbf{k})/[2E_\ell(\mathbf{k})]
$ 
is the zero temperature pair wave function.
In the BCS limit, $\xi_{\ell}^{{\rm pair}}$ is much larger than 
the interparticle distance $k_{\rm F}^{-1}$ since the Cooper pairs are weakly bound.
Furthermore, for $\mu_\ell < 0$, we expect that $\xi_{\ell}^{{\rm pair}}$ is a decreasing 
function of interaction for any $\ell$, 
since Cooper pairs become more tightly bound as the interaction increases.
Next, we compare  $(\xi^{\rm GL}_{\ell,m_\ell})_i$ and  $\xi^{\rm pair}_{\ell}$
for $s$-wave ($\ell = 0$) and $p$-wave ($\ell = 1$) states.

\begin{figure} [htb]
\centerline{\scalebox{0.67}{\includegraphics{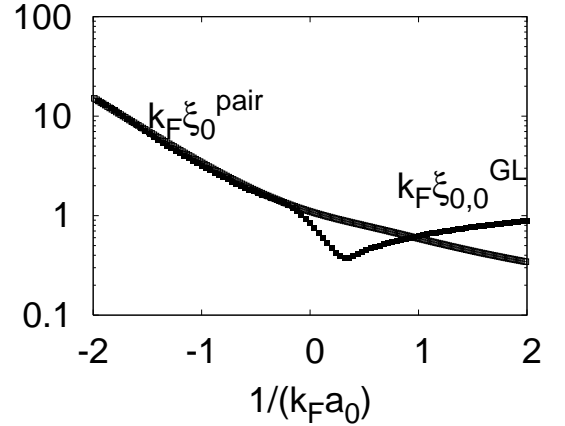}}}
\caption{\label{fig:swave.pairsize} Plots of GL coherence length
$k_{\rm F} \xi_{0,0}^{\rm{GL}}$ (solid squares), and zero temperature Cooper pair size
$k_{\rm F} \xi_{0,0}^{\rm{pair}}$ (hollow squares) versus interaction strength $1/(k_{\rm F} a_0)$ at 
$T = T_{{\rm c},0}$ for $k_0 \approx 200k_{\rm F}$.
}
\end{figure}

In Fig.~{\ref{fig:swave.pairsize}}, a comparison between $(\xi_{0,0}^{{\rm GL}})_i$
and $\xi_{0}^{{\rm pair}}$ is shown for $s$-wave ($\ell = 0, m_\ell = 0$).
$\xi_{0}^{\rm pair}$ changes from 
$
k_{\rm F} \xi_0^{\rm pair} = [e^{\gamma}/\sqrt{2}\pi] (\epsilon_{\rm F}/T_{c,0})
$
in the BCS limit to
$
k_{\rm F} \xi_0^{\rm pair} = [\epsilon_{\rm F}/(2|\mu_0|)]^{1/2} = k_{\rm F}a_0/\sqrt{2}
$
in the BEC limit as the interaction increases.
Here $\gamma \approx 0.577$ is the Euler's constant.
Furthermore, when $\mu_0 = 0$, we obtain 
$
k_{\rm F} \xi_0^{\rm pair} = \sqrt{7}[\Gamma^2(1/4)/\sqrt{\pi}]^{1/3}/4 \approx 1.29,
$
where $\Gamma(x)$ is the Gamma function.
Notice that, $\xi_{0}^{\rm pair}$ is continuous at $\mu_0 = 0$, 
and monotonically decreasing function of $1/(k_{\rm F}a_0)$ with a limiting 
value controlled by $a_0$ in strong coupling.
However, $(\xi_{0,0}^{\rm GL})_i$ is a non-monotonic function of $1/(k_{\rm F}a_0)$
having a minimum around $1/(k_{\rm F}a_0) \approx 0.32$ ($\mu_0 = 0$). 
It changes from
$
k_{\rm F}(\xi^{\rm GL}_{0,0})_i = [7\zeta(3)/(12\pi^2)]^{1/2} (\epsilon_{\rm F}/T_{{\rm c},0})
$
in the BCS to
$
k_{\rm F}(\xi^{\rm GL}_{0,0})_i = [\pi/(8k_{\rm F}a_0)]^{1/2}
$
in the BEC limit as the coupling increases,
where $\zeta(x)$ is the Zeta function.
Notice that, $(\xi_{0,0}^{\rm GL})_i$ grows as $1/\sqrt{k_{\rm F}a_0}$ 
in strong coupling limit.

\begin{figure} [htb]
\centerline{\scalebox{0.67}{\includegraphics{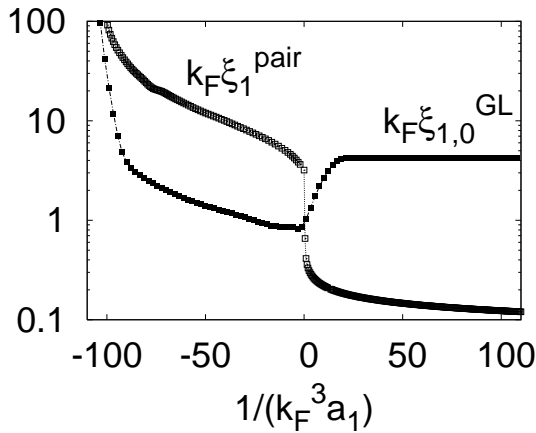}}}
\caption{\label{fig:pwave.pairsize} Plots of GL coherence length
$k_{\rm F} \xi_{1,0}^{\rm{GL}}$ (solid squares), and zero temperature Cooper pair size
$k_{\rm F} \xi_{1,0}^{\rm{pair}}$ (hollow squares) versus interaction strength $1/(k_{\rm F}^3 a_1)$ at 
$T = T_{{\rm c},1}$ for $k_0 \approx 200k_{\rm F}$.
}
\end{figure}

In Fig.~{\ref{fig:pwave.pairsize}}, a comparison between $(\xi_{1,0}^{{\rm GL}})_z$
and $\xi_{1}^{{\rm pair}}$ is shown for $p$-wave ($\ell = 1, m_\ell = 0$).
Notice that, $\xi_{1}^{\rm pair}$ is nonanalytic at $\mu_1 = 0$, 
and is a monotonically decreasing function of $1/(k_{\rm F}^3 a_1)$ with a limiting 
value controlled by $k_{\rm F}/k_0$ in strong coupling.
This nonanalytic behavior is associated with the
change in $E_1(\mathbf{k})$ from gapless (with line nodes) 
in the BCS to fully gapped in the BEC side.
However, $(\xi_{1,0}^{\rm GL})_z$ is a non-monotonic function of $1/(k_{\rm F}^3 a_1)$
having a minimum around $1/(k_{\rm F}^3 a_1) \approx 0.02$ ($\mu_1 = 0$).
It changes from 
$
k_{\rm F}(\xi^{\rm GL}_{1,0})_x = k_{\rm F}(\xi^{\rm GL}_{1,0})_y = 
k_{\rm F}(\xi^{\rm GL}_{1,0})_z/\sqrt{3} = [7\zeta(3)/(20\pi^2)]^{1/2} (\epsilon_{\rm F}/T_{{\rm c},1})
$
in the BCS to 
$
k_{\rm F}(\xi^{\rm GL}_{1,0})_i = [\pi k_0/(36k_{\rm F})]^{1/2}
$ 
in the BEC limit as the coupling increases.
Notice that, $\xi_{1,0}^{\rm GL}$ saturates in strong coupling limit 
reflecting the finite range of interactions.

It is important to emphasize that $(\xi_{\ell, m_\ell}^{\rm GL})_z$ shown in 
Figs.~(\ref{fig:swave.pairsize}) and (\ref{fig:pwave.pairsize}) is only qualitative in the 
intermediate regime around unitarity $1/(k_{\rm F}^{2\ell + 1} a_\ell) = 0$
since our theory is not strictly applicable in that region.

\section{Conclusions}
\label{sec:conclusions}

In this manuscript, we extended the $s$-wave ($\ell = 0$) functional
integral formalism to finite angular momentum $\ell$ including two 
hyperfine states (THS) pseudo-spin singlet
and single hyperfine states (SHS) pseudo-spin triplet channels.
We analyzed analytically superfluid properties of a dilute Fermi gas in the ground state ($T = 0$) 
and near critical temperatures ($T \approx T_{{\rm c},\ell}$) from weak coupling (BCS) to strong coupling
(BEC) as a function of scattering parameter ($a_\ell$) for arbitrary $\ell$.
However, we presented numerical results only for THS $s$-wave and SHS $p$-wave
symmetries which may be relevant for current experiments involving atomic Fermi gases.
The main results of our paper are as follows.

First, we analyzed the low energy scattering amplitude within a T-matrix approach.
We found that bound states occur only when $a_\ell > 0$ for any $\ell$.
The energy of the bound states $E_{{\rm b},\ell}$ involves only 
the scattering parameter $a_0$ for $\ell = 0$.
However, another parameter related to the interaction range $1/k_0$
is necessary to characterize $E_{{\rm b},\ell}$ for $\ell \ne 0$.
Therefore, all superfluid properties for $\ell \ne 0$ depend strongly on $k_0$
and $a_\ell$, while for $\ell = 0$ they depend strongly only on $a_0$ but weakly on $k_0$.

Second, we discussed the order parameter, chemical potential,
quasiparticle excitations, momentum distribution, atomic compressibility, 
ground state energy, collective modes and average Cooper pair size at $T = 0$.
There we showed that the evolution from BCS to BEC is just a crossover for $\ell = 0$, 
while the same evolution for $\ell \ne 0$ exhibits a quantum phase transition
characterized by a gapless superfluid on the BCS side to a fully gapped superfluid on the BEC side. 
This transition is a many body effect and takes place exactly when chemical
potential $\mu_{\ell \ne 0}$ crosses the bottom of the fermion band ($\mu_{\ell \ne 0} = 0$), 
and is best reflected as non-analytic behavior in the ground state atomic compressibility, 
momentum distribution and average Cooper pair size.

Third, we discussed the critical temperature, chemical potential,
and the number of unbound, scattering and bound fermions at $T = T_{{\rm c},\ell}$.
We found that the critical BEC temperature is the highest for $\ell = 0$.
We also derived the time-dependent Ginzburg-Landau functional (TDGL) near $T_{{\rm c},\ell}$
and extracted the Ginzburg-Landau (GL) coherence length and time.
We recovered the usual TDGL equation for BCS superfluids in the weak coupling limit,
whereas in the strong coupling limit we recovered the Gross-Pitaevskii (GP) equation
for a weakly interacting dilute Bose gas.
The TDGL equation exhibits anisotropic coherence lengths for $\ell \ne 0$ which
become isotropic only in the BEC limit, in sharp contrast to the $\ell = 0$ case,
where the coherence length is isotropic for all couplings.
Furthermore, the GL time is a complex number with a larger imaginary component 
for $\mu_\ell > 0$ reflecting the decay of Cooper pairs into the two particle
continuum. However, for $\mu_\ell < 0$ the imaginary component vanishes and Cooper pairs
become stable above $T_{{\rm c},\ell}$.

In summary, the BCS to BEC evolution in higher angular momentum ($\ell \ne 0$) states 
exhibit quantum phase transitions and is much richer than in conventional $\ell = 0$ 
$s$-wave systems, where there is only a crossover. 
These $\ell \ne 0$ states might be found not only in atomic Fermi gases, 
but also in nuclear (pairing in nuclei), astrophysics (neutron stars) 
and condensed matter (high-$T_c$ and organic superconductors) systems.

\section{Acknowledgement}
\label{acknowledgement}

We would like to thank NSF (DMR-0304380) for financial support.

\section{APPENDIX}
\label{sec:appendix}
\subsection{Expansion Coefficients at $T = 0$}
\label{sec:appa}

From the rotated fluctuation matrix $\widetilde{\mathbf{F}}_{\ell}^{-1}(q)$
expressed in the amplitude-phase basis, we can obtain the expansion
coefficients necessary to calculate the collective modes
at $T = 0$. In the long wavelength $(|\mathbf{q}| \to 0)$, and low
frequency limit $(w \to 0)$ the condition
\begin{equation}
\{ w ,\frac{q_i q_j}{2M} \} \ll \min\{2E_\ell(\mathbf{k})\},
\end{equation}
is used. 
While there is no Landau damping and a well defined expansion is 
possible for $\ell = 0$ case for all couplings,
extra care is necessary for $\ell \ne 0$ when $\mu_\ell > 0$
since Landau damping is present.

In all the expressions below we use the following simplifying notation 
$
\dot{\xi}_\ell^{i} = \partial\xi_\ell (\mathbf{k} + \mathbf{q}/2)/\partial q_i,
$
$
\ddot{\xi}_\ell^{i,j} = \partial \xi_\ell (\mathbf{k} + \mathbf{q}/2)/ (\partial q_i \partial q_j),
$
$
\dot{\Delta}_\ell^{i} = \partial \Delta_\ell(\mathbf{k} + \mathbf{q}/2)/\partial q_i
$
and
$
\ddot{\Delta}_\ell^{i,j} = \partial^2 \Delta_\ell(\mathbf{k} + \mathbf{q}/2)/(\partial q_i \partial q_j),
$
which are evaluated at $\mathbf{q} = 0$.
 
The coefficients necessary to obtain the matrix element 
$(\widetilde{\mathbf{F}}_{\ell,m_\ell,m_\ell'}^{-1})_{11}$ are
\begin{eqnarray}
A_{\ell,m_\ell,m_\ell'}  &=& \frac{\delta_{m_\ell,m_\ell'}}{4\pi {\cal V}^{-1} \lambda_\ell} \nonumber \\
&-& \sum_{\mathbf{k}}\frac{\xi_\ell^2} {2E_\ell^3} 
\Gamma_\ell^2(k) Y_{\ell,m_\ell}(\widehat{\mathbf{k}}) Y_{\ell,m_\ell'}^*(\widehat{\mathbf{k}}), 
\end{eqnarray}
corresponding to the $(\mathbf{q} = 0, w = 0)$ term,
\begin{eqnarray}
C_{\ell,m_\ell,m_\ell'}^{i,j} & = & \sum_{\mathbf{k}}\frac{\xi_\ell(\mathbf{k})}{4E_\ell^7(\mathbf{k})}
\big\lbrace
\ddot{\xi}_\ell^{i,j} E_\ell^2(\mathbf{k}) 
\left[\xi_\ell^2(\mathbf{k}) - 2\Delta_\ell^2(\mathbf{k})\right] \nonumber \\
&+& 3\ddot{\Delta}_\ell^{i,j} E_\ell^2(\mathbf{k}) \xi_\ell(\mathbf{k}) \Delta_\ell(\mathbf{k})  
+ 5\dot{\xi}_\ell^i \dot{\xi}_\ell^j \Delta_\ell^2(\mathbf{k}) \xi_\ell^2(\mathbf{k}) \nonumber \\
&+& \dot{\Delta}_\ell^i \dot{\Delta}_\ell^j \xi_\ell(\mathbf{k}) 
\left[\xi_\ell^2(\mathbf{k}) - 4\Delta_\ell^2(\mathbf{k})\right] \nonumber \\
&+& (\dot{\xi}_\ell^i \dot{\Delta}_\ell^j + \dot{\xi}_\ell^j \dot{\Delta}_\ell^i) 
\Delta_\ell(\mathbf{k}) 
\left[2\Delta_\ell^2(\mathbf{k}) - 3\xi_\ell^2(\mathbf{k})\right]
\big\rbrace \nonumber \\
&& \Gamma_\ell^2(k) Y_{\ell,m_\ell}(\widehat{\mathbf{k}}) Y_{\ell,m_\ell'}^*(\widehat{\mathbf{k}}),
\end{eqnarray}
corresponding to the $q_i q_j$ term, and
\begin{equation}
D_{\ell,m_\ell,m_\ell'} = \sum_{\mathbf{k}}\frac{\xi_\ell^2(\mathbf{k})}{8E_\ell^5(\mathbf{k})}
\Gamma_\ell^2(k) Y_{\ell,m_\ell}(\widehat{\mathbf{k}}) Y_{\ell,m_\ell'}^*(\widehat{\mathbf{k}}), 
\end{equation}
corresponding to the $w^2$ term.
Here $\delta_{m_\ell,m_\ell'}$ is the Kronecker delta.

The coefficients necessary to obtain the matrix element
$(\widetilde{\mathbf{F}}_{\ell,m_\ell,m_\ell'}^{-1})_{22}$ are
\begin{eqnarray}
P_{\ell,m_\ell,m_\ell'} &=& \frac{\delta_{m_\ell,m_\ell'}}{4\pi{\cal V}^{-1}\lambda_\ell} \nonumber \\
&-& \sum_{\mathbf{k}}\frac{1}{2E_\ell(\mathbf{k})}
\Gamma_\ell^2(k) Y_{\ell,m_\ell}(\widehat{\mathbf{k}}) Y_{\ell,m_\ell'}^*(\widehat{\mathbf{k}}),
\end{eqnarray}
corresponding to the $(\mathbf{q} = 0, w = 0)$ term,
\begin{eqnarray}
Q_{\ell,m_\ell,m_\ell'}^{i,j} & = & \sum_{\mathbf{k}}\frac{1}{4E_\ell^5(\mathbf{k})}
\big\lbrace
\ddot{\xi}_\ell^{i,j} E_\ell^2(\mathbf{k}) \xi_\ell(\mathbf{k}) \nonumber \\
&+& \ddot{\Delta}_\ell^{i,j} E_\ell^2(\mathbf{k}) \Delta_\ell(\mathbf{k}) \nonumber \\
&+& 3\dot{\xi}_\ell^i \dot{\xi}_\ell^j \Delta_\ell^2(\mathbf{k})
+ 3\dot{\Delta}_\ell^i \dot{\Delta}_\ell^j \xi_\ell^2(\mathbf{k}) \nonumber \\
&-& 3(\dot{\xi}_\ell^i \dot{\Delta}_\ell^j + \dot{\xi}_\ell^j \dot{\Delta}_\ell^i) \xi_\ell(\mathbf{k}) \Delta_\ell(\mathbf{k}) 
\big\rbrace \nonumber \\
&& \Gamma_\ell^2(k) Y_{\ell,m_\ell}(\widehat{\mathbf{k}})
Y_{\ell,m_\ell'}^*(\widehat{\mathbf{k}}),
\end{eqnarray}
corresponding to the $q_i q_j$ term, and
\begin{equation}
R_{\ell,m_\ell,m_\ell'} = \sum_{\mathbf{k}}\frac{1}{8E_\ell^3(\mathbf{k})}
\Gamma_\ell^2(k) Y_{\ell,m_\ell}(\widehat{\mathbf{k}}) Y_{\ell,m_\ell'}^*(\widehat{\mathbf{k}}), 
\end{equation}
corresponding to the $w^2$ term.

The coefficients necessary to obtain the matrix element 
$(\widetilde{\mathbf{F}}_{\ell,m_\ell,m_\ell'}^{-1})_{12}$ is
\begin{equation}
B_{\ell,m_\ell,m_\ell'} = \sum_{\mathbf{k}}\frac{\xi_\ell(\mathbf{k})}{4E_\ell^3(\mathbf{k})}
\Gamma_\ell^2(k) Y_{\ell,m_\ell}(\widehat{\mathbf{k}}) Y_{\ell,m_\ell'}^*(\widehat{\mathbf{k}}),  
\end{equation}
corresponding to the $w$ term.

\subsection{Expansion Coefficients at $T = T_{{\rm c}, \ell}$}
\label{sec:appb}

In this section, we perform a small $\mathbf{q}$ and $iv_j \to w + i0^+$ expansion near $T_{{\rm c},\ell}$,
where we assumed that the fluctuation field $\Lambda_{\ell,m_\ell}(\mathbf{x}, t)$ 
is a slowly varying function of $\mathbf{x}$ and $t$.

The zeroth order coefficient $L^{-1}_{\ell,m_\ell,m'_\ell}(0,0)$ is diagonal in
$m_\ell$ and $m'_\ell$, and is given by
\begin{equation}
a_{\ell,m_\ell,m'_\ell} = \frac{\delta_{m_\ell,m'_\ell}}{4\pi}
\left[ \frac{{\cal V}}{\lambda_\ell} - \sum_{\mathbf{k}} 
\frac{{\cal X}_\ell(\mathbf{k})}{2\xi_\ell(\mathbf{k})} \Gamma_\ell^2(k) \right],
\end{equation}
where ${\cal X}_\ell(\mathbf{k}) = \tanh{\left[\beta\xi_\ell(\mathbf{k})/2\right]}$.
The second order coefficient $M\partial^2L^{-1}_{\ell,m_\ell,m'_\ell}(\mathbf{q},0)/(\partial q_i \partial q_j)$
evaluated at $\mathbf{q} = 0$ is given by 
\begin{eqnarray}
c_{\ell,m_\ell,m'_\ell}^{i,j} &=& \frac{1}{4\pi} \sum_{\mathbf{k}}\big\lbrace 
\left[\frac{{\cal X}_\ell(\mathbf{k})}{8\xi_\ell^2(\mathbf{k})} 
- \frac{\beta {\cal Y}_\ell(\mathbf{k})}{16\xi_\ell(\mathbf{k})} \right]\delta_{m_\ell,m'_\ell} \delta_{i,j} \nonumber \\
&+& \alpha_{\ell,m_\ell,m'_\ell}^{i,j}
\frac{\beta^2 k^2 {\cal X}_\ell(\mathbf{k}) {\cal Y}_\ell(\mathbf{k})}{16M \xi_\ell(\mathbf{k})} \big\rbrace\Gamma_\ell^2(k),
\end{eqnarray}
where ${\cal Y}_\ell(\mathbf{k}) = \rm{sech}^2[\beta\xi_\ell(\mathbf{k})/2]$ and the angular average
\begin{equation}
\alpha_{\ell,m_\ell,m'_\ell}^{i,j} 
= \int d\widehat{\mathbf{k}} \widehat{k}_i \widehat{k}_j 
Y_{\ell,m_\ell}(\widehat{\mathbf{k}}) Y_{\ell,m'_\ell}^*(\widehat{\mathbf{k}}).
\end{equation}
Here, $d\widehat{\mathbf{k}} = \sin(\theta_\mathbf{k}) d\theta_\mathbf{k} d\phi_\mathbf{k}$,
$\widehat{k}_x = \sin(\theta_\mathbf{k})\cos(\phi_{\mathbf{k}})$,
$\widehat{k}_y = \sin(\theta_\mathbf{k})\sin(\phi_{\mathbf{k}})$ and
$\widehat{k}_z = \cos(\theta_\mathbf{k})$.
In general, $\alpha_{\ell,m_\ell,m'_\ell}^{i,j}$ is a fourth order tensor for fixed $\ell$.
However, in the particular case where only one of the spherical harmonics 
$Y_{\ell,m_\ell}(\widehat{\mathbf{k}})$
is dominant and characterizes the order parameter, 
$\alpha_{\ell,m_\ell,m_\ell'}^{i,j} = \alpha_{\ell,m_\ell,m_\ell}^{i,j} \delta_{m_\ell,m_\ell'}$
is diagonal in $m_\ell$ and $m_\ell'$. 
In this case, we use Gaunt coefficients~\cite{arfken} to show that
$\alpha_{\ell,m_\ell,m_\ell}^{i,j}$ is also diagonal in $i$ and $j$ leading to
$\alpha_{\ell,m_\ell,m_\ell'}^{i,j} = \alpha_{\ell,m_\ell,m_\ell}^{i,i} \delta_{m_\ell,m_\ell'} \delta_{i,j}$.

The coefficient of fourth order term is approximated at 
$q_n = 0$, and given by
\begin{eqnarray}
b_{\ell,\{m_{\ell_n}\}}(0) &=& \frac{\gamma_{\ell,\{m_{\ell_n}\}}}{4\pi} \nonumber \\ 
&& \sum_{\mathbf{k}} \left[ \frac{{\cal X}_\ell(\mathbf{k})}{4\xi_\ell^3(\mathbf{k})} 
- \frac{\beta {\cal Y}_\ell(\mathbf{k})}{8\xi_\ell^2(\mathbf{k})} \right]
\Gamma_\ell^4(k),
\end{eqnarray}
where the angular average
\begin{eqnarray}
\gamma_{\ell,\{m_{\ell_n}\}} &=& \int d\widehat{\mathbf{k}}
Y_{\ell,m_{\ell_1}}(\widehat{\mathbf{k}}) Y_{\ell,m_{\ell_2}}^*(\widehat{\mathbf{k}}) \nonumber \\
&& \hskip 1 cm Y_{\ell,m_{\ell_3}}(\widehat{\mathbf{k}}) Y_{\ell,m_{\ell_4}}^*(\widehat{\mathbf{k}}).
\end{eqnarray}

To extract the time-dependence, we expand
$Q_{\ell,m_\ell,m'_\ell}(iv_j) = L^{-1}_{\ell.m_\ell,m'_\ell}(\mathbf{q} = 0, iv_j) 
- L^{-1}_{\ell,m_\ell,m'_\ell}(0,0)$ in powers of $w$
after the analytic continuation $iv_j \to w + i0^+$.
We use the relation $(x \pm i0^+)^{-1} = {\cal P}(1/x) \mp i\pi \delta(x)$,
where ${\cal P}$ is the principal value and $\delta(x)$ is the Delta function,
to obtain
\begin{eqnarray}
Q_{\ell,m_\ell,m'_\ell}(iv_j) = 
- \frac{\delta_{m_\ell,m'_\ell}}{4\pi} 
\left[ \sum_{\mathbf{k}} \frac{{\cal X}_\ell(\mathbf{k})}{4\xi_\ell^2(\mathbf{k})}\Gamma_\ell^2(k) \right. \nonumber \\
 \left. - i\pi \sum_{\mathbf{k}} {\cal X}_\ell(\mathbf{k}) \delta[2\xi_\ell(\mathbf{k}) - w] \Gamma_\ell^2(k) 
\right]
\end{eqnarray}
Keeping only the first order terms in $w$ leads to 
$Q_{\ell,m_\ell,m'_\ell}(w + i0^+) = - d_{\ell,m_\ell,m'_\ell} w + ...$, where 
\begin{eqnarray}
d_{\ell,m_\ell,m'_\ell} &=& \frac{\delta_{m_\ell,m'_\ell}}{4\pi} 
\left[ \sum_{\mathbf{k}} \frac{{\cal X}_\ell(\mathbf{k})}{4\xi_\ell^2(\mathbf{k})}\Gamma_\ell^2(k) \right. \nonumber \\
&+& \left. i\frac{\pi\beta}{8}N(\epsilon_{\rm F}) \sqrt{\frac{\mu_\ell}{\epsilon_{\rm F}}} 
\Gamma_\ell^2(\mu_\ell) \Theta(\mu_\ell)
\right]
\end{eqnarray}
is also diagonal in $m_\ell$ and $m'_\ell$.
Here 
$
N(\epsilon_{\rm F}) = M{\cal V}k_{\rm F}/(2\pi^2)
$
is the density of states per spin at the Fermi energy,
$
\Gamma_\ell^2(x) = (\epsilon_0 x^\ell)/(\epsilon_0 + x)^{\ell + 1}
$
is the interaction symmetry in terms of energy 
and $\Theta(x)$ is the Heaviside function.


\begin{thebibliography}{99}
\bibitem{regal2} C. A. Regal, M. Greiner, and D. S. Jin, Phys. Rev. Lett. \textbf{92}, 040403 (2004).
\bibitem{greiner} M. Greiner, C. A. Regal, and D. S. Jin, Nature (London) \textbf{426}, 537 (2003).
\bibitem{hulet} K. E. Strecker, G. B. Partridge, and R. G. Hulet, Phys. Rev. Lett. \textbf{91}, 080406 (2003).
\bibitem{litium1} M. W. Zwierlein, C. A. Stan, C. H. Schunck, S. M. F. Raupach, S. Gupta, Z. Hadzibabic, and W. Ketterle, Phys. Rev. Lett. \textbf{91}, 250401 (2003).
\bibitem{litium2} T. Bourdel, L. Khaykovich, J. Cubizolles, J. Zhang, F. Chevy, M. Teichmann, L. Tarruell, S. J. J. M. F. Kokkelmans, C. Salomon, Phys. Rev. Lett. \textbf{93}, 050401 (2004).
\bibitem{litium3} M. Bartenstein, A. Altmeyer, S. Riedl, S. Jochim, C. Chin, J. Hecker Denschlag, and R. Grimm, Phys. Rev. Lett. \textbf{92}, 120401 (2004).
\bibitem{kinast} J. Kinast, S. L. Hemmer, M. E. Gehm, A. Turlapov and J. E. Thomas, Phys. Rev. Lett. \textbf{92}, 150402 (2004).
\bibitem{mitvortex} M. W. Zwierlein, J. R. Abo-Shaeer, A. Schirotzek, C. H. Schunck, and W. Ketterle, Nature \textbf{435}, 1047 (2005).
\bibitem{eagles} D. M. Eagles, Phys. Rev. \textbf{186}, 456 (1969).
\bibitem{leggett} A. J. Leggett, in \textit{Modern Trends in the Theory of Condensed Matter}, edited by A. Peralski and R. Przystawa (Springer-Verlag, Berlin 1980).
\bibitem{nsr} P. Nozieres and S. Schmitt-Rink, J. Low. Temp. Phys. \textbf{59}, 195 (1985).
\bibitem{carlos} C. A. R. S{\'a} de Melo, M. Randeria, and J. R. Engelbrecht, Phys. Rev. Lett. \textbf{71}, 3202 (1993).
\bibitem{jan} J. R. Engelbrecht, M. Randeria, and C. A. R. S{\'a} de Melo, Phys. Rev. B \textbf{55}, 15153 (1997).
\bibitem{perali} A. Perali, P. Pieri, L. Pisani, and G. C. Strinati, Phys. Rev. Lett. \textbf{92}, 220404 (2004).
\bibitem{griffin-trap} Y. Ohashi and A. Griffin, Phys. Rev. A \textbf{67}, 033603 (2003). 
\bibitem{holland} M. Holland, S. J. J. M. F. Kokkelmans, M. L. Chiofalo, and R. Walser, Phys. Rev. Lett. \textbf{87}, 120406 (2001). 
\bibitem{timmermans} E. Timmermans, K. Furuya, P. W. Milonni, and A. K. Kerman, Phys. Lett. A \textbf{285}, 228 (2001)
\bibitem{griffin-fb} Y. Ohashi and A. Griffin, Phys. Rev. Lett. \textbf{89} , 130402 (2002).
\bibitem{regal3} C. A. Regal, C. Ticknor, J. L. Bohn, and D. S. Jin, Phys. Rev. Lett. \textbf{90}, 053201 (2003).
\bibitem{ticknor} C. Ticknor, C. A. Regal, D. S. Jin, and J. L. Bohn, Phys. Rev. A \textbf{69}, 042712 (2004)
\bibitem{zhang} J. Zhang, E. G. M. van Kempen, T. Bourdel, L. Khaykovich, J. Cubizolles, F. Chevy, M. Teichmann, L. Tarruell, S. J. J. M. F. Kokkelmans, and C. Salomon, Phys. Rev. A \textbf{70}, 030702 (2004).
\bibitem{schunck} C. H. Schunck, M. W. Zwierlein, C. A. Stan, S. M. F. Raupach, W. Ketterle, A. Simoni, E. Tiesinga, C. J. Williams, and P. S. Julienne, Phys. Rev. A \textbf{71}, 045601 (2005).
\bibitem{gunter} K. G\"unter, T. St\"oferle, H. Moritz, M. K\"ohl, and T. Esslinger, Phys. Rev. Lett. \textbf{95}, 230401 (2005). 
\bibitem{john} J. L. Bohn, Phys. Rev. A \textbf{61}, 053409 (2000).
\bibitem{gurarie}  V. Gurarie, L. Radzihovsky, and A. V. Andreev, Phys. Rev. Lett. \textbf{94}, 230403 (2005).
\bibitem{skyip} Chi-Ho Cheng and S.-K. Yip, Phys. Rev. Lett. \textbf{95}, 070404 (2005).
\bibitem{borkowski} L. S. Borkowski and C. A. R. S{\'a} de Melo, cond-mat/9810370.
\bibitem{duncan} R. D. Duncan and C. A. R. S{\'a} de Melo, Phys. Rev B \textbf{62}, 9675 (2000).
\bibitem{annett} Jorge Quintanilla, Balazs L. Gyorffy, James F. Annett, and Jonathan P. Wallington, Phys. Rev. B \textbf{66}, 214526 (2002).
\bibitem{hertog} B. C. den Hertog, Phys. Reb. B \textbf{60}, 559 (1999).
\bibitem{andrenacci} N. Andrenacci, A. Perali, P. Pieri, and G. C. Strinati, Phys. Rev. B \textbf{60}, 12410 (1999).
\bibitem{tlho} Tin-Lun Ho and Roberto B. Diener,  Phys. Rev. Lett. \textbf{94}, 090402 (2005).
\bibitem{botelho1}  S. S. Botelho and C. A. R. S{\'a} de Melo, cond-mat/0409368.
\bibitem{iskin-lattice} M. Iskin and C. A. R. S{\'a} de Melo, Phys. Rev. B \textbf{72}, 224513 (2005).
\bibitem{botelho-pwave} S. S. Botelho and C. A. R. S{\'a} de Melo, J. Low Temp. Phys. \textbf{140}, 409 (2005).
\bibitem{ohashi} Y. Ohashi, Phys. Rev. Lett. \textbf{94}, 050403 (2005).
\bibitem{iskinprl} M. Iskin and C.A.R. S{\'a} de Melo, Phys. Rev. Lett. \textbf{96}, 040402 (2006).
\bibitem{volovik} G. E. Volovik, cond-mat/0601372.
\bibitem{landau} L. D. Landau and E. M. Lifshitz, {\it Quantum Mechanics}, Permagon, Oxford (1994). 
\bibitem{leggett-review} A. J. Leggett, Rev. Mod. Phys. 47, 331 (1975).
\bibitem{pieri} P. Pieri, L. Pisani and G. C. Strinati, Phys. Rev. B. \textbf{70}, 094508 (2004).
\bibitem{jin-correlations} M. Greiner, C. A. Regal, J. T. Stewart, and D. S. Jin, Phys. Rev. Lett. \textbf{94}, 110401 (2005). 
\bibitem{ealtman} E. Altman, E. Demler, and M. Lukin, Phys. Rev. A \textbf{70}, 013603 (2004).
\bibitem{reif} F. Reif, \textit{Fundamentals of Statistical and Thermal Physics}, Chap. 8, McGraw-Hill, Tokyo (1981).
\bibitem{volovik-book} G. E. Volovik, {\it Exotic Properties of Superfluid $^3{\rm He}$}, World Scientific, Singapore (1992).
\bibitem{lifshitz} I. M. Lifshitz, Sov. Phys. JETP \textbf{11}, 1130 (1960). [Zh. Eksp. Teor. Fiz. \textbf{38}, 1569 (1960)].
\bibitem{abrikosov} A. A. Abrikosov, {\it Fundamental Theory of Metals}, pages 111-114, North-Holland  (1988).
\bibitem{pieri-a} P. Pieri and G. C. Strinati, Phys. Rev. B \textbf{61}, 15370 (2000).
\bibitem{gora} D. S. Petrov, C. Salomon, and G. V. Shlyapnikov, Phys. Rev. Lett. \textbf{93}, 090404 (2004).
\bibitem{kagan} I. V. Brodsky, A. V. Klaptsov, M. Yu. Kagan, R. Combescot, X. Leyronas, cond-mat/0507240.
\bibitem{gurarie-a} J. Levinsen, V. Gurarie, cond-mat/0510672.
\bibitem{giorgini} G. E. Astrakharchik, J. Boronat, J. Casulleras, and S. Giorgini, Phys. Rev. Lett. \textbf{93}, 200404 (2004).
\bibitem{arfken} G. B. Arfken and H. J. Weber, {\it Mathematical methods for physicists}, Harcourt Academic Press, New York (2001).
\end{thebibliography}
\end{document}